\documentclass[pra,aps,longbibliography,showpacs,groupedaddress,superscriptaddress,twocolumn,toc=flat,nofootinbib]{revtex4-1}

\usepackage{appendix}
\usepackage{graphicx}
\usepackage{latexsym}
\usepackage{amssymb}
\usepackage{amsmath}
\usepackage{amsfonts}
\usepackage{upgreek}
\usepackage{subfigure}
\usepackage{bm}
\usepackage{bbold}
\usepackage{verbatim}
\usepackage[unicode=true,
 bookmarks=false,
 breaklinks=false,pdfborder={0 0 1},backref=false,colorlinks=true]
 {hyperref}
\hypersetup{
 linkcolor=[rgb]{0,0,1},citecolor=[rgb]{0,0,1},urlcolor=[rgb]{0,0,1}}
\usepackage{multirow}
\usepackage{color}
\usepackage{comment}
\usepackage{xcolor}
\usepackage{soul}
\usepackage{tikz}
\usepackage{tabularx}
\usepackage{comment}
\usepackage{siunitx}
\usepackage{enumerate}

\usepackage{braket}

\begin{document}

\title{Quantum simulation of fermionic non-Abelian lattice gauge theories \\ in $(2+1)$D with built-in gauge protection}

\author{Gaia De Paciani}
\email{gaia.paciani@lmu.de}
\affiliation{Department of Physics and Arnold Sommerfeld Center for Theoretical Physics (ASC), Ludwig-Maximilians-Universit\"at M\"unchen, Theresienstr. 37, M\"unchen D-80333, Germany}
\affiliation{Munich Center for Quantum Science and Technology (MCQST), Schellingstr. 4, M\"unchen D-80799, Germany}
\author{Lukas Homeier}
\affiliation{JILA and Department of Physics, University of Colorado, Boulder, CO, 80309, USA}
\affiliation{Center for Theory of Quantum Matter, University of Colorado, Boulder, CO, 80309, USA}
\affiliation{Department of Physics and Arnold Sommerfeld Center for Theoretical Physics (ASC), Ludwig-Maximilians-Universit\"at M\"unchen, Theresienstr. 37, M\"unchen D-80333, Germany}
\affiliation{Munich Center for Quantum Science and Technology (MCQST), Schellingstr. 4, M\"unchen D-80799, Germany}
\author{Jad C. Halimeh}
\affiliation{Max Planck Institute of Quantum Optics, Garching, Germany}
\affiliation{Department of Physics and Arnold Sommerfeld Center for Theoretical Physics (ASC), Ludwig-Maximilians-Universit\"at M\"unchen, Theresienstr. 37, M\"unchen D-80333, Germany}
\affiliation{Munich Center for Quantum Science and Technology (MCQST), Schellingstr. 4, M\"unchen D-80799, Germany}
\author{Monika Aidelsburger}
\affiliation{Max Planck Institute of Quantum Optics, Garching, Germany}
\affiliation{Department of Physics and Arnold Sommerfeld Center for Theoretical Physics (ASC), Ludwig-Maximilians-Universit\"at M\"unchen, Theresienstr. 37, M\"unchen D-80333, Germany}
\affiliation{Munich Center for Quantum Science and Technology (MCQST), Schellingstr. 4, M\"unchen D-80799, Germany}
\author{Fabian Grusdt}
\affiliation{Department of Physics and Arnold Sommerfeld Center for Theoretical Physics (ASC), Ludwig-Maximilians-Universit\"at M\"unchen, Theresienstr. 37, M\"unchen D-80333, Germany}
\affiliation{Munich Center for Quantum Science and Technology (MCQST), Schellingstr. 4, M\"unchen D-80799, Germany}

\date{\today}
\begin{abstract} 
Recent advancements in the field of quantum simulation have significantly expanded the potential for applications, particularly in the context of lattice gauge theories (LGTs). Maintaining gauge invariance throughout a simulation remains a central challenge, especially for large-scale non-Abelian LGTs with dynamical matter, which are particularly complex in terms of engineering for experiments. Gauge-symmetry breaking is inevitable in established rishon-based schemes for alkaline-earth-like atoms (AELAs) and controlling the magnitude of its effect is an open challenge. Here, we first construct a minimal model to quantum simulate non-Abelian LGTs ensuring that the gauge constraints are met and explicitly derive their unambiguous non-Abelian nature. Second, we present a proposal for a novel gauge protection scheme using native interactions in AELAs enabling the simulation of toy models of non-Abelian $U(2)$ LGTs with dynamical fermionic matter in $(2+1)$ dimensions on large scales. 
Due to the simplicity of the gauge protection mechanism, based on a Zeeman shift in combination with superexchange interactions, our scheme can be naturally included in other rishon-based quantum simulation protocols. Third, we extend our approach to a fully scalable, hybrid digital-analog simulator for $U(N)$ LGTs based on Rydberg AELA with variable rishon number. The proposed general mechanism for gauge protection provides a promising path towards the long-awaited simulation of non-Abelian LGTs relevant to particle physics.
\end{abstract}
\maketitle

\section{Introduction} \label{sec:introduction}
Over the past century, the development of gauge theories has provided a comprehensive framework for the description of strongly interacting systems, including the Standard Model of particle physics as well as emergent effective theories in condensed matter systems~\cite{Weinberg_book, Zee_book, Fradkin2013, YangMills1954}.
Despite the numerous advancements, ranging from quantum spin liquids to high-$T_C$ superconductivity and lattice quantum chromodynamics (QCD)~\cite{Savary2016, Balents2010, Senthil2000, kleinert_book, Sachdev2019, Auerbach_book}, gauge theories still pose significant challenges, particularly in addressing non-perturbative phenomena and at high (fermionic) matter density~\cite{Borla2022}. 

The structure of LGTs~\cite{Kogut1975} allows us to apply classical numerical simulations, such as Monte Carlo~\cite{Gazit2017, Creutz1983, Bender2023} or tensor network methods~\cite{Banuls2020, magnifico2024, Borla2022}, due to (i) the spatially discretized formulation and (ii) the ability to formulate models with finite-dimensional link Hilbert spaces~\cite{Wiese_review}. However, large-scale numerical studies in- and out-of-equilibrium remain challenging at strong coupling and beyond $(1+1)$D. 
The development of quantum technologies has spurred significant growth in the field of quantum simulation of LGTs, with numerous theoretical proposals and promising experimental demonstrations emerging to validate these approaches~\cite{Halimeh2025, Dalmonte_review,Zohar_review, aidelsburger2021cold, zohar2021quantum, Bauer2023, Bernien2017, gyawali2024, gonzalezcuadra2024, DiMeglio2024, Mil2020}. Today, a variety of platforms, both analog and digital, are being employed as quantum simulators, including ultracold atoms in optical lattices~\cite{Yang2020, Zhou2022, Surace2020, Wang2023, Schweizer2019, Zhang2025, Su2023, Zhu2024, Kapit2011, Buechler2005, Zohar2011}, Rydberg atoms~\cite{feldmeier2024, Weimer2010}, superconducting qubits~\cite{Marcos2013, Marcos2014, Hayata2024, Ge2022}, and trapped ions~\cite{martinez2016real, Muschik2017, Nguyen2022, Kokail2019}, including numerous Abelian~\cite{Klco2018,  Mil2020, Nguyen2022, Mildenberger2025, Wang2021, charles2023, surace2023, cochran2024, de2024} and non-Abelian proposals~\cite{Atas2021, banerjee2013, Klco2020, surace2024, halimeh2024}.

The long-standing goal of achieving large-scale quantum simulations of non-Abelian (LGTs) beyond~$(1+1)$D remains a formidable challenge, since the required local symmetries do not arise naturally but have to be engineered. One main challenge are gauge breaking terms that need to be suppressed over experimental timescales, before quantum devices can reliably and quantitatively predict outcomes in complex gauge theories such as QCD. Hence, the development of methods in toy models with non-Abelian gauge structure constitutes a significant pathway towards robust large-scale quantum simulations, allowing to study qualitative features and suitable probes of non-Abelian LGTs.
The construction of such models is generally more difficult than in the Abelian cases since multiple Gauss's laws with non-commuting generators have to be fulfilled.
Moreover, large-scale systems in dimensions greater than $(1+1)$D are often susceptible to gauge-breaking errors, which can undermine the accuracy of simulations.
This challenge has been tackled from various perspectives, leading to a range of proposed solutions, including the attempt to eliminate gauge-noninvariant subspaces~\cite{Zohar2019, Bernien2017, martinez2016real, Surace2020} or the energetic protection against gauge-breaking errors; the latter involves implementing a gauge protection mechanism, designed to constrain the simulation's dynamics to the gauge-invariant sector of the Hilbert space~\cite{halimeh2020reliability, halimeh2022gaugeprotection, halimeh2022stabilizinggaugetheoriesquantum, tagliacozzo2013, halimeh2024, banerjee2013}.

In this work, we propose a scheme for the quantum simulation of non-Abelian $U(N)$ LGTs in $(2+1)$D with dynamical, fermionic matter in a platform of ultracold atoms.
Previous approaches adopted a \textit{bottom-up} strategy, consisting in engineering microscopic processes that fulfil non-Abelian symmetries ensuring that gauge-violating terms remain sufficiently controlled~\cite{banerjee2013, tagliacozzo2013, Zohar2013coldatom}. Here, we develop a general \textit{top-down} approach to engineer gauge-invariant couplings of matter while maintaining the non-Abelian gauge symmetry of the system, enforcing a gauge-protection scheme.  In the limit of strong gauge protection, we show that the dynamics is governed by a tunable effective non-Abelian $U(N)$ LGT with timescales realistic for current experiments.

The aim of our work is to extend previous models, proposed and explored in Refs.~\cite{banerjee2013, surace2023, surace2024}, by considering explicit gauge protection mechanisms that enforce the gauge invariance of the simulation.
To this end, we propose two schemes with increasing complexity of the emergent non-Abelian LGT based on a combination of optical lattices with (i) two-body $SU(N)$-invariant Hubbard interactions in AELAs as well as (ii) periodically driven Rydberg atoms.
These schemes are scalable and gauge-invariant by construction and hence pave the way for the large-scale quantum simulation of non-Abelian LGTs with gauge-matter couplings.

The paper is structured as follows. In Sec.~\ref{sec:lattice-formalism}, we review the rishon formalism of the quantum link model we consider, specify the Gauss's laws of the theory and describe the general mechanism of our proposed scheme. In Sec.~\ref{sec:model}, we describe the Hamiltonian that allows us to protect against gauge-breaking terms and we derive the gauge-invariant effective theory. Then, in Sec.~\ref{sec:minimal-model}, we study a minimal experimentally relevant setup and demonstrate that our proposed scheme contains dynamics intrinsic to a non-Abelian gauge structure. In Sec.~\ref{sec:experimental-impl} and~\ref{sec:experimental-impl Rydberg}, we propose two experimental protocols to implement our scheme in quantum simulators based on AELAs and on a hybrid Rydberg platform. We conclude with a discussion and summary in Sec.~\ref{sec:summary}.
\section{Lattice formalism} \label{sec:lattice-formalism}
The implementation of a LGT in a quantum simulator requires the truncation of a typically infinite-dimensional local link Hilbert space to a finite-dimensional Hilbert space. The so-called rishon formulation of quantum link models~\cite{Brower1999, Chan_Wiese_1997, Wiese_review, Bar2001} is an efficient and quantum-simulator-friendly way to represent non-Abelian LGTs with finite-dimensional link Hilbert spaces while maintaining the underlying local symmetry. 

\textbf{\textit{Rishon formulation}---}
Rishons are anticommuting fermionic operators used to represent the link degrees-of-freedom, for instance, the electric and gauge fields in a $U(1)$ LGT. Rather than a realistic physical field, they work as mathematical tools to describe the Hilbert space of a color charged gauge field; hereby configurations in the rishon Hilbert space can be identified uniquely as a configuration in the gauge theory.

\begin{figure}[t!]
\centering
\includegraphics[width=\linewidth]{ 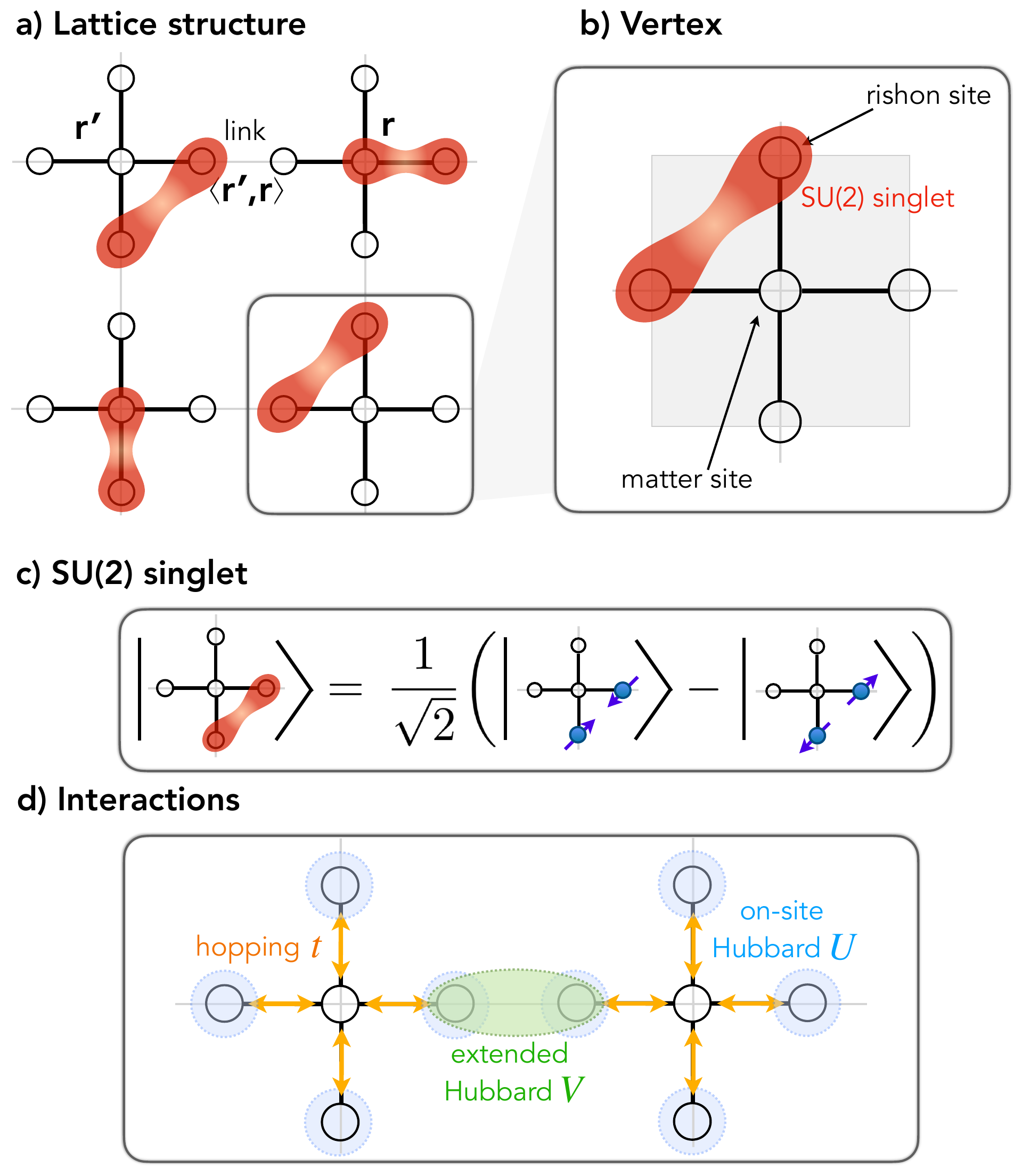}
\caption{\textbf{Rishon-formulation of non-Abelian $U(N)$ quantum-link models.} \textbf{a)} We illustrate the general lattice structure of the rishon model on a $(2+1)$D square geometry showing an allowed singlet covering for the case $N=2$. \textbf{b)} The Gauss's law constraints enforce total color singlets on each vertex as well as a conserved number of rishons~$\mathcal{N}$ per link; here we consider $\mathcal{N}=1$ for an $SU(2)$ gauge structure, where the color singlet is given by a spin-$1/2$ singlet, shown in \textbf{c)}. \textbf{d)} To energetically stabilize the rishon number constraint, we propose to use $SU(N)$-invariant on-site Hubbard interactions $U$ on rishon sites (blue) and extended Hubbard interactions $V$ (green) between rishon sites. The weak tunneling~$t$ of fermions within a vertex induces the perturbative gauge-invariant dynamics. }
\label{fig: Lattice structure and interactions}
\end{figure}

For the scope of our analysis, we consider a square lattice with vertices~$\mathbf{r}$, see Fig.~\ref{fig: Lattice structure and interactions}a.
The links are denoted with $\langle \mathbf{r},\mathbf{r'} \rangle$ and are located between neighboring vertices $\mathbf{r}$ and $\mathbf{r'}$. 
Each link is associated with two rishon sites, and hence one vertex consists of a central matter site and the adjacent four rishon sites on the connecting links, see Fig.~\ref{fig: Lattice structure and interactions}b. 

We define the generators of the $SU(N)$ LGT as~\cite{Brower1999}
\begin{equation}
    \hat{G}^a_{\mathbf{r}}=  \sum_{\alpha, \beta = 1}^{N} \bigg[ \hat{\psi}^{\alpha \dagger}_{\mathbf{r}} \, \lambda^a_{\alpha \beta} \, \hat{\psi}^{\beta}_{\mathbf{r}} \,+\, \sum_{\mathbf{k}} \sum_{\mathbf{v} = \pm \mathbf{k}} \hat{c}^{\alpha \dagger}_{\mathbf{r}, \mathbf{v}} \, \lambda^a_{\alpha \beta} \, \hat{c}^{\beta}_{\mathbf{r}, \mathbf{v}} \bigg]
\end{equation}
and that of the $U(1)$ LGT as~\cite{Brower1999}
\begin{equation} \label{eq: gauss laws}
    \hat{G}_{\mathbf{r}} =  \sum_{\alpha = 1}^{N} \bigg[ \hat{\psi}_{\mathbf{r}}^{\alpha \dagger} \, \hat{\psi}_{\mathbf{r}}^{\alpha}  \, - \, \sum_{\mathbf{k}} \big(\hat{E}_{\mathbf{r}, \mathbf{r} + \mathbf{k}}^{\alpha} - \hat{E}_{\mathbf{r} - \mathbf{k}, \mathbf{r}}^{\alpha} \big) - q_{\mathbf{r}}^0 \, \bigg],
\end{equation}
where $\hat{\psi}^{\alpha \dagger}_{\mathbf{r}}$ is the fermionic matter operator with color $\alpha, \beta=1,...,N$ at site $\mathbf{r}$; $\hat{c}^{\alpha \dagger}_{\mathbf{r} \pm \mathbf{k}}$ is the fermionic rishon operator acting on the rishon site adjacent to vertex~$\mathbf{r}$ in the direction $\mathbf{k}=(1 ,0), (0, 1)$; $\lambda^a_{\alpha \beta}$ are the $N \times N$ matrices that represent the $a= 1,...,(N^2 - 1)$ generators of the corresponding $SU(N)$ Lie Algebra~\cite{Brower1999}. The constant $q_{\mathbf{r}}^0$ depends on the choice of the vacuum sector and corresponds to a local background charge.
Further, the $U(1)$ electric field associated with color~$\alpha$ is defined as
\begin{equation} \label{eq: electric field}
    \hat{E}_{\mathbf{r},\mathbf{r} + \mathbf{k}}^{\alpha} = \frac{1}{2} \big( \, \hat{c}^{\alpha \dagger}_{\mathbf{r} + \mathbf{k}, \, - \, \mathbf{k}}\hat{c}_{\mathbf{r} + \mathbf{k}, - \mathbf{k}}^{\alpha} - \hat{c}^{\alpha \dagger}_{\mathbf{r}, + \mathbf{k}}\hat{c}^{\alpha}_{\mathbf{r}, + \mathbf{k}} \, \big),
\end{equation}
which measures the population imbalance of rishons with color $\alpha$ on a link $\langle \mathbf{r},\mathbf{r} + \mathbf{k} \rangle$.

In the rishon model, two constraints are imposed in the gauge-invariant subspace: (i) The total number~$\mathcal{N}$ of rishons per link has to be conserved, where
\begin{equation}
    \mathcal{N} =  \sum_{\alpha} \big( \, \hat{c}^{\alpha \dagger}_{\mathbf{r'}, \, - \, \mathbf{k}}\hat{c}_{\mathbf{r'}, - \mathbf{k}}^{\alpha} + \hat{c}^{\alpha \dagger}_{\mathbf{r}, + \mathbf{k}}\hat{c}^{\alpha}_{\mathbf{r}, + \mathbf{k}} \, \big)
\end{equation}
with $\mathbf{r'} = \mathbf{r} + \mathbf{k}$ and (ii) the non-Abelian $SU(N)$ Gauss's laws enforce local color singlets on each vertex
\begin{equation}    \label{eq: gauss law su(n)}
    \hat{G}^a_{\mathbf{r}} \vert\varphi\rangle=0 \,\,\, \forall a,\mathbf{r}.
\end{equation} 
In addition, we can impose a $U(1)$~conservation law
\begin{equation} \label{eq: gauss law u(1)}
    \hat{G}_{\mathbf{r}}\vert\varphi\rangle=0 \,\,\, \forall \mathbf{r}
\end{equation} 
to obtain the enlarged gauge group~$U(N)=SU(N) \times U(1)$. Consequently, any reliable quantum simulation of the non-Abelian LGT we aim to construct has to obey dynamics governed by the above constraints.

\section{Model and energetic gauge protection} \label{sec:model}
In this article, we develop two closely related schemes for simulating $(2+1)$D non-Abelian LGTs. Both start by ensuring the gauge constraints are satisfied, by introducing large gauge-protection energy scales adequate for the underlying interactions. Subsequently, we introduce perturbations that explicitly break the above constraints, but -- as we derive -- these terms induce (i) effective and manifestly gauge-invariant couplings, and (ii) superexchange terms that protect against subleading gauge-breaking terms, hence providing a scalable scheme for quantum simulators.
In our \textit{top-down} approach, where gauge-invariant dynamics is generated perturbatively, the minimal coupling term is the dominant one in the effective gauge theory and it describes the gauge-invariant tunneling of $U(N)$~gauge charges coupled to the color charged gauge fields, as we show in this section. We note that in this approach it is inherently difficult to engineer 4-body plaquette terms \cite{Wang2025}. Other potentially complementary schemes for strong, non-perturbative plaquette interactions have been proposed in the contest of $U(1)$ LGTs and toric-code simulations, see e.g. Refs.~\cite{feldmeier2024, homeier2021, Dai2017, Buechler2005}.
We point out that existing methods like the linear gauge protection allow to stabilize large system simulations of Abelian theories~\cite{Halimeh2020single_bodies}, but they are not sufficient for the implementation of non-Abelian symmetries and therefore need to be replaced with different strategies.

The general approach we consider has been introduced and initially studied by Banerjee \textit{et al.}~\cite{banerjee2013} and Surace \textit{et al.}~\cite{surace2024}. However, as we explicitly show in Sec.~\ref{sec:experimental-impl}, any experimentally \textit{realistic} large-scale setup has to be protected against gauge-symmetry breaking errors, which arise from inherent atomic scattering processes in $SU(N)$~Hubbard models. Such gauge-breaking terms were not included in the analysis of Ref.~\cite{banerjee2013} and only partially suppressed in the recent scheme of Ref.~\cite{surace2024}. 

In our proposed scheme, we start with the construction of a Hamiltonian that enforces the correct rishon number by energetically isolating states with the desired number from the unphysical ones: this energy protection term is set to be the most relevant scale of the system Hamiltonian. 
Additionally, the singlet protection is obtained by suppressing spin-exchange interactions between neighboring vertices using a combination of a strong external (magnetic) field and weaker spin-exchange interactions that appear at lower order. In this way, the resulting Hamiltonian can be made inherently gauge invariant.
Subsequently, the dynamics is promoted by additional subdominant terms, such as intra-vertex tunnelling, to achieve second-order dynamical processes within the gauge-invariant sector, through virtual gauge-symmetry breaking.
In the analysis below, we consider sectors with one rishon per link, but our scheme generalizes to different rishon numbers as well.
Our scheme is complementary but fundamentally different from the one proposed in Ref.~\cite{halimeh2024}: here, the rishon-number constraint is actively, perturbatively broken and the spin-exchange mechanism, that contributes as main dynamical protection strategy in Ref.~\cite{halimeh2024}, here appears as a lower-order protection term.

\textbf{\textit{Rishon-number protection mechanism}---}
Let us define a microscopic model which energetically separates states with different rishon numbers in order to realize the constraints discussed in Sec.~\ref{sec:lattice-formalism}. In particular, we introduce the Hamiltonian
\begin{equation} \label{eq:Basic Ham}
\begin{split}
    \hat{H}_0 & = V \sum_{\alpha \beta}\sum_{\mathbf{r}, \mathbf{k}}  \hat{n}_{\mathbf{r},\mathbf{k}}^{\alpha} \hat{n}_{\mathbf{r} + \mathbf{k}, - \mathbf{k}}^{\beta} + U \sum_{\alpha< \beta}\sum_{\mathbf{r}, \mathbf{k}} \sum_{\mathbf{v}=\pm \mathbf{k}} \hat{n}_{\mathbf{r}, \mathbf{v}}^{\alpha} \hat{n}_{\mathbf{r}, \mathbf{v}}^{\beta} \\
    &+ \delta \sum_{\alpha}\sum_{\mathbf{r}} \hat{n}_{\mathbf{r},m}^{\alpha} + U_m \sum_{\alpha < \beta}\sum_{\mathbf{r}} \hat{n}_{\mathbf{r},m}^{\alpha}\hat{n}_{\mathbf{r},m}^{\beta}
\end{split}
\end{equation}
where we used the notation $\hat{n}_{\mathbf{r},m}^{\alpha} = \hat{\psi}_{\mathbf{r}}^{\alpha \dagger} \, \hat{\psi}_{\mathbf{r}}^{\alpha}$ and  $\hat{n}_{\mathbf{r}, \mathbf{k}}^{\alpha} = \hat{c}_{\mathbf{r},\mathbf{k}}^{\alpha \dagger} \, \hat{c}_{\mathbf{r},\mathbf{k}}^{\alpha}$.
The first term $\propto V$ is an $SU(N)$-invariant extended Hubbard interaction (Fig.~\ref{fig: Lattice structure and interactions}d), which acts as an interaction between two rishons at opposite sides of a link. 
The second term $\propto U$ is an on-site $SU(N)$-invariant Hubbard interaction between rishons.
Further, we introduce a chemical potential on the matter site $\propto \delta$ that will allow us later to tune the resonance condition for the effective hopping terms~\cite{surace2023}. The Hamiltonian separates states with different rishon numbers through energy terms proportional to $V$ and $\delta$: this allows to energetically split the Hilbert space into sectors with well-defined rishon numbers per link.
The term $\propto U_m$ is an additional on-site Hubbard interaction between matter excitations.
The absence of hopping and spin-exchange interactions between adjacent vertices makes this idealized model Eq.~\eqref{eq:Basic Ham} inherently gauge invariant~\cite{banerjee2013}. When going beyond one-rishon per link simulations, we additionally need to impose the condition $V\approx U$ in order to maintain the energy resonance between states in the gauge-invariant sector, e.g., when considering a model with two rishons per link, a state with two rishon excitations localized on one side (rishon site) of the link must be near resonant with a state with one rishon excitation on each side of the link.

Thus far, the Hamiltonian~\eqref{eq:Basic Ham} does not have any dynamics. Hence, a state initialized in the target rishon-number sector and with total spin singlet per vertex remains static and gauge-invariant. As we argue next, an intra-vertex perturbation~$t$, that does not couple to the spin (or color) on neighboring vertices but that breaks the rishon number constraint, induces gauge-invariant dynamics~\cite{banerjee2013,surace2023}. In the limit of weak perturbation~$\vert t \vert \ll \delta,V$, the eigenstates are still approximately described by well-defined rishon number sectors~$\mathcal{N}$, such that the effective model satisfies the required gauge constraints up to controlled errors~\cite{halimeh2022gaugeprotection, halimeh2022stabilizinggaugetheoriesquantum}, which we analyse numerically for some realistic cases in Sec.~\ref{sec:experimental-impl}.

This approach requires the energy of the rishon number sectors to depend non-linearly on the rishon number~$\mathcal{N}$ to avoid a resonance between e.g. a gauge-breaking state characterized by two rishon excitations on the same link and zero rishon excitations on another link and a gauge-invariant state correctly characterized by two links with one rishon excitation each. We point out that the Hamiltonian we introduce in Eq.\eqref{eq:Basic Ham} allows us to stabilize any rishon number sector with~$\mathcal{N} \leq 2N$ and therefore can be considered a general scheme to explore gauge theories with increasing number of rishons per link.

\textbf{\textit{Gauge-invariant effective theory}---} 
Next, we introduce a perturbative Hamiltonian term
\begin{equation} \label{eq: general intra-vertex hopping}
    \hat{H}_t = -t \sum_{\alpha}\sum_{\mathbf{r}, \mathbf{k}} \sum_{\mathbf{v}=\pm\mathbf{k}} \bigl( \hat{c}_{\mathbf{r}, \mathbf{v}}^{\alpha \dagger} \, \hat{\psi}_{\mathbf{r}}^{\alpha} + \mathrm{h.c.} \bigr),
\end{equation}  
which describes tunneling of amplitude~$t$ between nearest-neighbor sites within a vertex, see Fig.~\ref{fig: Lattice structure and interactions}d; hence it preserves the total color singlet constraint in each vertex~\cite{banerjee2013}.
First, we consider the resulting effective model, at second order $\mathcal{O}(t^2)$, for a scheme with exactly one singlet per vertex and one rishon per link.
In this case, the effective interactions describe correlated \textit{intra-vertex} processes that conserve the total number of particles on each vertex, 
\begin{equation}
    \hat{N}_{\mathbf{r}}=\sum_{\alpha} \biggr(\hat{n}_{\mathbf{r},m}^{\alpha} + \sum_{\mathbf{k}} \sum_{\mathbf{v}=\pm \mathbf{k}} \hat{n}_{\mathbf{r}, \mathbf{v}}^{\alpha} \biggl),
\end{equation} 
imposing the additional $U(1)$~gauge constraint.

In the limit of~$\vert t \vert \ll V$, we perform a Schrieffer-Wolff transformation and obtain a gauge-invariant hopping given by~\cite{surace2023}
\begin{equation}
    \hat{H}_{\mathrm{eff}}^t = -t_{\mathrm{eff}} \sum_{\alpha \beta}\sum_{\mathbf{r}, \mathbf{k}}  \bigl(  \hat{\psi}_{\mathbf{r}}^{\alpha \dagger} \hat{c}_{\mathbf{r}, \mathbf{k}}^{\alpha}\hat{c}_{\mathbf{r + k, -\mathbf{k}}}^{\beta \dagger} \hat{\psi}_{\mathbf{r+k}}^{\beta} + \mathrm{h.c.} \bigr),
\end{equation}
with effective tunneling amplitude
\begin{equation}\label{Form: General Effective Hopping}
    t_{\mathrm{eff}} = t^2 \biggl( \frac{1}{\delta - V} + \frac{1}{-\delta} \biggr).
\end{equation}
For $\vert t \vert \ll \delta < V$, the coefficient scales $\propto t^2/\delta$, and we achieve gauge-invariant dynamics in the energy-protected manifold. 
This approach similarly allows one to derive the manifestly gauge-invariant tunnelings for any filling and rishon-number sector, enabling the implementation of theories beyond the simplest instance of a rishon link model treated above.

\textbf{\textit{Gauge-breaking interactions}---} 
So far we followed the strategies proposed in earlier works~\cite{banerjee2013,surace2023} working with an idealized unperturbed Hamiltonian $\hat{H}_0$ that does not violate gauge invariance. However, the model in Eq.~\eqref{eq:Basic Ham} is fine-tuned to a gauge-invariant point in the following sense: it requires strong inter-vertex density-density interactions $\propto V$ which commute with the local $SU(N)$ Gauss's laws of both involved vertices. In a realistic experimental setup, using e.g. AELAs, we may indeed assume that local interactions, such as the $V$-term in Eq.~\eqref{eq:Basic Ham} are $SU(N)$-invariant to a very good approximation, but the interactions commute with the total $SU(N)$ spin formed by the two involved rishon sites together and therefore couple particles of adjacent vertices.

Hence, another term in the microscopic Hamiltonian that we have to consider in any generic, i.e., non fine-tuned, situation involves spin-exchange processes between neighboring rishon sites~\footnote{In the specific case of AELAs, this term is only non-zero for interactions between two different orbitals, otherwise it vanishes due to the $SU(N)$ symmetric nature of the interactions.}
\begin{equation}    \label{eq: exc in general theory}
     \hat{H}_{\mathrm{ex}} = V_{\mathrm{ex}} \sum_{\alpha \beta}\sum_{\mathbf{r}, \mathbf{k}}   \hat{c}_{\mathbf{r},\mathbf{k}}^{\alpha \dagger}\hat{c}_{\mathbf{r+k},-\mathbf{k}}^{\beta \dagger}\hat{c}_{\mathbf{r},\mathbf{k}}^{\beta}\hat{c}_{\mathbf{r+k},-\mathbf{k}}^{\alpha} .
\end{equation}
In particular, one should generally expect $\vert V_{\mathrm{ex}} \vert$ to be comparable in strength to the stabilizing interaction $V$ in Eq.~\eqref{eq:Basic Ham}, which we assume to be a large energy scale in the problem~\cite{scazza2014observation, Cappellini2014}.
In fact, Eq.~\eqref{eq: exc in general theory} does not commute with the $SU(N)$ Gauss's laws of the individual vertices $\mathbf{r}$ and $\mathbf{r'}$, because it enables spin-singlets initially defined on one vertex to traverse and extend across the link $\langle \mathbf{r},\mathbf{r'} \rangle$. This should be contrasted with the effect of similar spin-exchange interactions within one vertex that commute with the local $SU(N)$ Gauss's laws. Notably, the presence of $SU(N)$ spin-exchange interactions was not considered in the seminal paper \cite{banerjee2013}, defining the main source of $SU(N)$ gauge-symmetry breaking of that scheme (the $U(1)$ Gauss's law is not affected by this, however \cite{surace2023}). The key goal of our present work is to include such leading gauge-breaking terms and develop a scheme to suppress and control these processes.

\textbf{\textit{Gauge-protection through external field}---}
As an initial strategy for the mitigation of the aforementioned processes, we propose to utilize an external field to guarantee that the color composition of each vertex does not change. The field is intended to introduce a different energy offset for each colored excitation. Any variation of the color composition on a vertex will bring the state out of resonance, suppressing it energetically.

We formalize this idea by adding the following gauge-protection term, with strength $H$, to the Hamiltonian,
\begin{equation}    \label{eq: H protection ext field}
      \hat{H}_{\mathrm{prot}}=H\sum_{\alpha} \gamma_{\alpha}^{\mathbf{r}} \biggl(\hat{n}_{\mathbf{r}, m}^{\alpha} + \sum_{\mathbf{k}} \sum_{\mathbf{v}=\pm\mathbf{k}} \hat{n}_{\mathbf{r}, \mathbf{v}}^{\alpha} \biggr) .
\end{equation}
This corresponds to a shift by $H\gamma_{\alpha}^{\mathbf{r}}$ of all fermions with color $\alpha$, matter and rishon, on vertex $\mathbf{r}$. By choosing
\begin{equation}    \label{eq: condition ext field}
      \vert (\gamma_{\alpha}^{\mathbf{r}} - \gamma_{\alpha}^{\mathbf{r'}}) - (\gamma_{\beta}^{\mathbf{r}} - \gamma_{\beta}^{\mathbf{r'}})\vert  >  0 \quad \text{for} \quad\alpha \neq \beta
\end{equation}
on neighboring vertices $\mathbf{r}$ and $\mathbf{r+k}$, spin-exchange processes within rishon links can be suppressed: the effect of exchanging colors $\alpha$ and $\beta$ on vertices $\mathbf{r}$, $\mathbf{r+k}$ by applying the corresponding term $\propto \hat{c}_{\mathbf{r},\mathbf{k}}^{\alpha \dagger}\hat{c}_{\mathbf{r+k},-\mathbf{k}}^{\beta \dagger}\hat{c}_{\mathbf{r},\mathbf{k}}^{\beta}\hat{c}_{\mathbf{r+k},-\mathbf{k}}^{\alpha}$ in Eq.~\eqref{eq: exc in general theory}, leads to a change of energy $\Delta H_{\mathrm{prot}} = H (\gamma_{\beta}^{\mathbf{r}} - \gamma_{\beta}^{\mathbf{r'}}) - H (\gamma_{\alpha}^{\mathbf{r}} - \gamma_{\alpha}^{\mathbf{r'}})$ for $\alpha \neq \beta$. 
Hence for $\vert H \vert \gg \vert V_{\mathrm{ex}} \vert$ color-changing spin-exchange is strongly suppressed. In Sec.~\ref{sec:experimental-impl} below we will discuss a concrete experimental realization of this gauge-protection term.

\begin{figure}[t]
    \centering  
    \includegraphics[width=\linewidth]{ 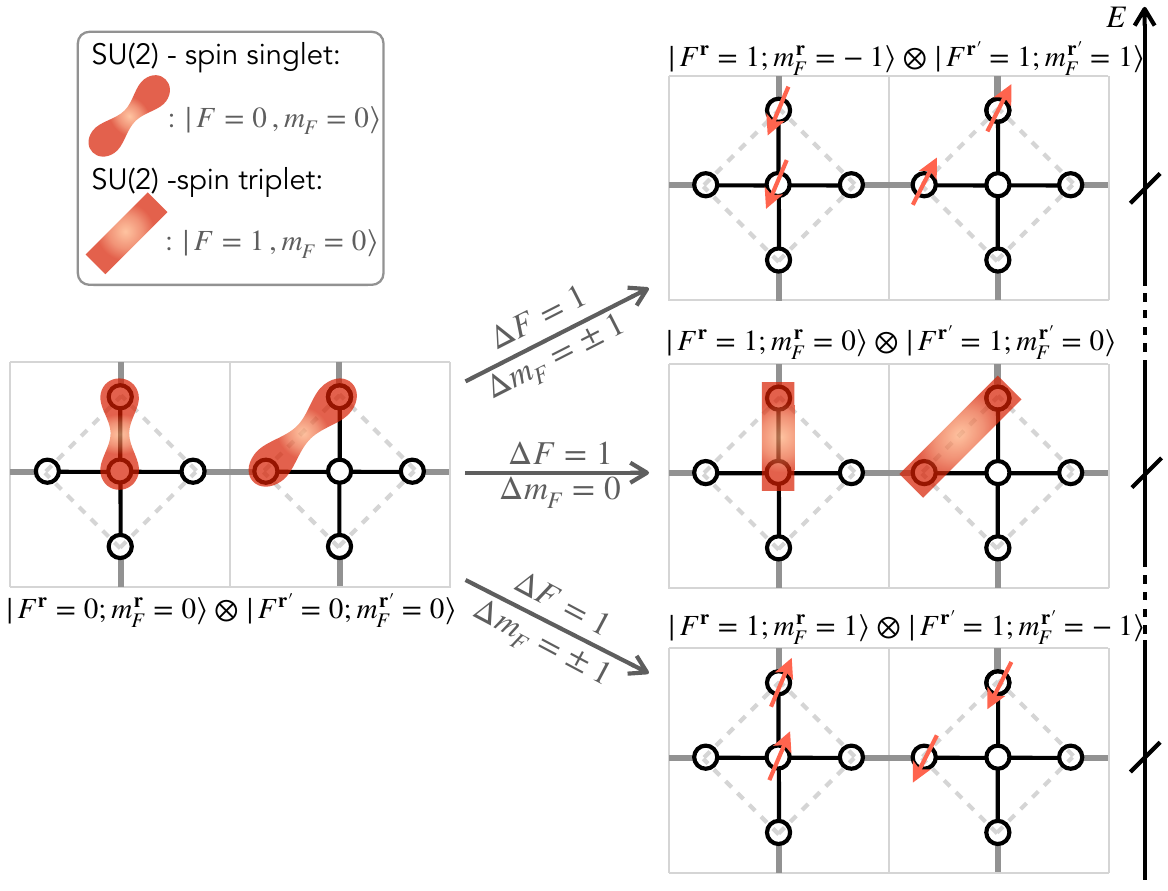}
    \caption{\textbf{Selection rules.} We show the selection rules for a system of two neighboring vertices. The spin-exchange interaction $V_{\mathrm{ex}}$ couples the gauge-invariant states with $\ket{F^{\mathbf{r}}=0;m_F^{\mathbf{r}}=0}\otimes \ket{F^{\mathbf{r'}}=0;m_F^{\mathbf{r'}}=0}$ to gauge-breaking states $\ket{F^{\mathbf{r}}=1;m_F^{\mathbf{r}}=0}\otimes \ket{F^{\mathbf{r'}}=1;m_F^{\mathbf{r'}}=0}$ (represented with the red squares) and $\ket{F^{\mathbf{r}}=1;m_F^{\mathbf{r}}=\pm1}\otimes \ket{F^{\mathbf{r'}}=1;m_F^{\mathbf{r'}}\mp1}$ (indicated by aligned spins). The introduction of an external, vertex-dependent field, Eq.~\eqref{eq: H protection ext field}, allows to energetically separate the states with $m_F=\pm1$ from those with $m_F=\pm0$ but does not protect from oscillations into the manifolds of states with $\ket{F^{\mathbf{r}}=1;m_F^{\mathbf{r}}=0}\otimes \ket{F^{\mathbf{r'}}=1;m_F^{\mathbf{r'}}=0}$.}
    \label{fig: selection rules} 
\end{figure}

Although the gauge-protection term above suppresses changes of the color configurations, it does not fully protect the $SU(N)$ Gauss's law in Eq.~\eqref{eq: gauss law su(n)}. To understand this, we consider an $SU(2)$ subgroup of $SU(N)$ obtained by focusing on two colors $\alpha \neq \beta$, and derive the selection rules associated with the spin-exchange interaction Eq.~\eqref{eq: exc in general theory}. We can label states in the $SU(2)$ subgroup by their total nuclear spin $F^{\mathbf{r}}$ and their spin-projection $m_F^{\mathbf{r}}= -F^{\mathbf{r}},...,F^{\mathbf{r}}$ quantum numbers, defined separately on each vertex $\mathbf{r}$ and $\mathbf{r'}$. In the gauge-invariant subspace, $F^{\mathbf{r}}=F^{\mathbf{r'}}=0$ since total color singlets on the vertices also realize singlets of the $SU(2)$ subgroup. The linear gauge protection term~\cite{halimeh2022gaugeprotection} of Eq.\eqref{eq: H protection ext field} only suppresses $m_F$-changing processes ($\Delta m_F^{\mathbf{r}} = \pm 1, \Delta m_F^{\mathbf{r}}+\Delta m_F^{\mathbf{r'}}=0$), illustrated in the top and bottom rows of Fig.~\ref{fig: selection rules}. 

In order to understand which other ($m_F^{\mathbf{r}} = m_F^{\mathbf{r'}} = 0$) states the $F_{\mathrm{tot}}=0$ singlets $\ket{F^{\mathbf{r}}=0;m_F^{\mathbf{r}}=0}\otimes\ket{F^{\mathbf{r'}}=0;m_F^{\mathbf{r'}}=0}$ (Fig.~\ref{fig: selection rules} left) are coupled by $\hat{T}_{\alpha \beta}^{(1)\otimes (1)} = V_{\mathrm{ex}} \, \hat{c}_{\mathbf{r},\mathbf{k}}^{\alpha \dagger}\hat{c}_{\mathbf{r+k},-\mathbf{k}}^{\beta \dagger}\hat{c}_{\mathbf{r},\mathbf{k}}^{\beta}\hat{c}_{\mathbf{r+k},-\mathbf{k}}^{\alpha}$, we note that the spin-exchange interaction is globally $SU(N)$-invariant; i.e., it conserves the total $SU(N)$ spin on both vertices $\mathbf{r}$, $\mathbf{r'}$ combined and, by extension, the same is true for the $SU(2)$ subgroup we consider. In principle, this symmetry consideration allows couplings to states with $F^{\mathbf{r}}=F^{\mathbf{r'}}=F$, which can combine to $F_{\mathrm{tot}}=0$. However, since $\hat{T}_{\alpha \beta}^{(1)\otimes (1) }$ acts as a rank-1 tensor operator on each vertex ($\mathbf{r}$ or $\mathbf{r'}$), the Wigner-Eckart theorem implies a further selection rule (besides $\Delta m_F^{\mathrm{tot}}=0$),
\begin{equation}
    \Delta F^{\mathbf{r}}, \Delta F^{\mathbf{r'}} = 0,1.
\end{equation}
Hence, the only gauge-noninvariant states reached by $V_{\mathrm{ex}}$ are $\ket{F^{\mathbf{r}}=1;m_F^{\mathbf{r}}=0}\otimes \ket{F^{\mathbf{r'}}=1;m_F^{\mathbf{r'}}=0}$, illustrated in the center row of Fig.~\ref{fig: selection rules}.
Below, in Sec.~\ref{sec:experimental-impl}, we will show for a realistic experimental setting that coupling matrix elements to the triplet states $\ket{F^{\mathbf{r}}=1;m_F^{\mathbf{r}}=0}\otimes \ket{F^{\mathbf{r'}}=1;m_F^{\mathbf{r'}}=0}$ are perturbative in $t^2 / (\delta - V)$, in contrast to the $\Delta m_F = \pm 1$ transitions which are of order $V_{\mathrm{ex}}$ and thus require strong gauge protection, $\vert H\vert > \vert V_{\mathrm{ex}} \vert$. This, in turn, allows to design an efficient gauge protection mechanism based on weak spin-exchange processes $\propto J \sim t^2/U$, which we introduce below. 

\section{Minimal model for non-Abelian lattice gauge theory } \label{sec:minimal-model}
So far, we focused on obtaining the gauge-invariant hopping terms, however, to confirm that our model captures inherently non-Abelian features, we must ensure that the scheme does not map to an Abelian $U(1)$ or $\mathbb{Z}_2$~LGT, or simpler models, e.g., because the model has too many constraints such that the dynamics becomes effectively Abelian. Indeed, we find that $U(2)$ \textit{one-singlet-per-vertex} ($\hat{N}_{\mathbf{r}} = 2$) models can be trivially mapped onto a $U(1)$~QLM, as we derive in this section. In addition, our analysis allows us to determine the inherently non-Abelian sectors of one-rishon models, which can be targeted experimentally by the choice of the initial state.

To identify the Abelian component in $U(2)$ one-rishon models, we first establish a one-to-one correspondence to a $U(1)$~QLM. If such a mapping can be applied, this implies that the spin states of the particles are irrelevant in the underlying non-Abelian LGT. As a result, the $U(2)$~LGT reduces to a $U(1)$~LGT with background charges. In order to find the simplest non-Abelian instance of one-rishon ($\mathcal{N} = 1$) models, we subsequently extend to \textit{multiple-singlet-per-vertex} ($\hat{N}_{\mathbf{r}} > 2$) models -- i.e., higher particle fillings at the vertices -- and consider different initializations, until we reach a minimal model with components that are beyond Abelian LGTs.

\textbf{\textit{Mapping to an Abelian model---}}
\begin{figure*}[t!]
\centering
\includegraphics[width=\textwidth]{ 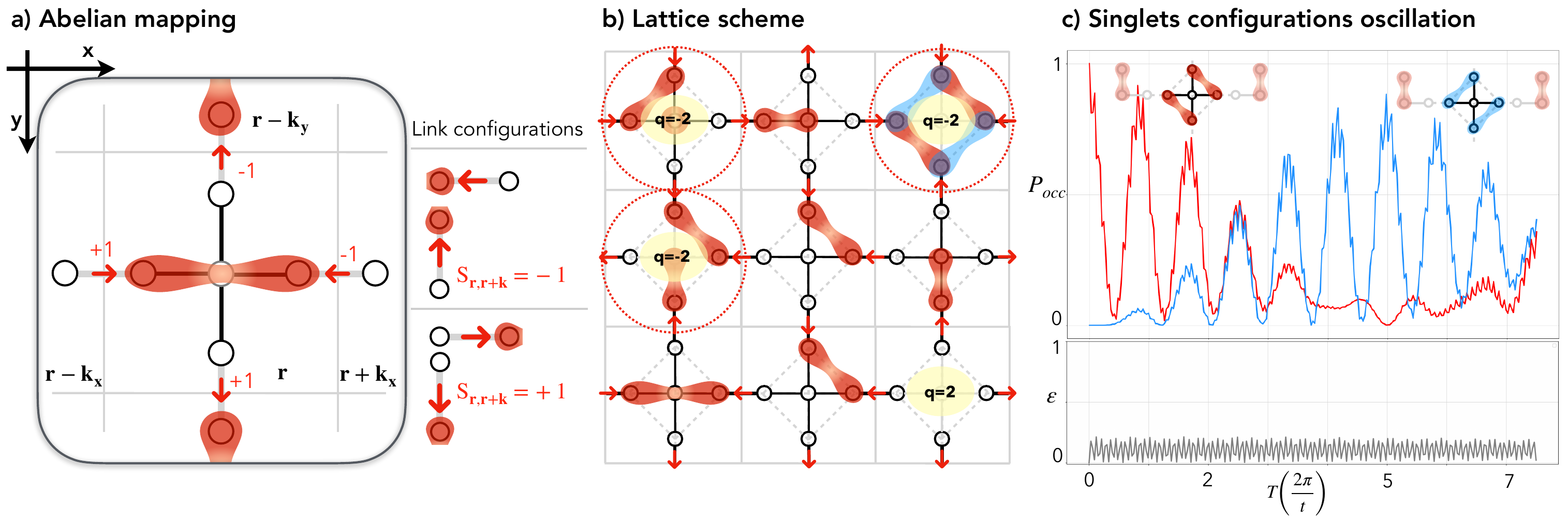}
\caption{\textbf{Mapping of $U(2)$ to $U(1)$~LGT and minimal non-Abelian setup: $U(2)$ with one rishon per link.} \textbf{a)} We illustrate the mapping that identifies the links' state through the operator $\hat{\mathrm{S}}_{\mathbf{r}, \mathbf{r+k}}$. The mapping to a $U(1)$ theory is valid for models with one rishon per link and up to one singlet configuration per vertex. The values of the link variable are determined by the position of the occupied rishon site with respect to the link, as illustrated. The link can be described through a binary variable $\mathrm{S}_{\mathbf{r}, \mathbf{r+k}} = \pm 1$. \textbf{b)} We show an example of lattice initialization for a non-Abelian simulation: at least one vertex is occupied by more than one singlet. The vertices circled in red are filled with more than one singlet and characterize the non-Abelian nature of the scheme. In the top-right vertex we show in blue an additional gauge-invariant singlet covering including components orthogonal to the red covering. \textbf{c)} We show the exact diagonalization results of the model discussed in the main text. Top: During the time evolution, the system, which is initialized in the state colored in red, evolves into the state in blue, with different singlet covering. Since the two states are not orthogonal, in the plot we show the evolution of the orthonormalized states (i.e. the blue state is projected in the space orthogonal to the red one). Three different frequencies of oscillation are noticeable: the fastest oscillations are due to the occupation of states with a wrong number of rishon excitation, the slowest ones represent the effective evolution of the state from the red to the blue singlet covering, while the intermediate oscillations appear when the system is in a gauge-invariant state with some matter excitation in the central vertex (e.g. when one particle or an entire singlet resides in the matter site). Bottom: The grey plot shows the occupation $\varepsilon$ of states with the wrong rishon-number and indeed match the frequency of the fastest oscillations in the top plot.}
\label{fig: Abelian and non Abelian}
\end{figure*}
The mapping to a $U(1)$~LGT that we attempt to make defines an orientation on each link establishing a one-to-one correspondence to a $U(1)$~electric field $\hat{E}_{\mathbf{r},\mathbf{r'}} =\sum_\alpha \hat{E}_{\mathbf{r},\mathbf{r'}}^{\alpha}$, as defined in Eq.~\eqref{eq: electric field}. In Fig.~\ref{fig: Abelian and non Abelian}a, we illustrate the fundamental elements of the mapping: we consider a vertex filled with up to one singlet and we impose the constraint of one rishon per link. The mapping consists in assigning a specific orientation to the links between two vertices $\langle \mathbf{r},\mathbf{r'} \rangle$ depending on the occupation of the corresponding rishon sites. In particular, we choose the~\textquotedblleft orientation\textquotedblright~of the link to point towards the occupied site, and we define the operator $\hat{\mathrm{S}}_{\mathbf{r}, \mathbf{r+k}} = \mathrm{sign}(\hat{E}_{\mathbf{r},\mathbf{r'}})$, with $\mathbf{r'} = \mathbf{r}+\mathbf{k}$ as before. 
Therefore, when the right rishon of a horizontal link is occupied, the link variable will accordingly be positive (i.e. $\hat{\mathrm{S}}_{\mathbf{r}, \mathbf{r+k}} =+1$) while the link variable will be negative when the left rishon is occupied (i.e. $\hat{\mathrm{S}}_{\mathbf{r}, \mathbf{r+k}} =-1$), see Fig.~\ref{fig: Abelian and non Abelian}a.
The same applies to a vertical link: when the down rishon is occupied the link variable is positive, while it is negative when the upper rishon is occupied. 

In this model, the $U(1)$~Gauss's law is defined as
$\hat{G}_{\mathbf{r}} = \sum_{\mathbf{k}} \bigl( \hat{E}_{\mathbf{r}, \mathbf{r}+\mathbf{k}} - \hat{E}_{\mathbf{r}-\mathbf{k}, \mathbf{r}} \bigr) - \hat{\psi}_{\mathbf{r}}^{\dagger}\hat{\psi}_{\mathbf{r}}-q_{\mathbf{r}}^0$ where the local background charge $q_{\mathbf{r}}^0$ is assigned accordingly in each vertex. In the example of Fig.~\ref{fig: Abelian and non Abelian}a, the background charge is $q_{\mathbf{r}}^0 = 0$, while the charge would be $q_{\mathbf{r}}^0 = 2$ if the vertex was empty. Since a $U(1)$ constraint is present (i.e. the singlets cannot hop from one vertex to the other), see Eq.~\eqref{eq: gauss law u(1)}, the background charge is also conserved.
In Appendix~\ref{supp: Mapping to Abelian models}, we show that this way a one-to-one correspondence between \textit{one-singlet-per-vertex} and the $U(1)$~QLM is established.

\textbf{\textit{Non-Abelian model---}}
The minimal non-Abelian model is obtained by initializing the lattice with at least one vertex having more than one singlet. 
To be precise, we consider vertices with four matter/rishon particles, ($\hat{N}_{\mathbf{r}} = 4$); For these configurations Gauss's law implies that states formed by different (non-orthogonal) singlet coverings constitute gauge-invariant states, see Fig.~\ref{fig: Abelian and non Abelian}b. These states cannot be mapped to a $U(1)$~QLM, because the particle's position does not uniquely determine its color-state, hence these configurations are necessary to construct models beyond Abelian LGTs. 

Next, to confirm that the addition of these four-rishons/matter configurations enables to probe features inherent to non-Abelian LGTs in our proposed Hamiltonian, see Sec.~\ref{sec:model}, we show dynamical coupling between the different singlet configurations on such a vertex.
To this end, we employ exact diagonalization on a minimal model and show that the time evolution under the Hamiltonian~$\hat{H} = \hat{H}_0 + \hat{H}_t$ (see Eq.~\eqref{eq:Basic Ham} and Eq.~\eqref{eq: general intra-vertex hopping}) leads to oscillations between the two gauge-invariant singlet coverings on a vertex with four rishon/matter particles.
The minimal model is constituted by one central vertex and we additionally consider the  two neighboring vertices (e.g. left and right), which fulfil an ancillary role and are both occupied by one singlet, see Fig.~\ref{fig: Abelian and non Abelian}c. This situation defines a simple instance of a model with minimal non-Abelian dynamics, where the auxiliary sites are introduced to promote  matter hopping, and where the remaining two neighboring vertices (above and below) are disregarded for simplicity.

Here, we assume on-site (extended) Hubbard interactions of strength~$U=2\delta$ ($V=2\delta$) on all rishon sites (rishon links), which energetically protects the rishon number per link, see Eq.~\eqref{eq:Basic Ham}. The tunneling term~$\hat{H}_t$ of particles within a vertex is a weak perturbation and here we choose~$t =\delta/10$. An additional on-site Hubbard interaction $U_m=\tilde{U}$ is included on the matter site; this term helps to enhance the oscillation between the two singlet configurations. 

To probe the coupling between the two singlet configurations, we first construct a set of orthonormal states~$\{\ket{\varphi_1},\ket{\varphi_2}\}$, where~$\ket{\varphi_1}$ is the initial singlet state, see Fig.~\ref{fig: Abelian and non Abelian}c (red singlet covering) while ~$\ket{\varphi_2}$ represents the normalized projection of the second configuration of singlets (blue singlet covering) on the orthonormal space.
We initialize the system in the state~$\ket{\varphi_1}$ and compute the occupation of states~$\ket{\varphi_1}$ and ~$\ket{\varphi_2}$ under time evolution with Hamiltonian~$\hat{H}$ with time~$T$, see Fig.~\ref{fig: Abelian and non Abelian}c. By construction of the Hamiltonian (without the spin-exchange interactions between neighboring vertices discussed in Sec.~\ref{sec:model}), the Gauss's law constraints Eq.~\eqref{eq: gauss law su(n)} are conserved. Moreover we observe that the correct rishon number constraint is satisfied to a good approximation, as can be seen in the plot of the gauge-breaking states occupation $\varepsilon$ in Fig.~\ref{fig: Abelian and non Abelian}c. The errors $\varepsilon$ in the rishon number constraint are controlled by the perturbation $t/\delta$: while they decrease for weaker perturbation $t$; as does the effective tunnelling amplitude $t_{\mathrm{eff}}$.

The total evolution of the system is characterized by a superposition of multiple oscillations with three different characteristic frequencies: indeed, the system couples to different intermediate states which are not represented in the plot of Fig.~\ref{fig: Abelian and non Abelian}c. For example, the matter excitation on the right (or left) auxiliary vertex can perturbatively hop to the central vertex resulting in a gauge-invariant state that respects the rishon number constraint: the occupation of these states explains the oscillation of the system with intermediate frequency. The fastest oscillations correspond to a micromotion associated with the virtual occupation of the off-resonant states with the wrong rishon number and match the profile of $\varepsilon$. 
Finally, the evolution associated with the slowest frequency shows the oscillation between the red and blue singlet coverings which happens while the values of the link variables, shown as arrows in Fig.~\ref{fig: Abelian and non Abelian}b, do not change. The scale of the time evolution suggests that, with appropriate parameter tuning, this oscillation can be directly accessed in experiments.
Hence, for a given link configuration, there exist at least two different spin configurations participating in the dynamics and therefore the mapping to the $U(1)$~QLM is not bijective any more: this means that the local Hilbert space of our scheme is larger than the Hilbert space of a spin-1/2 $U(1)$ model, providing the desired non-Abelian dynamics, though it does not ensure that the low-energy effective field theory of this system is ultimately non-Abelian. 
\section{Experimental implementation 1: AELAs in optical lattices} \label{sec:experimental-impl}
To realize the microscopic model described in Sec.~\ref{sec:model} in an analog quantum simulation experiment, we propose two different schemes utilizing ultracold $SU(N)$~fermions \cite{scazza2014observation, gorshkov2010, banerjee2013, Zhang2014}.
In this section, we focus on $U(2)$ LGTs with rishon number $\mathcal{N}=1$ per link and fermionic spin-$1/2$ particles; in particular we establish a new gauge protection mechanism necessary to maintain gauge invariance in the presence of strong exchange interactions, Eq.~\eqref{eq: exc in general theory}, present in AELAs.
In the following Sec.~\ref{sec:experimental-impl Rydberg}, we propose a periodically-driven Rydberg scheme which allows the gauge-invariant implementation of more complicated theories beyond one-rishon per link models.
 
Implementing toy models of non-Abelian LGTs in $(2+1)$D constitutes one of the main goals in the quantum simulation of LGTs. 
Recently there have been numerous proposals to build schemes for the simulation of QLMs and non-Abelian gauge theories ~\cite{banerjee2013, Mezzacapo2015, Klco2020, GonzalezCuadra2022, surace2024, Gaz2025}: indeed, $SU(N)$~Hubbard models are generally suited to implement the microscopic model~\eqref{eq:Basic Ham}, however they require the engineering of additional strategies to fully ensure the protection of the gauge symmetry, as discussed in Sec.~\ref{sec:model}.
Previous works, such as Refs.~\cite{banerjee2013, surace2024} have tackled these challenges and suggested experimental proposals for engineering gauge-invariant Hamiltonians. In Banerjee et al.~\cite{banerjee2013} a model similar to Eq.~\eqref{eq:Basic Ham} has been considered, obtaining a scheme for the implementation of $U(N)$ and $SU(N)$ theories, but some terms of the Hamiltonian that are naturally present and give rise to problematic symmetry breaking interactions have not been included in the calculations. On the other hand, Surace et al.~\cite{surace2024} studied in great detail their proposed model, including imperfections of the underlying tight-binding Wannier states, but did not achieve full gauge invariance of the system. 
In our work, we attempt to consider all the relevant symmetry breaking interactions in the system and to find a set of strategies that leads to a full protection of the gauge symmetry. 

As we discussed in Sec.~\ref{sec:model} and explicitly show below, spin-exchange interactions in the form of Eq.~\eqref{eq: exc in general theory} naturally arise in AELAs between neighboring vertices with a strength that is on the order of the largest energy scale in the system.
To achieve the simulation of large-scale non-Abelian LGTs, we propose to implement a twofold gauge protection scheme to efficiently suppress gauge-breaking terms present in AELA $SU(N)$~Hubbard models. First, we implement linear gauge protection~\cite{halimeh2022gaugeprotection} realized by an external magnetic field that allows us to suppress oscillations to the states $\ket{F^{\mathbf{r}}=1;m^{\mathbf{r}}_F=\pm1}\otimes\ket{F^{\mathbf{r'}}=1;m^{\mathbf{r'}}_F=\mp1}$ (see Fig.~\ref{fig: selection rules}). Second, we derive the strength of the remaining gauge-breaking terms and we find that on-site Hubbard interactions on the matter sites are sufficient to fully protect the non-Abelian gauge constraints. 

Our starting point is the setup proposed in previous works that utilizes a rishon formulation to implement Abelian $U(1)$~\cite{surace2023} and non-Abelian~$U(N)$~LGTs~\cite{banerjee2013, surace2024}.
In particular, we consider ultracold fermionic AELAs distinguished through their electronic ground $\ket{g}$ (metastable $\ket{e}$) state and nuclear spin~$\vert m_{\mathrm{F}}=\pm \frac{1}{2}\rangle$ for $U(2)$~models; we note that our scheme can be generalized to $U(N)$~models.
Using tune-out optical lattices for the $\ket{g}$ and $\ket{e}$ states, we engineer an optical superlattice with $\ket{g}$ ($\ket{e}$) atoms on sublattice A (sublattice B) of a square lattice in $(2+1)$D.
\begin{figure*}[t!]
    \centering
    \includegraphics[width=\textwidth]{ 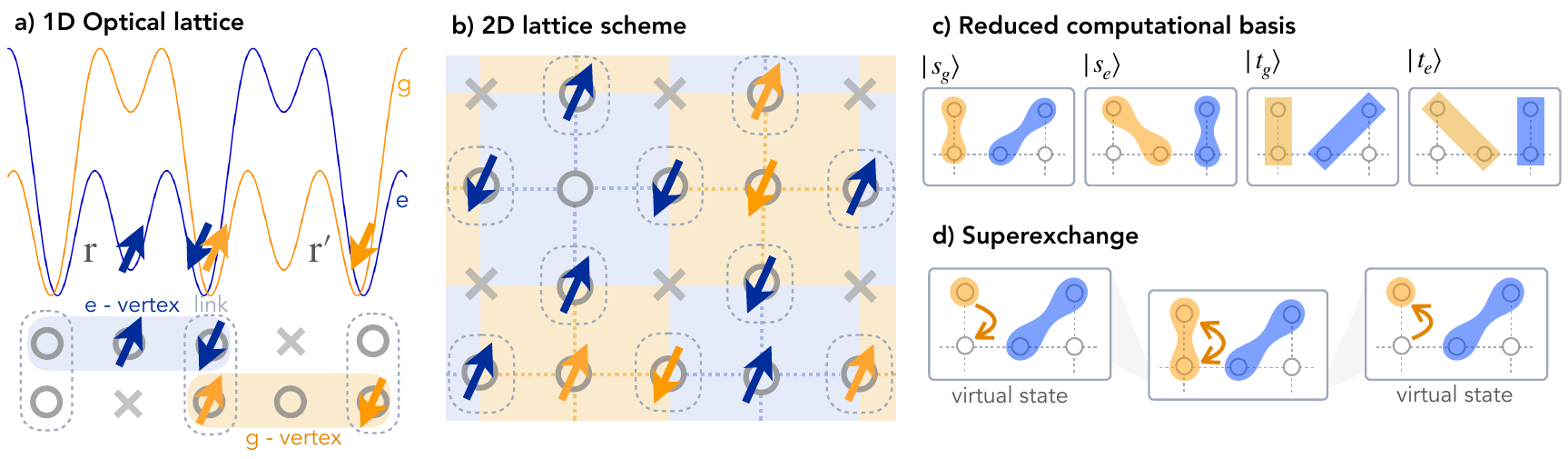}
    \caption{\textbf{AELA scheme with its reduced basis and superexchange mechanism.} \textbf{a)} 1D cut of a minimal building block: the orange line represents the optical lattice potential of atoms in the ground state ($g$) and the blue one of the atoms in the excited clock state ($e$). Along 1D, each vertex is given by a group of three wells that are of alternate ($g$ and $e$) nature and it is initialized to only contain spin singlets. The links are identified by the rishon sites, where the $g$ and $e$ wells overlap and where we realize interactions between atoms of the neighboring vertices. \textbf{b)} 2D generalization of the lattice scheme: each vertex is identified with blue or orange color and composed of one matter site and four adjacent rishon sites, for a total of five wells. The hopping of particles can only happen within the same vertex and it is shown with blue and orange dotted lines. \textbf{c)} The four $m^{\mathrm{tot}}_F = 0$ states that dominate the dynamics in the regime protected through the Zeeman splitting mechanism. \textbf{d)} The superexchange (here shown in the $g$-vertex) yields a correction to the total energy of the central state: the states in the left and right boxes are perturbatively occupied. Analogously, superexchange processes also occur in the $e$ vertices but have not been depicted here.}
    \label{fig: AELA and superexchange}
\end{figure*}

In Fig.~\ref{fig: AELA and superexchange}a, we illustrate a one-dimensional cut of a minimal building block: the two lattices have the same periodicity but are shifted relative to each other in space: the $\ket{g}$ and $\ket{e}$ lattice sites overlap only on the rishon sites allowing us to effectively generate the interactions between neighboring vertices shown in Fig.~\ref{fig: Lattice structure and interactions}d, without tunnelling between them; we refer to Ref.~\cite{surace2023} for an explicit calculation of the Wannier functions. Similarly, on the 2D square lattice the neighboring vertices overlap on the rishon sites.
Hence, each block is composed of five wells: a central matter site surrounded by four rishon sites, as illustrated in Fig.~\ref{fig: AELA and superexchange}b.
The total tight-binding discrete Hamiltonian of this system reads \cite{scazza2014observation, surace2023, surace2024}
\begin{equation} \label{main text: Hamiltonian total}
    \hat{H} = \hat{H}_g + \hat{H}_e + \hat{H}_{gg} + \hat{H}_{ee} + \hat{H}_{eg}.
\end{equation}

Here $\hat{H}_{\epsilon}$ ($\epsilon = g,e$) is the single-particle Hamiltonian in vertex $\epsilon$ that includes the chemical potential terms, proportional to $\delta^{\epsilon}$ in Eq.~\eqref{eq:Basic Ham}, and the intra-vertex  hopping proportional to $t^{\epsilon}$ in Eq.~\eqref{eq: general intra-vertex hopping}. In general, $t^{\epsilon}$ and $\delta^{\epsilon}$ can be tuned independently in this setting for $\epsilon = g,e$ states. $\hat{H}_{\epsilon \epsilon}$ is the on-site Hubbard interaction, proportional to $U_{\epsilon \epsilon}$, that takes the role of the on-site Hubbard interaction proportional to $U$ in Eq.~\eqref{eq:Basic Ham} and that will allow us to additionally implement a superexchange-based gauge protection mechanism.
Finally, $\hat{H}_{eg} = \hat{H}_V + \hat{H}_{\mathrm{ex}}$ describes the interaction between the $g$ and $e$ atoms in every rishon site and it is composed of two terms: the first one, $\hat{H}_V$, is an extended Hubbard term that arises when a $g$ and $e$ atom occupy the same rishon site and we identify it with the extended Hubbard term $\propto V$ in Eq.~\eqref{eq:Basic Ham}. The second term, that we indicate with $\hat{H}_{\mathrm{ex}}$, is defined as in Eq.~\eqref{eq: exc in general theory} and is proportional to $V_{\mathrm{ex}}$: it represents a spin-exchange interaction on the rishon sites, where the two lattices are spatially overlapping, and introduces gauge-breaking couplings, as explained in Sec.~\ref{sec:model}. For further discussion on the origin of the interaction terms $U_{\epsilon\epsilon}$, $V$, and $V_{\mathrm{ex}}$ we refer to Appendix~\ref{supp: effective theory}. In the next paragraphs we discuss our strategies for realizing gauge protection.

\textbf{\textit{Protection of gauge invariance}}---
After defining the general structure of the scheme at the level of the microscopic Hamiltonian, the goal is now to suppress gauge-breaking processes that are inevitable in the above described scheme due to the spin-exchange interaction of strength $V_{\mathrm{ex}}$ (see Eq.~\eqref{eq: exc in general theory}): this term introduces an energy splitting between the triplet and singlet states of two atoms located on the same rishon site and leads to the expansion of the spin singlets across a link, as discussed in Sec.~\ref{sec:model}.

To corroborate this picture, we study the microscopic Hamiltonian on a minimal building block comprised of two neighboring vertices, (Fig.~\ref{fig: AELA and superexchange}a). For a faithful quantum simulation, it is necessary to maintain gauge invariance, i.e., to remain in the total spin singlet sector $\ket{F^{\mathbf{r}}=0;m^{\mathbf{r}}_F=0}$ at each vertex~${\mathbf{r}}$, and suppress leakage into the triplet sector $\ket{F^{\mathbf{r}}=1;m^{\mathbf{r}}_F=0,\pm 1}$, as already explained in Sec.~\ref{sec:model} and shown in Fig.~\ref{fig: selection rules}. To this end, we propose a gauge protection scheme constituted by
\begin{enumerate}[(i)]
    \item linear Zeeman shifts suppressing singlet/triplet oscillation between neighboring vertices $\mathbf{r},\mathbf{r'}$, i.e. between states $\ket{F^{\mathbf{r}}=0;m^{\mathbf{r}}_F=0}\otimes\ket{F^{\mathbf{r'}}=0;m^{\mathbf{r'}}_F=0}$ and $\ket{F^{\mathbf{r}}=1;m^{\mathbf{r}}_F=\pm 1}\otimes\ket{F^{\mathbf{r'}}=1;m^{\mathbf{r'}}_F=\mp 1}$ with $m^{\mathbf{r}}_F = -m^{\mathbf{r}}_F$,
    \item superexchange processes within a vertex enabling protection~\cite{halimeh2024} against couplings between the $\ket{F^{\mathbf{r}}=0;m^{\mathbf{r}}_F=0}\otimes\ket{F^{\mathbf{r'}}=0;m^{\mathbf{r'}}_F=0}$ and $\ket{F^{\mathbf{r}}=1;m^{\mathbf{r}}_F= 0}\otimes\ket{F^{\mathbf{r'}}=1;m^{\mathbf{r'}}_F=0}$ manifold.
\end{enumerate}

The first protection mechanism is conveniently obtained by applying an external magnetic field~$B$: the atomic states $\epsilon = g,e$ are affected by Zeeman shifts $\Delta_Z = g_{\epsilon} \, m_F \, \mu_B \, B$, where $g_{\epsilon}$ is the corresponding Landé factor, $m_F$ the nuclear spin quantum number encoding color (spin) and $\mu_B$ is the Bohr magneton. Due to the non-zero difference $\delta g=g_e-g_g$ between the Landé factors of the $\ket{g}$ and $\ket{e}$ states, neighboring vertices experience different Zeeman shifts. 
This realizes the desired general gauge-protection term introduced in Sec.~\ref{sec:model}, Eq.~\eqref{eq: H protection ext field}, with the identification $H=\mu_B B$ and 
\begin{equation}    \label{eq: Zeeman shift}
     \gamma^{\mathbf{r}}_{\alpha} = m^{\alpha}_F \begin{cases} 
g_g & \mathbf{r}\in \text{$g$-vertices } \\
g_e & \mathbf{r}\in \text{$e$-vertices },
\end{cases}
\end{equation}
where $m^{\alpha}_F$ is the nuclear magnetic state corresponding to color $\alpha$. For $\alpha \equiv m^{\alpha}_F$ it follows that $\gamma^{\mathbf{r}}_{\alpha} - \gamma^{\mathbf{r'}}_{\alpha} = \pm \alpha \, \delta g \neq \gamma^{\mathbf{r}}_{\beta} - \gamma^{\mathbf{r'}}_{\beta}$ unless $\alpha = \beta$ with $\delta g = g_g -g_e$, i.e. condition~\eqref{eq: condition ext field} is satisfied. Hence, when $\vert \delta g \, \mu_B \, B \vert \gg \vert V_{\mathrm{ex}}\vert$, linear gauge-protection is achieved and $m_F$-changing spin-exchange processes are suppressed~\cite{scazza2014observation} for strong enough magnetic fields. Additionally, we assume to remain far away from Feshbach resonances, for simplicity, since near such Feshbach resonance modifications of the interaction strengths have to be accounted for~\cite{Bettermann2023, Pagano2015, Hofer2015}.

We note that the Zeeman splitting introduced here cannot protect against a class of processes where Gauss's law is violated on multiple vertices such that the contributions of the Zeeman energy sum up to zero and hence are coupled resonantly via a higher-order process. These weak higher-order processes are not captured by the minimal two vertex model considered. However, they are suppressed by the superexchange protection mechanism, introduced in the following paragraph.
Alternatively, the higher-order resonances can be efficiently suppressed by the introduction of a magnetic field gradient~\cite{Halimeh2022stabilizing, homeier2023Z2}.

As explained in Sec.~\ref{sec:model}, the linear Zeeman term cannot protect from oscillations between the $\ket{F^{\mathbf{r}}=0;m^{\mathbf{r}}_F=0}\otimes\ket{F^{\mathbf{r'}}=0;m^{\mathbf{r'}}_F=0}$ and $\ket{F^{\mathbf{r}}=1;m^{\mathbf{r}}_F=0}\otimes\ket{F^{\mathbf{r'}}=1;m^{\mathbf{r'}}_F=0}$ states and we therefore need an additional strategy.
To understand the origin of this problem, we consider the minimal model consisting of two vertices and explicitly derive how the effective matter tunneling from Eq.~\eqref{Form: General Effective Hopping} introduces perturbative gauge-breaking exchange interactions of type (ii) proportional to $V_{\mathrm{ex}}$. Concretely, we consider the configurations with one singlet on each vertex and those with one triplet on each vertex and calculate the effective couplings between these states. 

We find that the low-energy effective dynamics of the minimal model in the zero total magnetization sector ($m^{\mathrm{tot}}_F=0$) can be described through a four-dimensional Hilbert space spanned by the basis states (see Fig.~\ref{fig: AELA and superexchange}c)
\begin{equation}\label{eq: computational basis}
    \{ \vert s_g \rangle, \vert s_e \rangle, \vert t_g \rangle, \vert t_e \rangle \},
\end{equation}
where the subscript $g$ ($e$) indicates that the matter site in the vertex $g$ ($e$) is occupied and where $\vert s \rangle$ are the singlets states $\ket{F^{\mathbf{r}}=0;m^{\mathbf{r}}_F=0}\otimes\ket{F^{\mathbf{r'}}=0;m^{\mathbf{r'}}_F=0}$, while $\vert t \rangle$ are the triplets states $\ket{F^{\mathbf{r}}=1;m^{\mathbf{r}}_F=0}\otimes\ket{F^{\mathbf{r'}}=1;m^{\mathbf{r'}}_F=0}$.
Using the computational basis~\eqref{eq: computational basis} and assuming that the microscopic hopping $t=t^g=t^e$ and the chemical potential $\delta=\delta^g=\delta^e$ have the same values for $g$ and $e$ vertices, we can describe the system through the Hamiltonian
\begin{align}
\hat{H}_{\mathrm{eff}} = 
\begin{pmatrix} \label{mat: Heff4x4}
\delta + \delta_{\mathrm{eff}} & t_{\mathrm{eff}} & \lambda & \lambda \\
t_{\mathrm{eff}} & \delta + \delta_{\mathrm{eff}} & \lambda & \lambda \\ 
\lambda & \lambda & \delta + \delta_{\mathrm{eff}} & t_{\mathrm{eff}} \\
\lambda & \lambda & t_{\mathrm{eff}} & \delta + \delta_{\mathrm{eff}} \\
\end{pmatrix} .
\end{align}
From perturbation theory calculations (see more details in the Appendix~\ref{supp: effective theory}), we obtain the corrections to the chemical potential of the matter site $\delta_{\mathrm{eff}}=\mathcal{O}(t^2/\delta)$, the effective hopping within the singlets and triplets sectors $t_{\mathrm{eff}}$ similar to Eq.~\eqref{Form: General Effective Hopping}, and the coupling coefficient~$\lambda$ between the singlets and the triplets
\begin{equation} \label{eq: lambda g_e}
     \lambda = - \frac{t^2}{2(\delta - V)}\biggl(\frac{V_{ex}}{\delta - V + V_{ex}} \biggr);
\end{equation}
see the Tab.~\ref{tab: effective theory} in the Appendix for full expressions of $t_{\mathrm{eff}}$ and $\delta_{\mathrm{eff}}$.

In the limit $\vert t \vert \ll \delta \ll V$, the coupling between singlets and triplets is much weaker than the gauge-invariant hopping, $\vert \lambda \vert \ll t_{\mathrm{eff}}$, and appears as a perturbation to the effective gauge-invariant model. This is because in this limit $\lambda \sim t^2/V$ requires virtual two-rishon states whereas $t_{\mathrm{eff}} \sim t^2/\delta$ includes a contribution from a virtual zero-rishon state. In any case,  $\lambda$ leads to a resonant coupling between the triplet and singlet sectors.

Up to this point, the analysis did not include the contribution of superexchange processes within each vertex,  which can introduce a splitting between singlet and triplet states and thus serve as a gauge-protection mechanism. Moreover, we assumed the values of $\delta$ and $t$ to be the same on the $g$ and $e$ vertices. 
In the following, we distinguish between the on-site Hubbard interactions~$U_{gg}$ ($U_{ee}$) on the rishon and matter sites of $g$ ($e$) vertices, which yield intra-vertex superexchange interactions but cannot be tuned independently in a realistic experimental setting. 
In our scheme, we propose to use staggered potentials on the matter sites of the $g$- and $e$-vertices, i.e. $\delta^g \neq \delta^e$, such that the different contributions of the superexchange interaction are balanced and we obtain resonant gauge-protected dynamics. 

In Fig.~\ref{fig: AELA and superexchange}d, we illustrate the intra-vertex superexchange processes in the $g$-vertex. 
We distinguish the processes in the $g$- and $e$-vertices by introducing a state-dependent chemical potential~$\delta^{\epsilon}$ ($\epsilon=g,e$) on the matter site. 
The effective Hamiltonian describing the resulting superexchange~\cite{Trotzky2008, Duan2003} is given by
\begin{equation} \label{main text: superexchange operator}
    \hat{H}_J = \sum_{\mathbf{r}, \mathbf{v}=\pm\mathbf{k}} \frac{J_{\epsilon}}{2} \sum_{\alpha \not= \beta }\bigl( \hat{c}_{\mathbf{r,\mathbf{v}}}^{\alpha \dagger}\hat{c}_{\mathbf{r}, \mathbf{v}}^{\beta}\hat{\psi}_{\mathbf{r}}^{\beta \dagger}\hat{\psi}_{\mathbf{r}}^{\alpha} - \hat{n}_{\mathbf{r}, m}^{\alpha}\hat{n}_{\mathbf{r}, \mathbf{v}}^{\beta}  \bigr),
\end{equation}
where $\hat{c}_{\mathbf{r,\mathbf{v}}}^{\alpha}$ ($\hat{\psi}_{\mathbf{r}}^{\alpha}$) is the fermionic operator of the rishon (matter) site and
\begin{equation}
    J_{\epsilon}\,=\, \biggl( \frac{2 t^2}{U_{\epsilon\epsilon} + \delta^{\epsilon}} + \frac{2 t^2}{U_{\epsilon\epsilon} - \delta^{\epsilon}} \biggr).
\end{equation} 
This effective interaction can also be reformulated in terms of spin operators yielding the standard superexchange Hamiltonian
\begin{equation}
    \hat{H}_J = \sum_{\mathbf{r}, \mathbf{v}=\pm\mathbf{k}} J_{\epsilon} \biggl( \hat{\mathbf{F}}_{\mathbf{r}, m} \cdot \hat{\mathbf{F}}_{\mathbf{r}, \mathbf{v}} - \frac{1}{4} \hat{n}_{\mathbf{r}, m}\hat{n}_{\mathbf{r}, \mathbf{v}} \biggr).
\end{equation}

This interaction introduces an energy shift between the triplet and the singlet manifold, lowering singlet energy by $-J_\epsilon$. If $U_{gg}=U_{ee}$, these energy contributions would immediately allow us to separate the singlets and triplets into two well-defined energy sectors, as required for gauge protection,  but typically the atomic scattering interactions are different for $g$ and $e$ states, see Tab.~\ref{tab:scatt_lenghts}.
For example, in the fermionic isotope $^{173}$Yb, the superexchange energy scale is smaller for $e$ atoms than for $g$ atoms.

To mitigate the difference due to the unequal scattering lengths on $g$ and $e$ vertices, we propose to tune the chemical potentials on the matter sites of the two vertices $\delta^g$ and $\delta^e$ independently in such a way that the overall energy splitting between the two singlet configurations $\ket{s_g}$ and $\ket{s_e}$ vanishes, $\Delta^{s}=0$. Moreover, their values will be chosen such that a controlled energy separation $\vert \Delta^{s-t} \vert \gg \lambda$ from the gauge-breaking triplet states $\ket{t_g}$ and $\ket{t_e}$ is obtained. To this end, we repeat the analysis of the effective Hamiltonian in the basis \eqref{eq: computational basis} for $\delta^g \neq \delta^e$ and we obtain a modified version of Eq.~\eqref{mat: Heff4x4}, where the coefficients now depend on $\delta^g - \delta^e$. 

We write the local chemical potential $\delta^{\epsilon}$ in the microscopic Hamiltonian Eq.~\eqref{main text: Hamiltonian total} as 
\begin{equation}
    \delta^{\epsilon} = \delta + \tilde{\delta}^{\epsilon}, \quad \tilde{\delta}^{g}=-\tilde{\delta}^{e},\quad \delta=\frac{1}{2} (\delta^e + \delta^g ) , 
\end{equation}
where the small difference $\delta^e - \delta^g = \tilde{\delta}^{e} - \tilde{\delta}^{g}$ is expressed in terms of the small state-dependent detuning $\vert \tilde{\delta}^{\epsilon} \vert \ll \vert \delta \vert$; note that $\vert \delta \vert \gg t$ in order to conserve the correct rishon number per link, see Sec.~\ref{sec:model}.
The effective Hamiltonian~\eqref{mat: Heff4x4} then takes the form
\begin{align} \label{eq:effective_model_superexchange}
    \! \hat{H}_{\mathrm{eff}} \! = \! \begin{pmatrix} 
    \! \delta^g \!+ \! \delta^g_{\mathrm{eff}} \!-\!J_g & \! t_{\mathrm{eff}} & \! \lambda_{gg} & \! \lambda_{ge} \\
    \! t_{\mathrm{eff}} & \! \delta^e \!+\! \delta^e_{\mathrm{eff}} \!-\!J_e & \! \lambda_{ge} & \! \lambda_{ee} \\ 
    \! \lambda_{gg} & \! \lambda_{ge} & \! \delta^g \!+\! \delta^g_{\mathrm{eff}} & \! t_{\mathrm{eff}} \\
    \! \lambda_{ge} & \! \lambda_{ee} & \! t_{\mathrm{eff}} & \! \delta^e \! + \! \delta^e_{\mathrm{eff}}\\
    \end{pmatrix},
\end{align}
where $\vert \delta^{\epsilon}_{\mathrm{eff}}(t^{\epsilon}, \delta^{\epsilon}) \vert \ll \delta$ is a perturbative, dispersive energy shift $\mathcal{O}(t^2/\delta)$ and $t_{\mathrm{eff}} = t_{\mathrm{eff}}(t^{\epsilon}, \delta^{\epsilon})$ as well as $J_{\epsilon'} = J_{\epsilon'}(t^{\epsilon}, \delta^{\epsilon})$ and the off-diagonal couplings $\mathcal{O}(\lambda)$ depend on all microscopic parameters - explicit expressions are provided in Appendix~\ref{supp: effective theory}.

\begin{figure*}[t!]
    \centering
    \includegraphics[width=\textwidth]{ 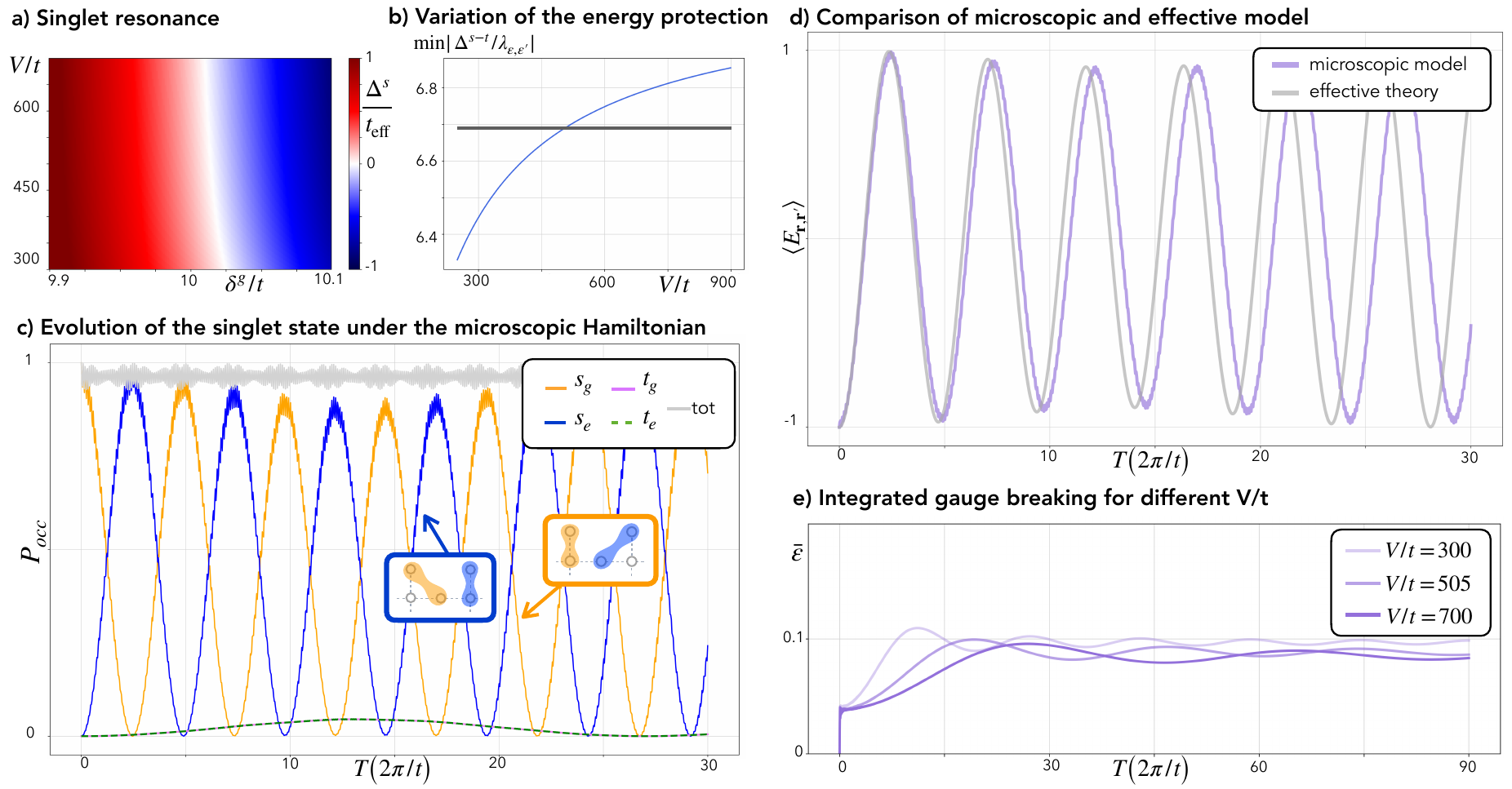}
    \caption{\textbf{Analysis of the superexchange-induced protection and time evolution of the full microscopic model.} \textbf{a)} The plot shows that for a fixed choice of $\delta^e=10\,t$ and variating values of $V/t$ (but we fix $V_{\mathrm{ex}}/V$, $U_{gg}/V$ and $U_{ee}/V$) we obtain the resonance of the singlets states, $\Delta^s=0$, for $\vert \delta^g/t \vert \approx 10$. \textbf{b)} We show the ratio between the protection energy $\Delta^{s-t}$ (i.e. the minimal energy difference between the triplets and singlets sector) and the singlet-triplet coupling coefficients $\lambda_{\epsilon, \epsilon'}$ for all combinations $\epsilon, \epsilon' = g,e$ varying values of $V/t$ (and fixed $V_{\mathrm{ex}}/V$, $U_{gg}/V$ and $U_{ee}/V$). The grey thick line corresponds to $V/t = 505$ that we use in the following analysis. \textbf{c)} We apply exact diagonalization to obtain the dynamics of an initial gauge-invariant state $\ket{s_g}$ (inset orange box) under the full microscopic model, i.e., including the gauge-breaking exchange interactions and the gauge protection strategies. The orange (blue) line denotes the population of the gauge-invariant state $\ket{s_g}$ ($\ket{s_e}$), whereas the magenta (green) curve shows the poppulation of the gauge-violating triplet state~$\ket{t_g}$ ($\ket{t_e}$). The grey curve shows the total occupation in the subspace spanned by $\ket{s_g}$, $\ket{s_e}$, $\ket{t_g}$and  $\ket{t_e}$; the remaining population is constituted by, e.g., high energy states with doublons.
    \textbf{d)} We compare the electric field $\langle \hat{E}_{\mathbf{r},\mathbf{r'}}\rangle$ in the microscopic (purple line) and effective (grey line) model, showing the emergence of an effective dynamics. \textbf{e)} We calculate the integrated gauge violation $\bar{\varepsilon}$ on long times for three values of $V/t$ and show that the gauge protection slightly improves when increasing it.}
    \label{fig: numeric_results}
\end{figure*}

The energy splitting between the two singlet states in Eq.~\eqref{eq:effective_model_superexchange} is given by $\Delta^s = (\delta^g + \delta^{g}_{\mathrm{eff}} -J_g) - (\delta^e + \delta^{e}_{\mathrm{eff}} -J_e)$. In order to obtain resonant, gauge-invariant dynamics we now determine $\delta^g$ and $\delta^e$ such that $\Delta^s = 0$. To this end, we set $t=t^e=t^g$, fix $\delta^e=10 \, t$, and find suitable parameters $\delta^g$ and $V$ such that $\Delta^s = 0$, see Fig.~\ref{fig: numeric_results}a. 
Here we keep $V_{\mathrm{ex}}/V = 790/1010$, $U_{gg}/V = 199/1010$ and $U_{ee}/V = 306/1010$ constant, at values corresponding to $^{173}$Yb, see Tab.~\ref{tab:scatt_lenghts}. Moreover, we assume the $g$ and $e$ optical potentials to be similar, in order to have equal $g$-$e$ Wannier orbitals and avoid their explicit calculation.
\begin{table}[b!]
\centering
\begin{tabular}{|c|c|c|c|}
\hline
$a_{gg}$ & $a_{ee}$ & $a_{eg}^-$ & $a_{eg}^+$ \\ \hline
$199.4a_{0}$  & $306.2a_{0}$  & $219.7a_{0}$ & $1800a_{0}$ \\\hline 
\end{tabular}
\caption{\textbf{Scattering lengths for $^{173}$Yb} in units of the Bohr radius~$a_0$~\cite{ediss18159}. The Hubbard interaction strength~$U_X \propto a_X$ depends on the two-particle scattering length with $X=gg, ee, eg^+, eg^-$. The absolute strength is determined by the Wannier function overlap between $g$ and $e$ sites. Throughout the main text, we define $V = (U_{eg}^+ + U_{eg}^-)/2$ and $V_{\mathrm{ex}} = (U_{eg}^+ - U_{eg}^-)/2$. }
\label{tab:scatt_lenghts}
\end{table}
For the points of ($\delta^g,V$) for which $\Delta^s = 0$, we then compute $\vert \Delta^{s-t}/ \lambda_{\epsilon, \epsilon'} \vert$ for all combinations $\epsilon, \epsilon' = g,e$ to estimate the efficiency of the superexchange gauge protection. Our results are shown in Fig.~\ref{fig: numeric_results}b where we observe that the atomic interactions in $^{173}$Yb naturally place the model in a gauge protected regime. The value of $\min \vert \Delta^{s-t}/ \lambda_{\epsilon, \epsilon'} \vert$, minimized over all combinations $\epsilon, \epsilon' = g,e$, increases for growing values of $V/t$. The dashed line shows $\min \vert \Delta^{s-t}/ \lambda_{\epsilon, \epsilon'} \vert = 6.6$ obtained for the reference values $V/t=505$ and $\delta^e/t=10$, which we use for the following time evolution, see Fig.~\ref{fig: numeric_results}c and~\ref{fig: numeric_results}d.

To validate our proposed gauge protection mechanism in a more realistic setting, we perform exact time evolution of the minimal toy model described in Fig.~\ref{fig: AELA and superexchange}c using the microscopic Hamiltonian Eq.~\eqref{main text: Hamiltonian total}.
We consider a system of two neighboring vertices, we set $t=t^e=t^g$ and $\delta^e=10 \, t$. We use $V_{\mathrm{ex}}/V = 790/1010$, $U_{gg}/V = 199/1010$ and $U_{ee}/V = 306/1010$ to be proportional to the scattering lengths in Tab.~\ref{tab:scatt_lenghts} and choose the value of $\delta^g$ determined perturbatively above to yield~$\Delta^s=0$.
The microscopic model is based on the Hamiltonian in Eq.~\eqref{main text: Hamiltonian total} and additionally includes the Zeeman splitting, Eqs.~\eqref{eq: condition ext field} and~\eqref{eq: Zeeman shift}. We use $\mu_B \, B \,  \delta g = 3V$ to guarantee sufficient protection.

The system is initialized in the state~$\ket{s_g}$, see Fig.~\ref{fig: numeric_results}c (inset), and we calculate the occupation of the low-energy states $\{ \vert s_g \rangle, \vert s_e \rangle, \vert t_g \rangle, \vert t_e \rangle \}$ from the full time-evolved state. Our results are shown in Fig.~\ref{fig: numeric_results}c. We observe that within this subspace, the singlet and triplet manifolds are well decoupled due to the introduction of the superexchange gauge protection. Hence, our simulation confirms that the system remains in a controlled regime with limited gauge breaking, consistent with the perturbative calculations.

We further compare the dynamics obtained from the microscopic time evolution to our effective model~\eqref{eq:effective_model_superexchange} 
and find excellent agreement, see Fig.~\ref{fig: numeric_results}d. For the parameters used in this simulation, the resulting characteristic timescale of the effective model is on the order of 5 tunneling times $2\pi/t$, realistic for current optical lattice experiments. We emphasize that the simulation includes the intrinsic gauge-breaking errors due to direct Hubbard exchange interactions~$V_{\mathrm{ex}}$. Therefore, our proposed gauge protection mechanism using linear Zeeman shifts and superexchange interactions allows us to still obtain an effective gauge-invariant $U(2)$ LGT with fermionic matter, as we demonstrate numerically and analytically.

To further study the efficiency of our gauge protection mechanism, we tune the parameter~$V/t$. Thus, we directly control the strength of gauge-breaking terms by increasing $V$, and consequently also the other interaction parameters, with respect to $t$ and $\delta^e$ that remain unchanged ($\delta^g$ is updated to maintain the singlet resonance as explained above).
In Fig.~\ref{fig: numeric_results}e, we show the long-time dynamics of the integrated gauge violation~$\bar{\varepsilon}(T) = \frac{1}{T}\int \varepsilon(\tilde{T}) d\tilde{T}$ ~\cite{Halimeh2020single_bodies}, where $\varepsilon(T)$ is the instantaneous gauge violation at time $T$ defined as the probability of occupation of gauge-breaking states normalized by its infinite-temperature value. Our results confirm that the gauge protection can be controlled by~$V/t$, although $\bar{\varepsilon}$ does not change much for the chosen parameters when $V$ is doubled - in agreement with the expectations of Fig.~\ref{fig: numeric_results}b. 
Additionally, Fig.~\ref{fig: numeric_results}e shows that when the ratio~$V/t$ is lowered, the gauge protection persists. 
The results obtained in this analysis on a single building block confirm that the intrinsic gauge protection mechanism is a promising starting point to achieve gauge-invariant dynamics in large-scale quantum simulations. Our scheme can be naturally implemented in an AELA setup since it relies on an external magnetic field, staggered chemical potentials, and on-site Hubbard interactions protecting it against both intrinsic and external gauge-breaking terms.

\textit{Limitations.}---
The scheme that we proposed offers promising results for the simulation of non-Abelian $U(2)$ LGTs, but it is important to recognize few inherent limitations that may impact its broader applicability and accuracy. These limitations arise mainly from the occurrence of atomic losses: while this phenomenon can actually appear during any two-body interaction, it is more pronounced for the $e$-$e$ interactions in the $e$ optical lattice. Atomic losses can clearly influence the stability of the scheme and impact the gauge invariance of the simulations, and generally affect the precision and scalability of the model in scenarios with larger fillings and higher atomic occupancies. For this reason, we propose to limit the particle occupation in the $e$-lattice to a maximum of one singlet (therefore two atoms) on each $e$-vertex. In Appendix~\ref{e-e losses} we check the occurrence of losses due to the $e$-$e$ and $e$-$g$ interactions on experimentally relevant timescales and find that this is acceptable for the range of parameters that we choose for our simulations.
Clearly, the constraint on the number of particles in the $e$-vertices generates a limitation for the implementation of more complex and interesting theories: to access a wider range of simulations, we therefore propose a second scheme, in the next section, which is based on Rydberg interactions and can ideally sustain $U(N)$ theories with higher rishon-number occupation.

\section{Experimental implementation 2: Rydberg platform} 
\begin{figure}[t!]
    \centering  
    \includegraphics[width=0.97\linewidth]{ 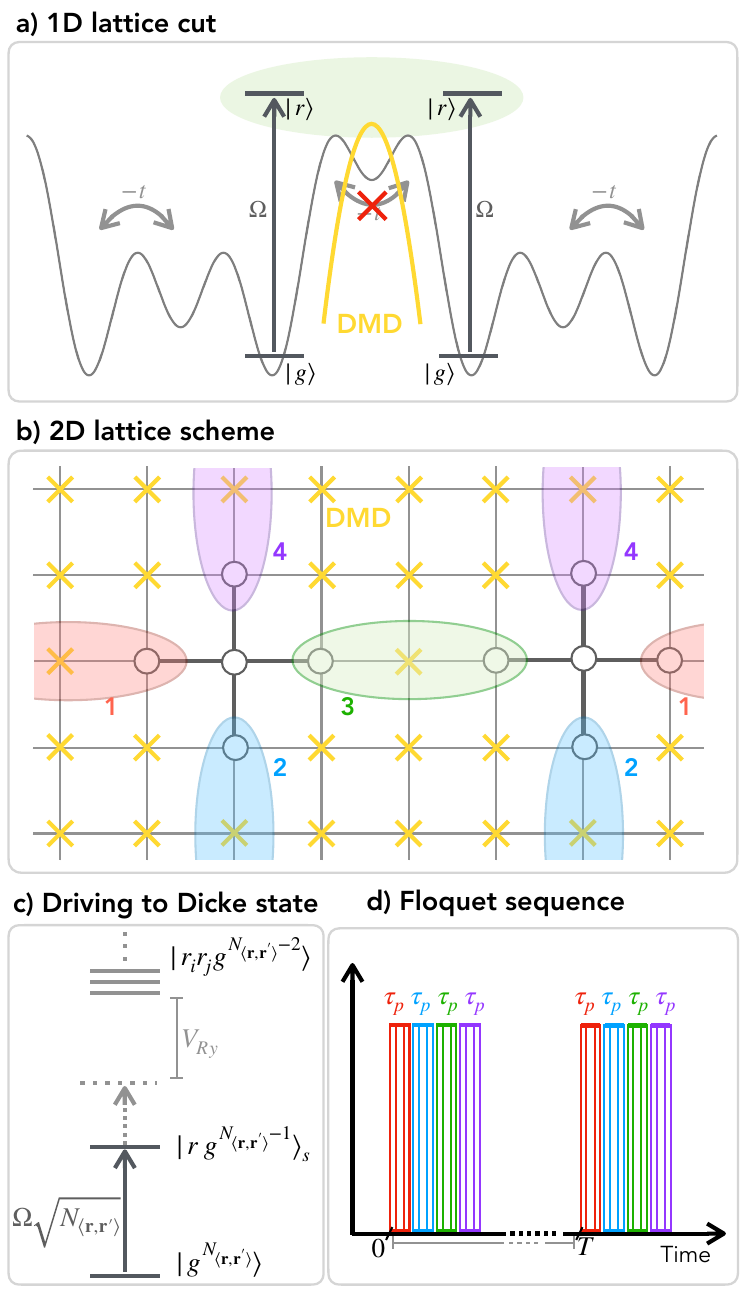}
    \caption{\textbf{Digital-analog hybrid Rydberg scheme} \textbf{a)} We show a 1D cut of the lattice setup: atoms in the rishon sites of a link are excited to the Rydberg states with Rabi frequency $\Omega$, they interact in a Rydberg-blockade regime, and are subsequently de-excited The hopping between different vertices is eliminated through optical potentials, e.g. using DMDs~\cite{Stuart_2018}. \textbf{b)} In the 2D lattice scheme we identify with different colours the four links of each vertex: we propose to excite the rishon sites of these links through four separated pulses in order to ensure that there are no interactions between rishon excitations of different links. \textbf{c)} A set of $N_{\langle \mathbf{r},\mathbf{r'} \rangle}$ atoms in the blockade regime undergoes a collective Rabi oscillation with frequency $\Omega\sqrt{N_{\langle \mathbf{r},\mathbf{r'} \rangle}}$, leading to a coupling between the ground state $\vert g^{N_{\langle \mathbf{r},\mathbf{r'} \rangle}} \rangle$ and the symmetric Dicke state $\vert r \,g^{N_{\langle \mathbf{r},\mathbf{r'} \rangle}-1} \rangle_s$. Our scheme utilizes the corresponding collective AC-Stark shifts $\propto \Omega\sqrt{N_{\langle \mathbf{r},\mathbf{r'} \rangle}}$ \textbf{d)} We show the Floquet sequence of period $T$ used in the scheme: initially, the four different pulses of duration $\tau_p$ act separately on the different links~--see \textbf{b)}--, then the systems evolves under its free Hamiltonian for the rest of the period $T$.
    }
    \label{fig: Rydberg general scheme}
\end{figure}

Next, we focus on models with more than one rishon per link~$\mathcal{N}>1$, i.e. non-Abelian $U(N)$ LGTs with a richer and more complex representation of the gauge group. 
When referring to the scheme as digital-analog hybrid implementation, we mean a combination of analogue time evolution with a digitized interaction gate: we propose to use a combination of a two-dimensional square lattice and optical tweezer arrays to engineer a periodically driven Floquet scheme~\cite{Santos2000,Nishad2023, Zhao2023, Weckesser2024}, designed to yield the desired gauge-invariant dynamics for AELAs. 
Specifically, the tweezers are employed to~\textquotedblleft cookie-cut\textquotedblright~some of the sites in the lattice, in order to avoid the hopping of particles within neighboring vertices, see Fig.~\ref{fig: Rydberg general scheme}a,b~\cite{Young2022, Stuart_2018}.
\label{sec:experimental-impl Rydberg}

In order to obtain a fully scalable scheme, in terms of color index $N$ and number of rishons per link $\mathcal{N}$, we pursue a different strategy from before. Our goal is to implement a generalization of the idealized $SU(N)$ invariant Hamiltonian, Eqs.~\eqref{eq:Basic Ham} and~\eqref{eq: general intra-vertex hopping}, fine-tuned as much as possible to be $SU(N)$-invariant. To this end, we propose to realize a generalized gauge-invariant multi-rishon interaction $\propto V$ using a Rydberg scheme that completely avoids gauge-symmetry breaking spin-exchange interactions $V_{\mathrm{ex}}$. Indeed, this second scheme works only with AELAs in their electronic ground state $\ket{g}$ with spatially non-overlapping rishon sites on each link, which furthermore completely eliminates the problem of $e$-$e$ losses, contributing to its full scalability. Since deviations from the perfect $SU(N)$ invariance of the new interactions $\propto V$ cannot be ruled out completely, we assume that residual $g$-$g$ superexchange interactions similar to those described in the previous scheme will serve as a gauge-protection mechanism. The main goal of this section is to introduce a fully scalable, conceptually new alternative scheme for implementing non-Abelian QLMs, and we will devote a detailed study of the efficiency of the gauge-protection scheme to future work.

\textbf{\textit{SU(N)-invariant Dicke interaction.}---}To obtain the interaction and the local addressability between neighboring vertices, see Eq.~\eqref{eq:Basic Ham}, necessary to achieve the gauge-invariant dynamics (see Sec.~\ref{sec:model}), we propose to use a sequence of fast excitations to the Rydberg states of the atoms in neighboring rishon sites. 
The goal of this scheme is to achieve an interaction term in the Hamiltonian that takes the role of $V$ in Eq.~\eqref{eq:Basic Ham} and depends on the total rishon number $\hat{N}_{\langle \mathbf{r},\mathbf{r'} \rangle}$ on a link $\langle \mathbf{r},\mathbf{r'} \rangle$ in a non-linear way:
\begin{equation}    \label{eq: omega term}
    \hat{H}_{\Omega} = \Omega \sum_{\langle \mathbf{r},\mathbf{r'} \rangle} f(\hat{N}_{\langle \mathbf{r},\mathbf{r'} \rangle}).
\end{equation}
This term, proportional to $\Omega$, plays a fundamental role for the scheme, serving two goals: it enforces the rishon-number conservation since it contributes an energy that is non-linear in $\hat{N}_{\langle \mathbf{r},\mathbf{r'} \rangle}$ and therefore implements the constraint $\hat{N}_{\langle \mathbf{r},\mathbf{r'} \rangle} = \mathcal{N}$ on all links separately for large enough $\Omega$. Moreover, this allows to obtain effective perturbative dynamics induced by a microscopic intra-vertex hopping $t$ -- as explained in Sec.~\ref{sec:model}; non-linearity is required to obtain a non-zero perturbative hopping coefficient. 

The Hamiltonian that we aim to implement in the scheme is therefore comprised of the following terms:
\begin{equation}    \label{eq: ideal Hamiltonian Rydberg}
    \hat{H} = \hat{H}_{t} + \hat{H}_{\delta} + \hat{H}_{\Omega},
\end{equation}
with $\hat{H}_t$ defined as in Eq.~\eqref{eq: general intra-vertex hopping}, $\hat{H}_{\Omega}$ as in Eq.~\eqref{eq: omega term} and $ \hat{H}_{\delta} = \delta \sum_{\alpha}\sum_{\mathbf{r}} \hat{n}_{\mathbf{r},m}^{\alpha}$; in the following analysis we will assume $U_{gg}=0$ for simplicity. This ideal Hamiltonian respects the gauge invariance requirements: indeed, the intra-vertex hopping $\hat{H}_{t}$ and the potential staggering $\hat{H}_{\delta}$ do not affect the gauge symmetry, while $\hat{H}_{\Omega}$ exclusively depends on the total number of rishon excitations of a link and not on the color of the particles, allowing to couple neighboring vertices while preserving the $SU(N)$ invariance on both vertices independently.

The main challenge consists in realising the interaction $f(\hat{N}_{\langle \mathbf{r},\mathbf{r'} \rangle})$ in a cold-atom platform.
We start considering a simple isolated link and we assume the distance between the rishons of such a link to be smaller than the Rydberg blockade radius: the general idea is to resonantly couple the ground state of the atomic system to the symmetric Dicke state with exactly one Rydberg excitation shared by all the atoms of the link.
In particular, we consider a system of $N_{\langle \mathbf{r},\mathbf{r'} \rangle}$ atoms in the two rishon sites of a link, whose ground states $\vert g \rangle_i$ are resonantly coupled to Rydberg states $\vert r \rangle_i$ with Rabi frequency $\Omega_i$, mediating van-der-Waals interactions, where $i=1,...,N_{\langle \mathbf{r},\mathbf{r'} \rangle}$ labels the atoms.
The strong Rydberg blockade (due to interactions $\propto C_6$) between two excited atoms prevents the simultaneous excitation of multiple particles to the states $\vert r_ir_j \,g^{N_{\langle \mathbf{r},\mathbf{r'} \rangle}-2} \rangle$, $\forall i\neq j$ (see Fig.~\ref{fig: Rydberg general scheme}c).
If we assume $\Omega_i\equiv \Omega$ $\forall i$, the system oscillates between the ground state $\vert g^{N_r} \rangle$ and the resonant symmetric Dicke-state $\vert r \,g^{N_{\langle \mathbf{r},\mathbf{r'} \rangle}-1} \rangle_s \equiv \frac{1}{\sqrt{N_{\langle \mathbf{r},\mathbf{r'} \rangle}}}\sum_{i}\vert r_i \,g^{N_{\langle \mathbf{r},\mathbf{r'} \rangle}-1} \rangle$ in an effective collective Rabi oscillation with frequency $\Omega^*=\Omega \sqrt{N_{\langle \mathbf{r},\mathbf{r'} \rangle}}$~\cite{Keating2016, Browaeys2020}.   
This reduced system can be diagonalized in the basis $ \{ \vert g^{N_{\langle \mathbf{r},\mathbf{r'} \rangle}} \rangle, \vert r \,g^{N_{\langle \mathbf{r},\mathbf{r'} \rangle}-1} \rangle_s\}$, and the instant eigenvalues are $E_{\pm}= \pm\Omega \sqrt{N_{\langle \mathbf{r},\mathbf{r'} \rangle}}$ with relative eigenvectors $\vert \varphi_{\pm} \rangle = \frac{1}{\sqrt{2}}(\vert g^{N_{\langle \mathbf{r},\mathbf{r'} \rangle}} \rangle \pm \vert r \,g^{N_{\langle \mathbf{r},\mathbf{r'} \rangle}-1} \rangle_s$). In our scheme, we want to exploit the non-linear dependence of these eigenenergies on $N_{\langle \mathbf{r},\mathbf{r'} \rangle}$ to implement the interaction $f(\hat{N}_{\langle \mathbf{r},\mathbf{r'} \rangle})$.

To maintain gauge invariance, spin-\textit{dependent} interactions between vertices have to be suppressed, see Sec.~\ref{sec:model}. 
While the amplitude of scattering interactions between AELAs in the electronic ground state are independent of the nuclear spin, this is in general not true for the pair-wise interactions between atoms excited to the Rydberg states~\cite{peper2024}. This implies that the coefficients $C_6^{ij}$ of the van-der-Waals interactions arising for the states $\vert r_ir_j \,g^{N_{\langle \mathbf{r},\mathbf{r'} \rangle}-2} \rangle$ depend on the spins of the excited atoms (i.e. if the atom of index $i$ has spin $\alpha$ and the atom $j$ has spin $\beta$, then $C_6^{ij}\equiv C_6^{\alpha \beta}$). 
Moreover, atoms in different spin states, once excited to their Rydberg states, are subjected to different hyperfine energy splitting~\cite{peper2024, Robicheaux2018} and the corresponding Rabi frequencies also depend on the spin, i.e., $\Omega_i \equiv \Omega_{\alpha}$ for the atom of index $i$ with spin $\alpha$. 
To ensure that the collective Rabi dynamics takes place, and to achieve the desired spin-independent interactions, our scheme is designed to bypass these issues: the Rydberg atoms on the links interact in a blockade regime, which requires that all $C_6^{\alpha \beta}$ are sufficiently large, such that their precise values become irrelevant for the effective interaction. Moreover, Rydberg states must be excited with a beam of multicolor lasers, which can be correctly tuned in order to achieve the same Rabi frequency for all spins, $\Omega_{\alpha} \equiv \Omega $, $\forall \alpha$~\cite{peper2024, Wu2022, Tang_2024} (see more details in Appendix~\ref{sec: Rydberg multicolor drive}).

\textbf{\textit{Floquet sequence.}---} To include the coherent dynamics within the vertices and to ensure that the collective Rabi oscillations are correctly applied on each link, we propose to utilize the following Floquet sequence:~(i) On each link, Rydberg states are quickly excited, held briefly, and de-excited for a total time $\tau_p$. We assume large enough Rabi frequencies, on a $\SI{}{\MHz}$ scale, such that $\tau_p$ can be on a $\SI{}{\micro s}$ scale - much shorter than a typical $\SI{}{m s}$ scale of a tunnelling time in the optical lattice;~(ii) In order to avoid interactions between Rydberg atoms in different links, we distribute the entire square lattice into four inequivalent types of links, as shown in Fig.~\ref{fig: Rydberg general scheme}b. This is done in such a way that a link of a given type only has links on other types of neighbours. Individual local Rydberg excitation and de-excitation pulses are then applied sequentially for each type of link.~(iii) After these four pulses, the system undergoes free dynamics described by the Hamiltonian
\begin{equation}
    \hat{H}_{\mathrm{free}} = -t \sum_{\alpha}\sum_{\mathbf{r}, \mathbf{k}} \sum_{\mathbf{v}=\pm\mathbf{k}} \bigl( \hat{c}_{\mathbf{r}, \mathbf{v}}^{\alpha \dagger} \, \hat{\psi}_{\mathbf{r}}^{\alpha} + \mathrm{h.c.} \bigr) + \delta \sum_{\alpha}\sum_{\mathbf{r}} \hat{n}_{\mathbf{r},m}^{\alpha},
\end{equation}
and the entire cycle is repeated after a determined Floquet period $T \gg \tau_{p}$, see Fig.~\ref{fig: Rydberg general scheme}d. This strategy addresses two main problems that would otherwise arise: firstly, the short duration of the excitation and the following de-excitation from the Rydberg level allows to reduce the probability of significant losses in the Rydberg-excited states since the experimental parameters and $\tau_p$ can be fine-tuned accordingly to achieve this; secondly, the application of four separated pulses allows to exclude the interaction between rishon sites that do not belong to the same link and further protects against losses, as the Rydberg atoms are significantly distant from each other. 

Each link $\langle \mathbf{r},\mathbf{r'} \rangle$ can now  be described as the ideal system introduced in the previous paragraph. The pulse applied on the link must guarantee a controlled evolution of the system. Therefore, we need to implement a fast but adiabatic process that allows the evolution from the ground state $\vert g^{N_{\langle \mathbf{r},\mathbf{r'} \rangle}} \rangle$ to its antisymmetric superposition $\ket{\varphi_-}$ with the Dicke state, which contributes with an energy equivalent to $-\Omega \sqrt{N_{\langle \mathbf{r},\mathbf{r'} \rangle}}$. Here we do not include a deep analysis of the adiabatic evolution but assume that the dynamical phase can be neglected or controlled through a spin echo sequence~\cite{Mitra2020, Jaksch2000, Conolly1989, Gregefalk2022}: we therefore approximate the time evolution operator of the pulse on the link $\langle \mathbf{r},\mathbf{r'} \rangle$  with $\hat{U}_{\langle \mathbf{r},\mathbf{r'} \rangle}= e^{i\Omega\sqrt{N_{\langle \mathbf{r},\mathbf{r'} \rangle}}\tau_{p}} \vert g^{N_{\langle \mathbf{r},\mathbf{r'} \rangle}} \rangle \langle g^{N_{\langle \mathbf{r},\mathbf{r'} \rangle}} \vert$, see more details in Appendix~\ref{SM sec: Adiabatic sweep}. 

We can define the total period of the Floquet sequence as $T = N_p\tau_p$, with $N_p \gg 1$ and the time-evolution operator over the Floquet period as $\hat{U}(T) = \hat{U}_{\mathrm{free}} \hat{U}_4 \hat{U}_3 \hat{U}_2 \hat{U}_1$, where $\hat{U}_{\mathrm{free}} = e^{-i\hat{H}_{\mathrm{free}}(N_p-4)\tau_{p}} \approx e^{-i\hat{H}_{\mathrm{free}}T}$~\cite{Goldman2014, Jaksch2000, Chinni2022}. In the limit of $\tau_p \ll T \ll 1/t$, with $t$ the hopping coefficient of the free Hamiltonian, we use the Trotter decomposition method to obtain the effective Hamiltonian of the periodically driven system $\hat{U}(T) \approx e^{-i\hat{H}_{\mathrm{Floq}}T}$ up to errors $\mathcal{O}(\tau_p^2)$, obtaining
\begin{equation}    \label{eq: hamiltonian floquet}
\begin{split}
    \hat{H}_{\mathrm{Floq}} \approx & -t \sum_{\alpha}\sum_{\mathbf{r}, \mathbf{k}} \sum_{\mathbf{v}=\pm\mathbf{k}} \bigl( \hat{c}_{\mathbf{r}, \mathbf{v}}^{\alpha \dagger} \, \hat{\psi}_{\mathbf{r}}^{\alpha} + \mathrm{h.c.} \bigr) \\
    & + \delta \sum_{\alpha}\sum_{\mathbf{r}} \hat{n}_{\mathbf{r},m}^{\alpha} - \frac{\Omega}{N_p} \sum_{\langle \mathbf{r},\mathbf{r'} \rangle} \sqrt{\hat{N}_{\langle \mathbf{r},\mathbf{r'} \rangle}},
\end{split}
\end{equation}
where $\hat{N}_{\langle \mathbf{r},\mathbf{r'} \rangle} = \sum_{\alpha \beta}\hat{n}_{\mathbf{r},\mathbf{k}}^{\alpha}+ \hat{n}_{\mathbf{r'}, - \mathbf{k}}^{\beta}$ is the total number of atoms on the rishon sites of the link $\langle \mathbf{r},\mathbf{r'} \rangle$.
This effective Hamiltonian has the same structure of the ideal one of Eq.~\eqref{eq: ideal Hamiltonian Rydberg} and therefore can be used to implement different types of non-Abelian dynamics.
The emerging term, $\hat{H}_{\Omega}$, plays the same role as the Hubbard ($U$) and extended Hubbard ($V$) interactions in Eq.~\eqref{eq:Basic Ham} with the difference that it is proportional to the square root of the number of rishons. Despite the different form, this term still fulfils the general requirement of non-linear proportionality to $N_{\langle \mathbf{r},\mathbf{r'} \rangle}$, necessary to enforce the correct rishon number $\mathcal{N}$ constraint, as explained in the previous paragraph.

\textbf{\textit{Effective theory.}---}
From Eq.~\eqref{eq: hamiltonian floquet} we can determine the effective hopping of the gauge-invariant perturbative dynamics for a fixed rishon number $\mathcal{N}$,
\begin{equation}
\begin{split}
    t_{\mathrm{eff}} = & \frac{t^2}{\delta +(\sqrt{\mathcal{N}} -\sqrt{\mathcal{N}+1})\frac{\Omega}{N_p}} \\
    &- \frac{t^2}{\delta +(\sqrt{\mathcal{N}-1}- \sqrt{\mathcal{N}})\frac{\Omega}{N_p}}.
\end{split}
\end{equation}
The coefficient is therefore influenced by the ratio $\frac{\Omega}{N_p \delta}$, which governs the scale of the dynamics.
Typically, we assume $\Omega$ to be on the order of $\mathcal{O}(\SI{}{\MHz})$ or $\mathcal{O}(\SI{10}{\MHz})$, while $t$ is on the order of $\mathcal{O}(\SI{}{\kHz})$: by choosing the values of $N_p$ and $\delta$ while respecting all the necessary requirements for the correctness of the scheme, we gain flexibility in tuning the effective hopping strength. This method is particularly efficient for small $\mathcal{N}$, leading to an effective dynamics observable on the timescales of the experiment: e.g. the strength of the effective hopping can be tuned to values between $\vert t_{\mathrm{eff}}\vert \approx \vert t\vert /10$ to $\vert t_{\mathrm{eff}}\vert \approx \vert t\vert /100$, assuming $\mathcal{N}=3$ or $\mathcal{N}=4$, $N_p\approx100$ and $\delta \approx \mathcal{O}(\SI{}{\kHz})$ or $\mathcal{O}(\SI{10}{\kHz})$.

Few comments on the effective implementation and the possible limitations of this platform are necessary. Atoms excited in their Rydberg states are inevitably affected by losses, through our Floquet scheme we aim to strongly limit the time that atoms spend in the excited states: if we assume the duration of the excitation pulses $\tau_p$ to be limited to a maximum of a fraction of a $\SI{}{\micro s}$~\cite{Levine2019}, and we estimate the theoretical lifetime of the atoms in the Rydberg states to be of the order of $\SI{1e2}{\micro s}$, we can expect that a sufficient number of Floquet cycles can be performed before observing experimentally relevant atomic losses. 
Moreover, since generally optical traps result in different trapping potentials for the ground and Rydberg states we expect heating and higher-band excitations; 
we note, however, that our proposal is based on AELAs, which offer the chance to realize magic traps for Rydberg atoms~\cite{Zhang2011, Ahlheit2025}, improving the reliability of the simulations.

\section{Summary and Outlook} 
\label{sec:summary}
Simulating non-Abelian LGTs with dynamical matter remains a central goal of contemporary quantum simulators~\cite{Wiese_review,Banuls2020,aidelsburger2021cold,Halimeh2025}. In this article we have analyzed a general approach to simulate (2+1)D non-Abelian $U(N)$ LGTs using AELAs in optical lattices. As a starting point we consider an AELA scheme previously proposed in the literature~\cite{banerjee2013}, which we show to be limited by gauge-symmetry breaking interactions. However, we demonstrate how the latter can be overcome by exploiting various protection terms in the microscopic Hamiltonian. Specifically, we have developed a comprehensive model that allows for the gauge-invariant hopping of matter while maintaining the system's gauge symmetry. The target non-Abelian LGT emerges as an effective theory within a protected region of the Hilbert space, where different energy terms enforce the separation between physical sectors -- with the correct rishon number and color charge -- and unphysical ones. We discussed a minimal experimentally relevant initialization of the lattice to achieve quantum simulations with genuinely non-Abelian character.

After outlining the general framework, we have proposed two specific experimental implementations of our scheme. The first makes use of AELAs in their atomic ground and metastable clock states and is closest to existing experiments~\cite{Darkwah_Oppong2022, Riegger2018}, but subject to several fundamental limitations. Among them are strong losses occurring when two metastable-excited clock state atoms occupy the same lattice site~\cite{scazza2014observation}. This leads to probably the most restrictive condition, namely, that two clock-state atoms cannot occupy the same site, which motivated us to study our first scheme only for cases with one rishon per link. Likewise, for this first scheme, it is advisable to work with only one singlet per clock-state vertex in order to limit undesired losses; but as we have demonstrated, this still allows for genuinely non-Abelian simulations if more than one singlet is allowed on ground-state vertices. For similar reasons, we also restricted our detailed analysis of the AELA scheme to the ${\rm U}(2)$ gauge group, although in principle larger groups ${\rm U}(N)$ with $N=3,4,..$ could be experimentally realized. The closeness of the first proposed scheme to realistic experimental setups and the built-in gauge-protection scheme we describe in this article nevertheless, makes it highly promising for performing large-scale quantum simulation studies of a restricted class of non-Abelian quantum link models.

Unless the strong collisional losses, which have only been measured experimentally in the recent years~\cite{scazza2014observation}, can somehow be overcome, we conclude that the scheme originally introduced by Banerjee et al.~\cite{banerjee2013}, on which our first proposal is also based, cannot be used to realize scalable quantum link models with tunable gauge-invariant couplings and more than one rishon per link~\cite{Wiese_review}. For this reason, in the last part of our article, we devised a digital-analog hybrid scheme that implements the required gauge-invariant couplings through Rydberg interactions in the blockade regime~\cite{Keating2016}. This allows to work only with ground-state atoms. As we have demonstrated, by fine-tuning all Rabi-couplings to the Rydberg states corresponding to different colors - realized through nuclear spin states - a fully scalable scheme for simulating non-Abelian quantum link models is obtained. We expect that gauge-symmetry breaking interactions are significantly weaker in this setup than in the first AELA scheme we analyzed, but a detailed study of the atomic losses in the excited states and the general stability of the scheme is necessary for a full experimental implementation and will be devoted to future work. 

Our work paves the way for a first experimental study of a large-scale non-Abelian quantum link model~\cite{Brower1999}. A next major step that will be necessary to explore the rich phase diagram of such models, in particular in the deconfined regime, is the implementation of strong plaquette terms~\cite{homeier2021,feldmeier2024,Meth2025}. In the schemes proposed here so far, only perturbatively small plaquette terms can be expected~\cite{Wang2025}, and we believe that further extensions of the digital-analog hybrid scheme we introduced are most promising for reaching this task. On the theoretical side, only few quantitative results exist on the detailed structure of the equilibrium phase diagrams of the resulting non-Abelian quantum link models~\cite{Banerjee2018}. In contrast to the field-theoretical models studied in particle physics, the quantum link models discussed here are subject to strong lattice and discretization effects due to the finite dimension of the link Hilbert space, which requires advanced numerical methods or quantum simulation platforms to reach a full quantitative description. Finally, quantum simulators offer new perspectives, such as studies of far-from equilibrium dynamics~\cite{Homeier_inprep, gyawali2024, schuhmacher2025, davoudi2025}.

\section*{Acknowledgements}
We are grateful for inspiring discussions with P. Preiss and F. Scazza and helpful comments from P. Weckesser on the Rydberg implementation scheme.
This work was supported by the QuantERA grant DYNAMITE, by the Deutsche Forschungsgemeinschaft (DFG, German Research Foundation) under project number 499183856. This project was funded within the QuantERA II Programme that has received funding from the European Union’s Horizon 2020 research and innovation programme under Grand Agreement No 101017733. This project has received funding from the European Research Council (ERC) under the European Union’s Horizon 2020 research and innovation programm (Grant Agreement no 948141) — ERC Starting Grant SimUcQuam. This project was funded by the Deutsche Forschungsgemeinschaft (DFG, German Research Foundation) under Germany's Excellence Strategy -- EXC-2111 -- 390814868.
G.D.P. aknowledges funding from the initiative Munich Quantum Valley and is supported by the Bavarian State Ministry of Science and the Arts through the Hightech Agenda Bayern Plus.
L.H. is supported by the Simons Collaboration on Ultra-Quantum Matter, which is a grant from the Simons Foundation (651440).
J.C.H. further acknowledges funding by the Max Planck Society and the European Research Council (ERC) under the European Union’s Horizon Europe research and innovation program (Grant Agreement No.~101165667)—ERC Starting Grant QuSiGauge.
M.A. further acknowledges funding under the Horizon Europe programme HORIZON-CL4-2022-QUANTUM-02-SGA via the project 101113690 (PASQuanS2.1).

\section*{Author contributions}
G.D.P. conducted the analytical and numerical simulations, G.D.P., L.H. and F.G conceived the conceptual ideas, with input from M.A. on the experimental realization and J.C.H on the theoretical scheme. L.H. and F.G supervised, all authors contributed to writing the manuscript.
\appendix 
\section{Mapping to Abelian Models} \label{supp: Mapping to Abelian models}
\begin{figure*}[t!]
\centering
\includegraphics[width=\textwidth]{ 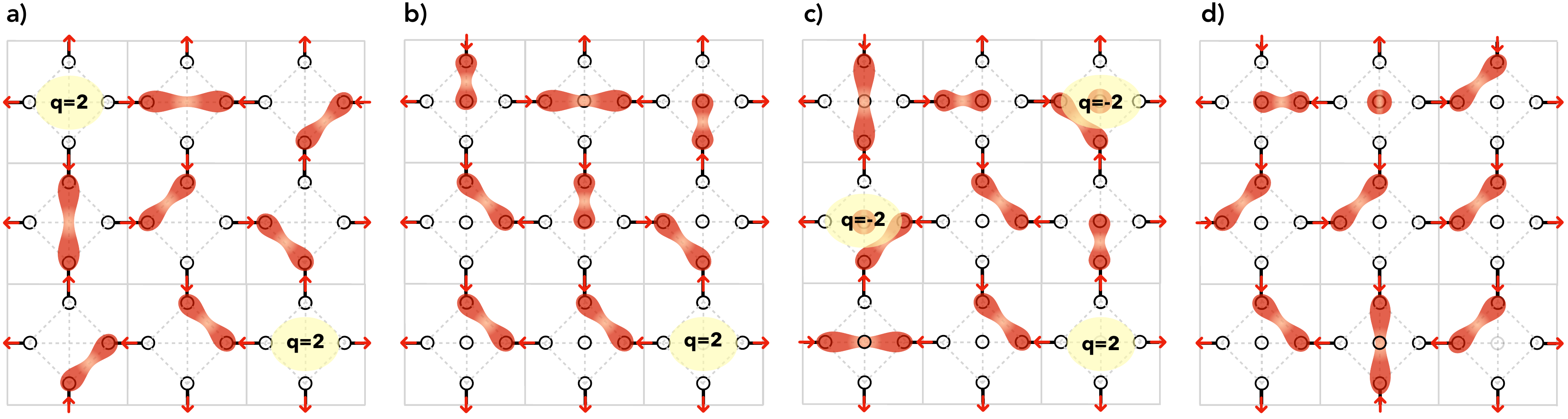}
\caption{\textbf{Abelian configurations.} 
We show examples of simple configurations that map to an Abelian theory, i.e.~configurations where the locations of the fermions as well as their spin state can be uniquely determined from the link variables (red arrows). \textbf{a)} The first scheme is characterized by the absence of matter sites, each vertex has one or zero singlets that correspond to a background charge of $q=0$ or $q=2$, respectively. \textbf{b)} In this case, we allow for matter excitations and each vertex contains one or zero singlets (background charge of $q=0$ or $q=2$). \textbf{c)} Here, each vertex can have either one or two singlets, but in the latter case one of the two singlets is constrained to the matter site and thus corresponds to a background charge of $q=-2$. \textbf{d)} Here we allow one or zero singlets per vertex including matter excitations with double occupancy. }
\label{fig: U(1) abelians}
\end{figure*}
In the main text, we have introduced a scheme to identify a discrete spin-variable of the links given the position of the particles in the lattice. Here we show some examples of configurations that map to an Abelian theory.
The first one has only one singlet per vertex and no matter excitations, as illustrated in Fig.~\ref{fig: U(1) abelians}a. Compared to the general discussion in the main text, here the central \textit{matter} site is absent.
Doublons are not permitted on any site, meaning the singlets at each vertex can be represented by an equivalent state consisting of two hard-core bosons. As a result, the atomic spin becomes irrelevant to the scheme's description.
We identify the vertices initialized without any atom with a background charge $q_{\mathbf{r}}^0 = 2$ in order to fulfil the Gauss's laws, i.e. $\sum_{i=x,y} \bigl( \hat{E}_{\mathbf{r}, \mathbf{r}+\mathbf{k}_i} - \hat{E}_{\mathbf{r}-\mathbf{k}_i, \mathbf{r}} \bigr) - \hat{\psi}_{\mathbf{r}}^{\dagger}\hat{\psi}_{\mathbf{r}} = 2 $.
Because of the local $U(1)$ symmetry of the system, the charge remains conserved as a static background charge.
The next case, shown in Fig.~\ref{fig: U(1) abelians}b, includes matter excitations in the vertices.
If all vertices are initialized to have only one or zero singlets per vertex, the entire lattice configuration can effectively be translated into an Abelian theory fully characterized by the spins on the links.
Another possible configuration allows two singlets on a vertex but we enforce a double occupancy on the matter site in that vertex (Fig.~\ref{fig: U(1) abelians}c). 
The vertices with two singlets would be associated with a background charge $q_{\mathbf{r}}^0 = -2$ as $\sum_{i=x,y} \bigl( \hat{E}_{\mathbf{r}, \mathbf{r}+\mathbf{k}_i} - \hat{E}_{\mathbf{r}-\mathbf{k}_i, \mathbf{r}} \bigr) - \hat{\psi}_{\mathbf{r}}^{\dagger}\hat{\psi}_{\mathbf{r}} = -2$.
The last configuration, shown in Fig~\ref{fig: U(1) abelians}d, has the constraint of one or zero singlets per vertex and allows for double occupancy on the matter site. In this case, only the empty vertices will have a background charge $q_{\mathbf{r}}^0 = 2$.
\section{Tight binding Hamiltonian and effective theory}\label{supp: effective theory}
In the main text, we introduced the tight binding discrete Hamiltonian as Eq.~\eqref{main text: Hamiltonian total}. Here we specify the individual components in detail. 
The single particle terms are defined as
\begin{equation} 
\label{support mat: H_eps}
    \hat{H}_{\epsilon} =  \sum_{\alpha} \sum_{\mathbf{r}_{\epsilon}, \mathbf{k}} \sum_{\mathbf{v}=\pm \mathbf{k}}  \bigg[-t^{\epsilon} \, \bigr( \hat{\psi}_{\mathbf{\mathbf{r}_{\epsilon}}}^{\alpha \dagger} \, \hat{c}_{\mathbf{\mathbf{r}_{\epsilon},\mathbf{v}}}^{\alpha} + \mathrm{h.c.} \bigl) + \, {\delta}^{\epsilon} \, \hat{\psi}_{\mathbf{r}_{\epsilon}}^{\alpha \dagger} \, \hat{\psi}_{\mathbf{r}_{\epsilon}}^{\alpha} \bigg],
\end{equation}
while the interaction terms read
\begin{equation} 
\label{support mat: H_epF_eps}
    \hat{H}_{\epsilon \epsilon} = \sum_{\alpha<\beta} \sum_{\mathbf{r}_{\epsilon}, \mathbf{k}} \sum_{\mathbf{v}=0,\pm \mathbf{k}} U_{\epsilon \epsilon} \hat{n}_{\mathbf{r}_{\epsilon}, \mathbf{v}}^{\alpha} \, \hat{n}_{\mathbf{r}_{\epsilon}, \mathbf{v}}^{\beta} ,
\end{equation}
\begin{equation} 
\label{support mat: H_g_e}
    \begin{split}
    \hat{H}_{ge} = &\sum_{\alpha,\beta} \sum_{\langle \mathbf{r}_g, \mathbf{r}_e \rangle} (V \, \hat{c}_{\mathbf{r}_g, \mathbf{k}}^{\alpha \dagger}\hat{c}_{\mathbf{r}_g, \mathbf{k}}^{\alpha}\hat{c}_{\mathbf{r}_e, -\mathbf{k}}^{\beta \dagger}\hat{c}_{\mathbf{r}_e, -\mathbf{k}}^{\beta}  \\ &+ V_{\mathrm{ex}} \, \hat{c}_{\mathbf{r}_g, -\mathbf{k}}^{\alpha \dagger}\hat{c}_{\mathbf{r}_e, \mathbf{k}}^{\beta \dagger}\hat{c}_{\mathbf{r}_g, -\mathbf{k}}^{\beta}\hat{c}_{\mathbf{r}_e, \mathbf{k}}^{\alpha}),
    \end{split}
\end{equation}
where we use $\hat{c}^{\alpha}_{\mathbf{r}_{\epsilon}, \mathrm{v}}$ ($\hat{c}_{\mathbf{r}_{\epsilon}, \alpha}^{\dagger}$) as generic lattice fermionic operators that create/annihilate an atom on a site described by a Wannier function centred at position $\mathbf{r}_{\epsilon}$ ($\epsilon = g,e$) and with spin ${\alpha}$; further we denote by $\hat{n}_{\mathbf{r}_{\epsilon}, \mathbf{v}=0}^{\alpha} = \hat{\psi}_{\mathbf{r}_{\epsilon}}^{\alpha \dagger} \hat{\psi}_{\mathbf{r}_{\epsilon}}^{\alpha}$ the matter density on vertex $\mathbf{r}_{\epsilon}$.
The strength of the two-body contact interactions $V$, $V_{\mathrm{ex}}$, $U_{gg}$ and $U_{ee}$ depend on the spatial shape and overlap of the Wannier functions of the lattice sites and are proportional to the interaction parameters $g_X = \frac{4\pi \hbar^2 a_X}{m}$, with $m$ the atomic mass and  $a_X$ the four scattering lengths of the interaction channels, where $X=gg, ee, eg^+, eg^-$, and where $V = (U_{eg}^+ + U_{eg}^-)/2$ and $V_{\mathrm{ex}} = (U_{eg}^+ - U_{eg}^-)/2$ are combination of the symmetric and antisymmetric channel strengths \cite{scazza2014observation, surace2023}.

Now we assume the first gauge-protection mechanism, i.e., the Zeeman protection, to be already present and from this microscopic Hamiltonian we can calculate the coefficients that characterize the perturbation theory, shown in Tab.~\ref{tab: effective theory}, where we did not include the superexchange-induced terms, as they are already shown in the main text. We limit the perturbative processes to those on the vertices included in the toy-model simulation: $t_{\mathrm{eff}}$ is the coupling coefficient between the two singlet (triplet) states $\ket{s_g}$ and $\ket{s_e}$ ($\ket{t_g}$ and $\ket{t_e}$) while $\lambda_{\epsilon \epsilon'}$ are the couplings between the singlet and triplet sectors and $\delta^{\epsilon}_{\mathrm{eff}} $ are the corrections to the total energy $\delta^{\epsilon}$. Since $t_{\mathrm{eff}}$ and $\lambda_{ge}$ couple states with different energy (at $0^{\mathrm{th}}$ order), we used the symmetric $2^{\mathrm{nd}}$ order matrix element expression.  
\begin{table*}[t!]
\centering
\begin{tabular}{m{0.5\textwidth} m{0.3\textwidth}} 
\hline
\centering
\textbf{Total contribution} & \textbf{Effective theory term} \\
\hline
\centering
\begin{equation} 
    t_{\mathrm{eff}} = -\frac{t^g t^e}{2} \sum_{\epsilon=g,e} \biggl(\frac{1}{\delta^{\epsilon}}  +\frac{1}{2} \biggl(\frac{1}{V-\delta^{\epsilon}} + 
     \frac{1}{V-V_{\mathrm{ex}}-\delta^{\epsilon}} \biggr) \biggr) \nonumber
\end{equation} 
&  \includegraphics[width=0.8\linewidth]{ 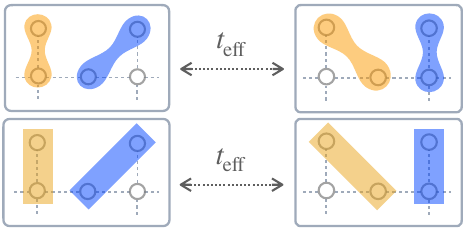} \\
\hline
\centering
\begin{equation} 
    \lambda_{\epsilon \epsilon} = -\frac{1}{2} \biggl( \frac{(t^{\epsilon})^2}{\delta^{\epsilon} - V} - \frac{(t^{\epsilon})^2}{\delta^{\epsilon} - V +V_{\mathrm{ex}}} \biggr) \nonumber 
\end{equation}
\begin{equation}
    \lambda_{ge} = -\frac{t^gt^e}{4} \sum_{\epsilon=g,e} \biggl( \frac{1}{\delta^{\epsilon} - V} - \frac{1}{\delta^{\epsilon} - V +V_{\mathrm{ex}}} \biggr) \nonumber
\end{equation}
& \includegraphics[width=0.8\linewidth]{ 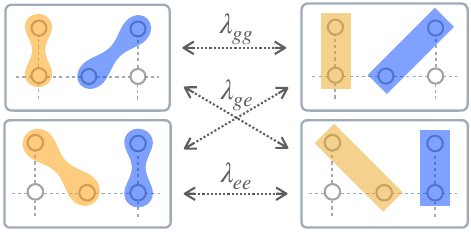} \\
\hline
\centering
\begin{equation} 
    \delta^e_{\mathrm{eff}} = -2\frac{(t^g)^2}{\delta^g} 
    -\frac{(t^e)^2}{2} \bigg( \frac{1}{V-\delta^e} +\frac{1}{V-V_{\mathrm{ex}}-\delta^e} \bigg) \nonumber
\end{equation}
& \quad \quad \quad \includegraphics[width=0.5\linewidth]{ 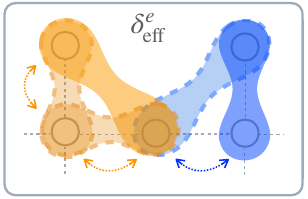} \\
\hline
\centering
\begin{equation} 
    \delta^g_{\mathrm{eff}} = -2\frac{(t^e)^2}{\delta^e} 
    -\frac{(t^g)^2}{2} \bigg( \frac{1}{V-\delta^g} +\frac{1}{V-V_{\mathrm{ex}}-\delta^g} \bigg) \nonumber
\end{equation}
& \quad \quad \quad\includegraphics[width=0.5\linewidth]{ 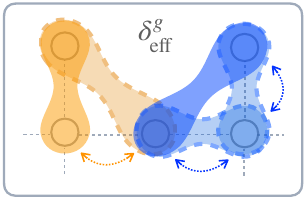} \\
\hline
\end{tabular}
\caption{\textbf{Effective model of the two-vertex building block.} We show the terms that contribute to the effective model in Eq.~\eqref{mat: Heff4x4}. The coefficients of the effective singlet hopping ($t_{\mathrm{eff}}$) and the coupling between singlet and triplet sectors ($\lambda_{ge}$) are calculated with corrected second-order matrix element expression in perturbation theory, since $\delta^g$ can be different from $\delta^e$. For reference, the values here represented are those used in Eq.~\eqref{eq:effective_model_superexchange}, while for Eq.~\eqref{mat: Heff4x4} it is sufficient to impose $t^e=t^g=t$ and $\delta^e=\delta^g=\delta$.} 
\label{tab: effective theory}
\end{table*}

In general, we can expect $\lambda_{gg} \approx \lambda_{ee} \approx \lambda_{ge}$: the error that would come from assuming them equal is negligible for the effective dynamics.
For example, if we consider $t^e=t^g=t$ and $\delta = \delta^e = 10t$, the difference between the coupling $\lambda_{gg}$ and $\lambda_{ee}$ can be approximated to $0$ when considering $\tilde{\delta}^{g}\ll\delta$:
\begin{equation}
    \begin{split}
        \lambda_{gg} &- \lambda_{ee} = - \frac{t^2}{2(\delta + \tilde{\delta}^g - V)}\biggl(\frac{V_{\mathrm{ex}}}{\delta + \tilde{\delta}^g - V + V_{\mathrm{ex}}}\biggr) \\ & + \frac{t^2}{2(\delta - V)}\biggl(\frac{V_{\mathrm{ex}}}{\delta  - V + V_{\mathrm{ex}}}\biggr)\\ & \approx \frac{t^2  V_{\mathrm{ex}} }{2(\delta - V) (\delta - V + V_{\mathrm{ex}}) }\biggl( \frac{\tilde{\delta}^g}{\delta - V} + \frac{\tilde{\delta}^g}{\delta - V + V_{\mathrm{ex}} }  \biggr) \\&\ll \lambda_{gg} , \lambda_{ee} .
    \end{split}
\end{equation}
\begin{figure*}[t]
    \centering
    \includegraphics[width=0.9\textwidth]{ 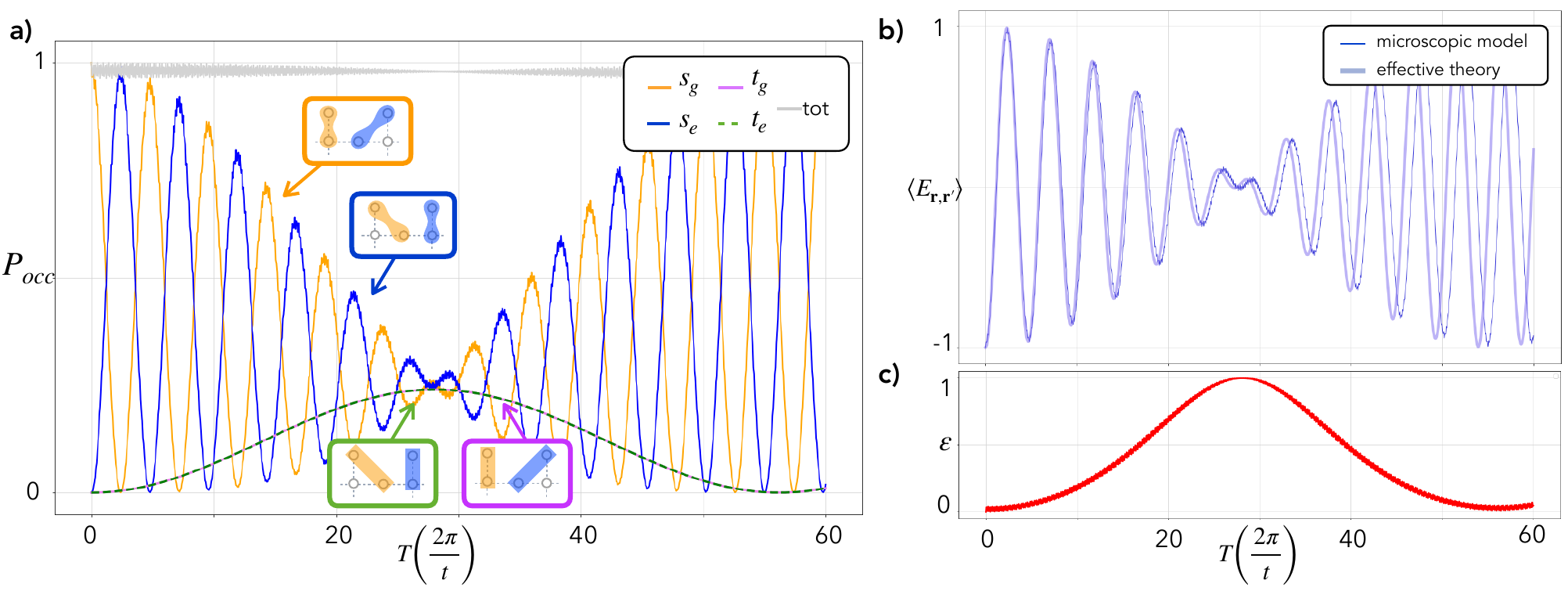}
    \caption{\textbf{Time evolution of the system without superexchange protection.} The plots show the time evolution of the two-vertex toy model presented in the main text: in this case the Zeeman-splitting mechanism is implemented and it protects from the occupation of triplet states $\ket{F^{\mathbf{r}}=1;m^{\mathbf{r}}_F=\pm1}\otimes\ket{F^{\mathbf{r'}}=1;m^{\mathbf{r'}}_F=\mp1}$, while we do not include the on-site Hubbard interactions; therefore there is no superexchange-induced protection. \textbf{a)} The first plots shows the evolution of populations in the initial state into a superposition of the singlet states (in orange and blue) and the triplet states (in green and purple). \textbf{b)} The second plot shows the evolution of the electric field in the microscopic (thin line) and effective (thick line) theory: it has a peculiar beat, which corresponds with the peak of the gauge violation $\varepsilon$, shown in red in \textbf{c)}.}
    \label{app_fig: no sup ex 1}
\end{figure*}
A similar comparison can be done with $\lambda_{ge}$.
In Sec.\ref{sec:experimental-impl}, we have mentioned how the spin dependence of the interactions between particles on the links introduces an oscillation between states in the singlet and triplet states with $m_F=0$.

From the microscopic model and considering the representation of these states in the Fock's basis, we now provide a more physical intuition for the non zero coupling $\lambda$ (see Tab.~\ref{tab: effective theory}) between the triplet and singlet sector. 
Only as a reference, we first write down the general form of the state with a singlet on both neighboring vertices,  $\ket{F^{\mathbf{r}}=0;m^{\mathbf{r}}_F=0}\otimes\ket{F^{\mathbf{r'}}=0;m^{\mathbf{r'}}_F=0}$:
\begin{equation}\label{eq: singlet in two}
    \begin{split}
        & \vert s \rangle = \vert s \rangle_g \otimes \vert s \rangle_e = \frac{\vert \uparrow \downarrow \rangle_g - \vert \downarrow \uparrow\rangle_g }{\sqrt{2}} \otimes \frac{\vert \uparrow \downarrow \rangle_e - \vert \downarrow \uparrow\rangle_e }{\sqrt{2}} \\ & \!=\! \frac{1}{2} \bigl( \vert \!\uparrow\!\textcolor{magenta}{\downarrow} \rangle_g\vert \textcolor{magenta}{\uparrow }\!\downarrow \rangle_e \!-\! \vert \! \uparrow \! \textcolor{cyan}{\downarrow} \rangle_g\vert\textcolor{cyan}{\downarrow} \! \uparrow\rangle_e \! -\vert \! \downarrow \! \textcolor{cyan}{\uparrow}\rangle_g\vert \textcolor{cyan}{\uparrow}\!\downarrow\rangle_e\!+\!\vert \! \downarrow \!\textcolor{magenta}{\uparrow}\rangle_g\vert \textcolor{magenta}{\downarrow} \! \uparrow \rangle_e \bigr),
    \end{split}
\end{equation}
and a state with spin triplets on both neighboring vertices,  $\ket{F^{\mathbf{r}}=1;m^{\mathbf{r}}_F=0}\otimes\ket{F^{\mathbf{r'}}=1;m^{\mathbf{r'}}_F=0}$:
\begin{equation}\label{eq: triplet in two}
    \begin{split}
        & \vert t \rangle = \vert t \rangle_g \otimes \vert t \rangle_e = \frac{\vert \uparrow \downarrow \rangle_g + \vert \downarrow \uparrow\rangle_g }{\sqrt{2}} \otimes \frac{\vert \uparrow \downarrow \rangle_e + \vert \downarrow \uparrow\rangle_e }{\sqrt{2}} \\ & \!=\! \frac{1}{2} \bigl( \vert \!\uparrow\!\textcolor{magenta}{\downarrow} \rangle_g\vert \textcolor{magenta}{\uparrow }\!\downarrow \rangle_e \!+\! \vert \! \uparrow \! \textcolor{cyan}{\downarrow} \rangle_g\vert\textcolor{cyan}{\downarrow} \! \uparrow\rangle_e \! +\vert \! \downarrow \! \textcolor{cyan}{\uparrow}\rangle_g\vert \textcolor{cyan}{\uparrow}\!\downarrow\rangle_e\!+\!\vert \! \downarrow \!\textcolor{magenta}{\uparrow}\rangle_g\vert \textcolor{magenta}{\downarrow} \! \uparrow \rangle_e \bigr),
    \end{split}
\end{equation}
where we coloured the spins that interact when both occupy the rishon sites, using blue for spin with the same state $m_F=\pm1/2$, and purple for those with opposite $m_F$. 
Indeed, the contribution of the atoms interacting in the rishon sites depends on their specific spin configuration: the term proportional to $V_{\mathrm{ex}}$ in Eq.~\ref{support mat: H_g_e} yields a spin exchange between two neighboring vertices when the interaction in the rishon site happens between two atoms with opposite $m_F=\pm1/2$ (purple in Eq.~\eqref{eq: singlet in two} and Eq.~\eqref{eq: triplet in two}); in that case, the Zeeman mechanism that we introduced energetically protects from the occupation of gauge-breaking states and the exchange interaction is therefore suppressed.
On the other hand, when the interaction in the rishon site happens between atoms with the same quantum number $m_F$ (blue in Eq.~\eqref{eq: singlet in two} and Eq.~\eqref{eq: triplet in two}), the term proportional to $V_{\mathrm{ex}}$ in Eq.~(\ref{support mat: H_g_e}) can be written as
\begin{equation}
    \begin{split}
    &V_{\mathrm{ex}} \sum_{\alpha,\beta} \! \bigl(\hat{c}_{\mathbf{r}_g, -\mathbf{k}}^{\alpha \dagger}\hat{c}_{\mathbf{r}_e, \mathbf{k}}^{\beta \dagger}\hat{c}_{\mathbf{r}_g, -\mathbf{k}}^{\beta}\hat{c}_{\mathbf{r}_e, \mathbf{k}}^{\alpha} \bigr)  \delta_{\beta}^{\alpha} \\ &=  V_{\mathrm{ex}} \sum_{\alpha }  \bigl(\hat{c}_{\mathbf{r}_g, -\mathbf{k}}^{\alpha \dagger}\hat{c}_{\mathbf{r}_e, \mathbf{k}}^{\alpha \dagger}\hat{c}_{\mathbf{r}_g, -\mathbf{k}}^{\alpha}\hat{c}_{\mathbf{r}_e, \mathbf{k}}^{\alpha} \bigr)  \\& \!  = \! -V_{\mathrm{ex}} \! \sum_{\alpha}  \! \bigl(\hat{c}_{\mathbf{r}_g, -\mathbf{k}}^{\alpha \dagger}\hat{c}_{\mathbf{r}_g, -\mathbf{k}}^{\alpha}\hat{c}_{\mathbf{r}_e, \mathbf{k}}^{\alpha \dagger} \hat{c}_{\mathbf{r}_e, \mathbf{k}}^{\alpha} \bigr) \!  = \! -V_{\mathrm{ex}} \! \sum_{\alpha} \, \hat{n}_{\mathbf{r}_g, -\mathbf{k}}^{\alpha} \hat{n}_{\mathbf{r}_e, \mathbf{k}}^{\alpha},
    \end{split}
\end{equation}
where we only considered the contribution on a single rishon.
This leads to a reformulated version of Eq.~(\ref{support mat: H_g_e}) that we can express as follows~\cite{gorshkov2010, Foss-Feig2010, Sotnikov2020}:
\begin{equation} 
\label{support mat: H_g_e rewritten}
    \begin{split}
    \hat{H}_{ge} = & \sum_{\langle \mathbf{r}_g, \mathbf{r}_e \rangle} \big[ V \sum_{\alpha \neq \beta}  (\, \hat{c}_{\mathbf{r}_g, \mathbf{k}}^{\alpha \dagger}\hat{c}_{\mathbf{r}_g, \mathbf{k}}^{\alpha}\hat{c}_{\mathbf{r}_e, -\mathbf{k}}^{\beta \dagger}\hat{c}_{\mathbf{r}_e, -\mathbf{k}}^{\beta})  \\ & +(V - 
     V_{\mathrm{ex}}) \sum_{\alpha} \, (\, \hat{c}_{\mathbf{r}_g, \mathbf{k}}^{\alpha \dagger}\hat{c}_{\mathbf{r}_g, \mathbf{k}}^{\alpha}\hat{c}_{\mathbf{r}_e, -\mathbf{k}}^{\alpha \dagger}\hat{c}_{\mathbf{r}_e, -\mathbf{k}}^{\alpha}) \\ & + 
     V_{\mathrm{ex}} \sum_{\alpha \neq \beta} \, (\hat{c}_{\mathbf{r}_g, -\mathbf{k}}^{\alpha \dagger}\hat{c}_{\mathbf{r}_e, \mathbf{k}}^{\beta \dagger}\hat{c}_{\mathbf{r}_g, -\mathbf{k}}^{\beta}\hat{c}_{\mathbf{r}_e, \mathbf{k}}^{\alpha}) \big],
    \end{split}
\end{equation}
where the term in the last line is effectively suppressed by the Zeeman protection. The term in the second line is proportional to $(V-V_{\mathrm{ex}}) = U_{eg}^-$, meaning that the atoms interact via the antisymmetric interaction channel.
The different energy of the intermediate states leads to a phase oscillation between singlet states and triplet states. 

In the limit $|t| \ll |\delta|$ we can define the two quantities
    $t_{\mathrm{eff}}^{\uparrow \downarrow} = - \frac{t^2}{\delta} + \frac{t^2}{(\delta - V )}$ 
    and
    $t_{\mathrm{eff}}^{\uparrow \uparrow} = - \frac{t^2}{\delta} + \frac{t^2}{(\delta - V +V_{\mathrm{ex}})}$,
which represent the perturbative hopping of the states components that interact in the rishon sites with spins of same $m_F$ ($t_{\mathrm{eff}}^{\uparrow \uparrow}$) or opposite $m_F$ ($t_{\mathrm{eff}}^{\uparrow \downarrow}$). The imbalance in the tunneling amplitudes causes a dephasing during the time evolution that leads to oscillations between the singlet state $\vert s \rangle$ and the state with a triplet on each vertex $\vert t \rangle$. We can also use these quantities to define the coupling $\lambda$ between these two sectors as $\lambda = (t_{\mathrm{eff}}^{\uparrow \uparrow} -t_{\mathrm{eff}}^{\uparrow \downarrow})/2$.
This analysis of Fig.~\ref{app_fig: no sup ex 1}, which does not include the superexchange interactions, shows that these couplings $\lambda$ quickly lead to accumulation of gauge-breaking errors as the four states of the basis \{$ \vert s_g \rangle, \vert s_e \rangle, \vert t_g \rangle, \vert t_e \rangle $\} are on resonance, the gauge-symmetry is easily broken by the evolution of the system. As shown in Fig.~\ref{fig: numeric_results} of the main text, the superexchange couplings lead to gauge-protection, overcoming the effect of $\lambda$.
\section{Atomic losses} \label{e-e losses}
The atomic two-body interactions can lead to losses in the system. When atoms interact with each other in the same lattice site, it is possible for some of them to be lost during the simulation. This phenomenon can affect any two-body interaction but it is particularly pronounced in the case of $e$-$e$ ones. 
Here, we want to estimate the contribution of such events to the scheme.
We start by considering the on-site pair lifetime, $\tau_{a_s}$, where $a_s$ indicates the scattering length of a generic two-body interaction, which we can define through the two-body loss rate coefficient $\beta_{a_s}$ using the relation \cite{ediss18159, garcia-ripoll2009dissipation}
\begin{figure*}[t!]
    \centering
    \includegraphics[width=0.9\linewidth]{ 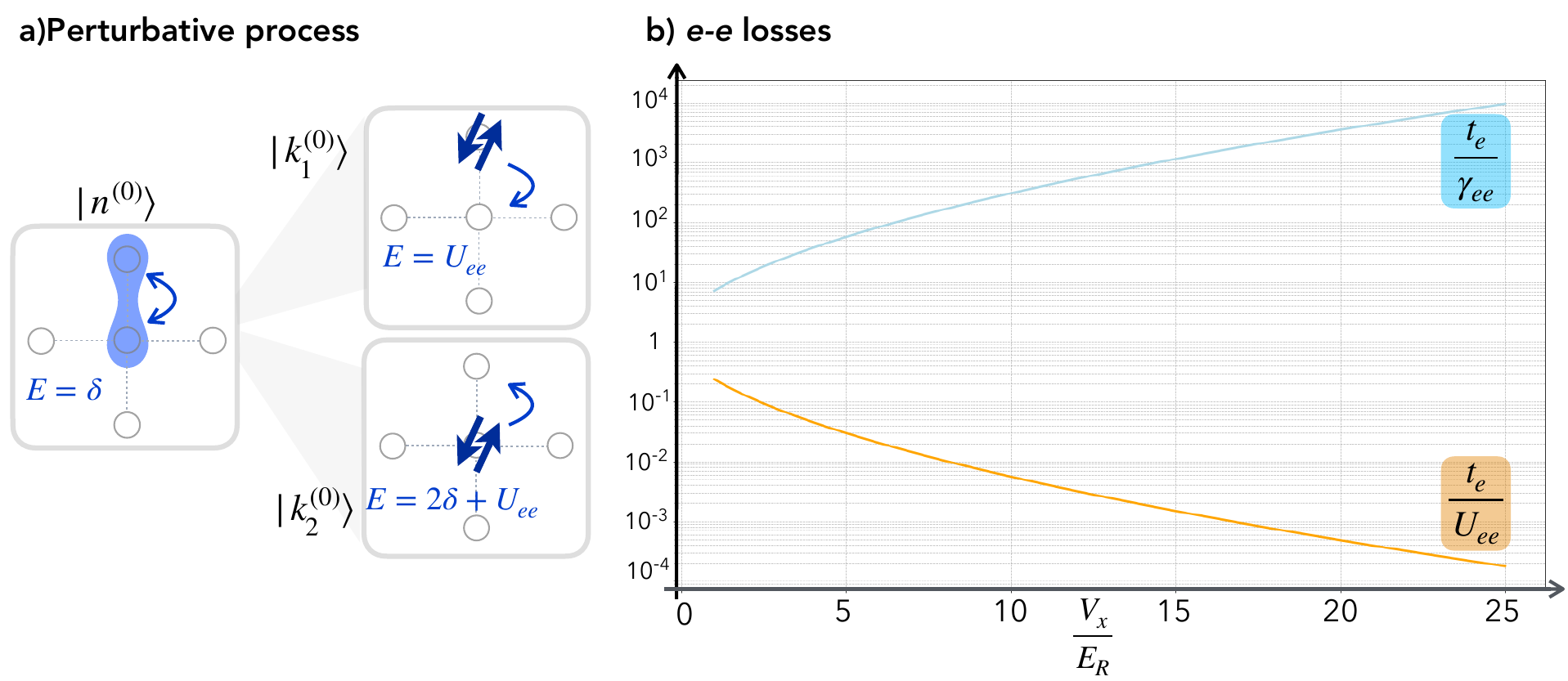}
    \caption{\textbf{Analysis for the e-e losses.} \textbf{a)} Here we show the superexchange process that contributes to a correction of the total energy of the initial singlet configuration $\vert n^{(0)} \rangle$: the two virtual states $\vert k_1^{(0)} \rangle$ and $\vert k_2^{(0)} \rangle$ are characterized by a two-body interaction that can lead to losses. \textbf{b)} In the figure, we show the average number of hoppings that we expect to be experimentally observable before the system loses a full atom in the $e$ state (blue plot). The orange plot is the ratio of the hopping coefficients over the interaction strength.}
    \label{fig: e-e losses scheme}
\end{figure*}
\begin{equation}
    (\tau_{a_s})^{-1} = \beta_{a_s} \int \, d\mathbf{r} \vert w(\mathbf{r})\vert ^4.
\end{equation}
We use this formula to calculate the loss-rate $\gamma_{ee}$, with $\beta_{ee}=\SI{2.2e-11}{\cubic\cm\per\second}$ \cite{ediss18159}:
\begin{equation}
    \gamma_{ee} = P_{ee}  (\tau_{a_{ee}})^{-1} 
\end{equation}
where we indicate with $P_{ee}$ the probability of double occupancy on the same $e$ lattice site.
Assuming that $t_e \ll \delta \ll U_{ee}$, we can estimate from the perturbative processes shown in Fig.~\ref{fig: e-e losses scheme}a 
\begin{equation}
    P_{ee} \approx 2 \biggl( \frac{t_e}{U_{ee}} \biggr) ^2,
\end{equation}
and we obtain
\begin{equation} \label{eq: gamma alpha alpha}
    \gamma_{ee} = P_{ee} (\tau_{a_{ee}})^{-1} = \frac{t_e^2}{U_{ee}^2} \, \, \beta_{ee} \int \vert w \vert^4.
\end{equation}
To avoid numerically evaluating the integrals with the Wannier functions we assume the $e$ lattice to be fairly deep and we use the harmonic approximation to calculate it.
In this case, we consider the approximation introduced in \cite{Jaksch1998, Duan2003} where the lattice structure is approximately described by the function
\begin{equation}
    V(\mathbf{r}) = \sum_{\mu} V_{\mu} \mathrm{sin}^2(x_{\mu}),
\end{equation}
where the index $\mu$ is referred to the spacial coordinates $\mu=x,y,z$.
The variables $V_{\mu}$ indicate the depth of the lattice in the three possible directions and are assumed to be large enough to allow a correct application of the approximation. The three-dimensional lattice can be treated as a two-dimensional one if the depth of one of the three variables is much bigger than the others, basically eliminating any kind of interaction and hopping in that direction.
First of all, the interaction and the hopping coefficients are calculated as
\begin{equation}
    t_{\mu} = - \int d\mathbf{r} \,\, w_{\mu}^*(\mathbf{r}) \left[-\frac{\hbar^2}{2m}\nabla^2 + V_{\mu}\right] w_{\alpha}(\mathbf{r} - \mathbf{r}),
\end{equation}
\begin{equation} \label{eq: U alpha alpha}
    U_{\alpha \alpha} = g_{\alpha \alpha } \int d\mathbf{r} \,\, \vert w_{\alpha}(\mathbf{r}) \vert^4,
\end{equation}
where we use the usual definition
\begin{equation}
    g_{\alpha \alpha} = \frac{4\pi \hbar^2 a_{\alpha \alpha}}{m}
\end{equation}
Moreover, we can use Eq.~\eqref{eq: U alpha alpha} and rewrite Eq.~\eqref{eq: gamma alpha alpha} as follows
\begin{equation} \label{eq: gamma mod}
    \gamma_{ee} = \frac{t_e^2}{U_{ee}} \, \, \frac{\beta_{ee}}{g_{ee}}.
\end{equation}
To apply the approximation we assume the Wannier functions to be expressed in a Gaussian-like form around the lattice site they refer to and we obtain \cite{Duan2003}
\begin{equation} \label{eq: Duan t}
    t_{e, \mu} \approx \biggl(\frac{4}{\sqrt{\pi}} \biggr) \, E_R \, \biggl( \frac{V_{\mu}}{E_R} \biggr)^{\frac{3}{4}} \, e^{-2\bigl( \frac{V_{\mu}}{E_R} \bigr)^{\frac{1}{2}}},
\end{equation}
\begin{equation} \label{eq: Duan U}
    U_{ee} \approx \biggl(\frac{8}{\pi} \biggr) \, \biggl(\frac{\pi \, a_{ee}}{a} \biggr) \, E_R \, \biggl( \frac{V_{x}}{E_R} \frac{V_{y}}{E_R} \frac{V_{z}}{E_R} \biggr)^{\frac{1}{4}}, 
\end{equation}
where $E_R$ is the recoil energy that depends on the lattice momentum
\begin{equation}
    E_R = \frac{\hbar^2 k^2}{2 m} = \frac{\hbar^2 \pi^2}{2 a m}. 
\end{equation}
For our calculations, we assume that $V_x/E_R=V_y/E_R \ll V_z/E_R$. Moreover, we focus on calculating the hopping in the $x$ (or $y$) direction, as ($t_x = t_y \gg t_z$ ).
Using these approximations, we substitute ~\eqref{eq: Duan t} and ~\eqref{eq: Duan U} in ~\eqref{eq: gamma mod} and we obtain
\begin{equation}
\begin{split}
    \gamma_{ee} &= \frac{\frac{16}{\pi} \, E_R \, \bigl( \frac{V_x}{E_R} \bigr)^{\frac{3}{2}} e^{-4 \bigl( \frac{V_{x}}{E_R} \bigr)^{\frac{1}{2}} } } {\frac{8}{\pi} \bigl(\frac{\pi \, a_{ee}}{a} \bigr) \, E_R \, \bigl( \frac{V_{x}}{E_R} \frac{V_{y}}{E_R} \frac{V_{z}}{E_R} \bigr)^{\frac{1}{4}} } \, \frac{\beta_{ee} m}{4 \pi \hbar^2 a_{ee}} \\& =  \frac{\pi}{4} \frac{V_x/E_R}{(V_z/E_R)^\frac{1}{4}} e^{-4 \bigl( \frac{V_{x}}{E_R} \bigr)^{\frac{1}{2}}} \frac{\beta_{ee}}{ a \, a_{ee} }, 
\end{split}
\end{equation}
which is correctly expressed in $\SI{}{\Hz}$.
We calculate the ratio $\gamma_{ee}/t_{e}$ to be able to estimate the average number of hopping processes that can happen before observing a loss
\begin{equation}
    \frac{\gamma_{ee}}{t_e} = \frac{t_e}{U_{ee}} \, \, \frac{\beta_{ee}}{g_{ee}} = \frac{\sqrt{\pi}}{8}  \bigl(\frac{V_x/E_R}{(V_z/E_R)}\bigr)^\frac{1}{4} e^{-2 \bigl( \frac{V_{x}}{E_R} \bigr)^{\frac{1}{2}}} \frac{\beta_{ee}}{ a \, a_{ee}^2 } \frac{1}{E_R}.
\end{equation}
From the same calculations we obtain the ratio $t_{e}/U_{ee}$
\begin{equation}
    \frac{t_e}{U_{ee}} = \frac{1}{2\sqrt{\pi}} \frac{a}{a_{ee}}  \bigl(\frac{V_x/E_R}{(V_z/E_R)}\bigr)^\frac{1}{4} e^{-2 \bigl( \frac{V_{x}}{E_R} \bigr)^{\frac{1}{2}}}.
\end{equation}
We plot these different ratios to find a specific range of the variable $V_x/E_R$ for which the losses remain controlled. We have used a typical value for $a \approx \SI{300}{\nano\meter}$ and we set $V_z/E_R = 150$.
The harmonic approximation fails to correctly reproduce the hopping amplitude for too small values of the lattice depth, and we assume it to be reliable for values $V/E_R > 5$.
In Fig.~\ref{fig: e-e losses scheme}b we observe that, when considering a lattice with ratio $t_e/U_{ee} \approx 1/100$ to $1/300$ (so when $V/E_R$ has values between $7$ and $9$), the ratio of particle decay in units of the hopping coefficient is set to be in a range that goes from $1/100$ to almost $1/300$, meaning that the decay is not particularly relevant. This particular range of $t_e/U_{ee}$ has been chosen because it is similar to the values used in the various analysis in the main text, where we considered $t/\delta \approx 1/10$ and $U_{ee}/\delta \approx 15$. From similar approximated calculation and considerations on the symmetry of the lattice, and using $\beta_{ge}=\SI{3.9e-13}{\cubic\cm\per\second}$ \cite{ediss18159}, we estimate that the corresponding value $\gamma_{ge}/t\approx \SI{e-2}\cdot \gamma_{ee}/t_e$ while $t_{e}/U_{ee}\approx 3\cdot t_{e}/U_{ee}$ making the $g$-$e$ losses even less relevant.
Because of these results, we assume the scheme to be reliable and experimentally implementable.
\section{Rydberg multicolor drive} \label{sec: Rydberg multicolor drive}
\begin{figure*}[t!]
    \centering  
    \includegraphics[width=0.9\textwidth]{ 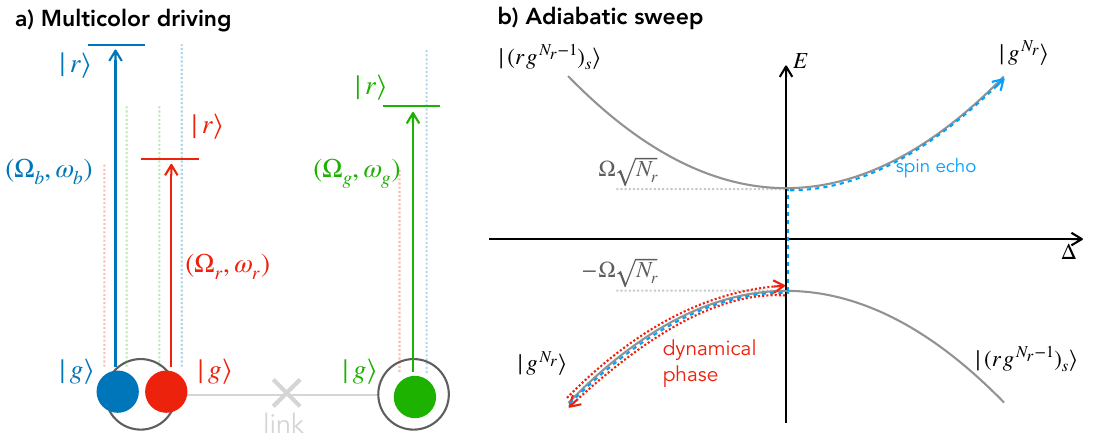}
    \caption{\textbf{Rydberg scheme} \textbf{a)} The Rydberg-excited states of atoms with different nuclear spin experience an hyperfine splitting that separate the energy levels. For this reason, we propose to use a multicolor-laser scheme to excite all the atoms in the link at the same time. Each laser must be tuned to yield the same Rabi frequency $\Omega$. Here we show the example of a link with three atoms of different color. \textbf{b)} We show the energy of the two eigenstates of an Hamiltonian describing a Rydberg dressing scheme for different values of the energy offset $\Delta$, to pick up the phase necessary to our scheme we assume to undergo an adiabatic evolution to the eigenenergy of the Hamiltonian with $\Delta=0$, hold the system in such state for a certain period, and then evolve the state back to the original ground state (red line). Alternatively, it is possible to use a spin echo procedure (blue line).}
    \label{fig: Rydberg SM}
\end{figure*}
In the the main text, we assumed that $N_{\langle \mathbf{r}, \mathbf{r'} \rangle}$ atoms in the rishon sites of a link are excited to the Rydberg states in a process that results in a collective Rabi oscillation between the ground state $\vert g^{N_{\langle \mathbf{r}, \mathbf{r'} \rangle}} \rangle$ and the symmetric Dicke-state $\vert r \,g^{N_{\langle \mathbf{r}, \mathbf{r'} \rangle}-1}_F \rangle$ with a frequency $\Omega^*=\sqrt{N_{\langle \mathbf{r}, \mathbf{r'} \rangle}}\Omega$.
The general requirement for the Dicke dynamics to be valid is that each atom must be excited to the Rydberg state with Rabi frequency $\Omega$ but, in our case, the excitation cannot be achieved using a monochromatic laser: the Rydberg states are split into different hyperfine levels, resulting in the excited states of atoms with different color degree of freedom to be out of resonance, see Fig.~\ref{fig: Rydberg SM}a.
We therefore propose to use of multi-chromatic lasers that allow to achieve excitations of the atoms in the ground state to the correct Rydberg states. The lasers frequencies and intensities must be accordingly tuned, in order to achieve the same Rabi frequency $\Omega$.
For clarity, we calculate the collective Rabi frequency for an easy example of a link with two atoms of different colors (e.g. the blue and red atoms in Fig.~\ref{fig: Rydberg SM}a).
We start defining the basis used, $ \{ \vert \textcolor{blue}{g}\textcolor{red}{g}\rangle, \vert \textcolor{blue}{r}\textcolor{red}{g}\rangle, \vert \textcolor{blue}{g}\textcolor{red}{r}\rangle, \vert \textcolor{blue}{r}\textcolor{red}{r}\rangle \}$, where $g$ is the ground state and $r$ is the Rydberg excited one and we assume the atoms to be separated by distance $R$.
The Hamiltonian of the system is then
\begin{align} \label{eq:multicolor_raw}
\hat{H} = \begin{pmatrix} 
0 & \textcolor{blue}{\Omega_b} & \textcolor{red}{\Omega_r} & 0 \\
\textcolor{blue}{\Omega_b} & 0 & 0 & \textcolor{red}{\Omega_r} \\ 
\textcolor{red}{\Omega_r} & 0 & 0 & \textcolor{blue}{\Omega_b} \\
0 & \textcolor{red}{\Omega_r} & \textcolor{blue}{\Omega_b} & C_6/R^6 \\
\end{pmatrix} ,
\end{align}
The states $\vert \textcolor{blue}{r}\textcolor{red}{g}\rangle, \vert \textcolor{blue}{g}\textcolor{red}{r}\rangle$ can be rotated to 
\begin{equation}
    \vert \varphi_ \pm \rangle = \frac{\Omega_r}{\sqrt{\Omega_r^2+\Omega_b^2}}\vert \textcolor{blue}{r}\textcolor{red}{g} \rangle \pm \frac{\Omega_b}{\sqrt{\Omega_r^2+\Omega_b^2}}\vert \textcolor{blue}{g}\textcolor{red}{r}\rangle
\end{equation}
Since $\vert \varphi_- \rangle$ is a "dark state", the Hamiltonian can be reduced using the basis $ \{ \vert \textcolor{blue}{g}\textcolor{red}{g}\rangle, \vert \varphi_+ \rangle,  \vert \textcolor{blue}{r}\textcolor{red}{r}\rangle \}$ yielding
\begin{align} \label{eq:multicolor_after}
\hat{H} = \begin{pmatrix} 
0 & \Omega^* & 0 \\
\Omega^* & 0 & \Omega^* \\ 
 0 & \Omega^* & C_6/R^6  \\
\end{pmatrix} ,
\end{align}
where $\Omega^* = \frac{2 \textcolor{red}{\Omega_r} \textcolor{blue}{\Omega_b}}{\sqrt{\textcolor{red}{\Omega_r}^2+\textcolor{blue}{\Omega_b}^2}}$ which is therefore dependent on the color of the particles in the link. Once the Rabi frequencies are tuned to be all equal to $\Omega$ for each one of the lasers, $\Omega^*$ reduces to $\sqrt{2}\Omega$ and the color-independence is achieved. The same process can be repeated for more atoms in each link. 
\section{Adiabatic sweep} \label{SM sec: Adiabatic sweep}
The adiabatic process used for the Rydberg scheme in the main text could be achieved through a fast Rydberg dressing sweep (see Fig.~\ref{fig: Rydberg SM}b):
 We define the Hamiltonian of the dressing to the Rydberg state as a function of a time dependent energy off-set $\Delta(t)$. The instantaneous eigenvalues are $E_{\pm}(t)= (-\Delta(t) \pm \sqrt{\Delta(t)^2 + 4\Omega^{*2}})/2$ with relative eigenvectors (non-normalised) $\vert \varphi_{+} \rangle = \vert g^{N_{\langle \mathbf{r}, \mathbf{r'} \rangle}} \rangle + \frac{E_+}{\Omega^*} \vert r \,g^{N_{\langle \mathbf{r}, \mathbf{r'} \rangle}-1} \rangle_s$ and $\vert \varphi_{-} \rangle = \frac{ \Delta (t) + E_- }{\Omega^*} \vert g^{N_{\langle \mathbf{r}, \mathbf{r'} \rangle}} \rangle + \vert r \,g^{N_{\langle \mathbf{r}, \mathbf{r'} \rangle}-1} \rangle_s$.
When $\Delta(t)=0$ we obtain the energy $E_-=\Omega^*= \sqrt{N_r}\Omega$ and the eigenstate is the antisymmetric superposition $\vert \varphi_{-} \rangle = \vert A \rangle = (\vert g^{N_{\langle \mathbf{r}, \mathbf{r'} \rangle}} \rangle - \vert r \,g^{N_{\langle \mathbf{r}, \mathbf{r'} \rangle}-1} \rangle_s)/\sqrt{2}$.

The total sequence is comprised of three steps: first the adiabatic evolution from $\vert g^{N_{\langle \mathbf{r}, \mathbf{r'} \rangle}} \rangle$ to $\vert A \rangle $ happens over the time $\tau_s$ and is associated to an evolution operator $\hat{U}_{\mathrm{fwd}}$, afterwards the Hamiltonian is not changed; then the state is held to remain in the eigenstate with energy $-\sqrt{N_{\langle \mathbf{r}, \mathbf{r'} \rangle}}\Omega$ ($\hat{U}_{\mathrm{hold}}$) for a time $\tau_h \gg\tau_s$; and finally it is evolved back through the same adiabatic path ($\hat{U}_{\mathrm{bwd}}$). The adiabatic evolution leads to the appearance of a dynamical $\theta = -\int_0^{\tau_s} E(t') dt'$ and Berry $\gamma$ phase, but the latter does not contribute, since the contributions from the forward and backward evolution sum up to zero. Therefore, we define the dynamical phase of the forward process as $\theta_{\mathrm{f}}$, that of the backward process as $\theta_{\mathrm{b}}$ and the total operator of the sweep as
\begin{equation}
\begin{split}
    &\hat{U}_{p} = \hat{U}_{\mathrm{bwd}}\hat{U}_{\mathrm{hold}}\hat{U}_{\mathrm{fwd}} \\&= e^{i\theta_{\mathrm{b}}} e^{i\theta_{\mathrm{f}}} e^{i\sqrt{N_{\langle \mathbf{r}, \mathbf{r'} \rangle}}\Omega\tau_h}\vert g^{N_{\langle \mathbf{r}, \mathbf{r'} \rangle}} \rangle\langle g^{_{\langle \mathbf{r}, \mathbf{r'} \rangle}} \vert
\end{split}
\end{equation}
In our analysis we assume that the contribution of the dynamical phases can be neglected or canceled if using a spin-echo procedure, see Fig.~\ref{fig: Rydberg SM}b. 
For reference we show its analytical form assuming that the energy offset of the dressing to vary as $\Delta(t)=-\Delta_0 + vt$ with $v = \Delta_0/\tau_s$ and $\Delta_0>0$
\begin{equation}
    \begin{split}
    &\theta_{\mathrm{f}} = -\int_0^{\tau_s} E_-(t') dt'= \\ &-\int_0^{\tau_s} \frac{-\Delta(t')}{2} - \frac{\sqrt{\Delta(t')^2 + 4\Omega^{*2}}}{2} dt'.
\end{split}
\end{equation}
An incomplete short evolution is not sufficient to guarantee the desired $\sqrt{N_{\langle \mathbf{r}, \mathbf{r'} \rangle}}$ term; indeed, we calculate the contribution of the dynamical phase on a short time $\tilde{\tau}\ll\tau_s$, therefore assuming that the state does not completely evolve to $\vert A \rangle$ (i.e. we assume the red dashed path in Fig.~\ref{fig: Rydberg SM}b to advance much less before turning back)
\begin{equation}
    -\int_0^{\tilde{\tau}} E_-(t') dt' \approx -\bigg[\frac{\Delta_0}{12} \biggl( \frac{\tilde{\tau}}{\tau_s} \biggr)^2 - \frac{N_{\langle \mathbf{r}, \mathbf{r'} \rangle}\Omega^{2}}{2\Delta_0} \bigg]\tilde{\tau}.
\end{equation}
The contribution of this integral is linear in $N_{\langle \mathbf{r}, \mathbf{r'}\rangle}$, not sufficient to guarantee the wanted gauge invariance.  
Holding the system in the state $\ket{A}$ for a time $\tau_h$, during which the Hamiltonian remains unchanged, also contributes to the tunability of the effective interaction resulting from the full Floquet scheme.  

\bibliographystyle{apsrev4-1}

\begin{thebibliography}{122}%
\makeatletter
\providecommand \@ifxundefined [1]{%
 \@ifx{#1\undefined}
}%
\providecommand \@ifnum [1]{%
 \ifnum #1\expandafter \@firstoftwo
 \else \expandafter \@secondoftwo
 \fi
}%
\providecommand \@ifx [1]{%
 \ifx #1\expandafter \@firstoftwo
 \else \expandafter \@secondoftwo
 \fi
}%
\providecommand \natexlab [1]{#1}%
\providecommand \enquote  [1]{``#1''}%
\providecommand \bibnamefont  [1]{#1}%
\providecommand \bibfnamefont [1]{#1}%
\providecommand \citenamefont [1]{#1}%
\providecommand \href@noop [0]{\@secondoftwo}%
\providecommand \href [0]{\begingroup \@sanitize@url \@href}%
\providecommand \@href[1]{\@@startlink{#1}\@@href}%
\providecommand \@@href[1]{\endgroup#1\@@endlink}%
\providecommand \@sanitize@url [0]{\catcode `\\12\catcode `\$12\catcode `\&12\catcode `\#12\catcode `\^12\catcode `\_12\catcode `\%12\relax}%
\providecommand \@@startlink[1]{}%
\providecommand \@@endlink[0]{}%
\providecommand \url  [0]{\begingroup\@sanitize@url \@url }%
\providecommand \@url [1]{\endgroup\@href {#1}{\urlprefix }}%
\providecommand \urlprefix  [0]{URL }%
\providecommand \Eprint [0]{\href }%
\providecommand \doibase [0]{http://dx.doi.org/}%
\providecommand \selectlanguage [0]{\@gobble}%
\providecommand \bibinfo  [0]{\@secondoftwo}%
\providecommand \bibfield  [0]{\@secondoftwo}%
\providecommand \translation [1]{[#1]}%
\providecommand \BibitemOpen [0]{}%
\providecommand \bibitemStop [0]{}%
\providecommand \bibitemNoStop [0]{.\EOS\space}%
\providecommand \EOS [0]{\spacefactor3000\relax}%
\providecommand \BibitemShut  [1]{\csname bibitem#1\endcsname}%
\let\auto@bib@innerbib\@empty
\bibitem [{\citenamefont {Weinberg}(1995)}]{Weinberg_book}%
  \BibitemOpen
  \bibfield  {author} {\bibinfo {author} {\bibfnamefont {S.}~\bibnamefont {Weinberg}},\ }\href {https://books.google.de/books?id=doeDB3_WLvwC} {\emph {\bibinfo {title} {The Quantum Theory of Fields}}},\ \bibinfo {series} {Quantum Theory of Fields, Vol. 2: Modern Applications}\ No.\ \bibinfo {number} {Bd. 1}\ (\bibinfo  {publisher} {Cambridge University Press},\ \bibinfo {year} {1995})\BibitemShut {NoStop}%
\bibitem [{\citenamefont {Zee}(2003)}]{Zee_book}%
  \BibitemOpen
  \bibfield  {author} {\bibinfo {author} {\bibfnamefont {A.}~\bibnamefont {Zee}},\ }\href {https://books.google.de/books?id=85G9QgAACAAJ} {\emph {\bibinfo {title} {Quantum Field Theory in a Nutshell}}}\ (\bibinfo  {publisher} {Princeton University Press},\ \bibinfo {year} {2003})\BibitemShut {NoStop}%
\bibitem [{\citenamefont {Fradkin}(2013)}]{Fradkin2013}%
  \BibitemOpen
  \bibfield  {author} {\bibinfo {author} {\bibfnamefont {E.}~\bibnamefont {Fradkin}},\ }\href {https://doi.org/10.1017/cbo9781139015509} {\emph {\bibinfo {title} {Field {T}heories of {C}ondensed {M}atter {P}hysics}}}\ (\bibinfo  {publisher} {Cambridge University Press},\ \bibinfo {year} {2013})\BibitemShut {NoStop}%
\bibitem [{\citenamefont {Yang}\ and\ \citenamefont {Mills}(1954)}]{YangMills1954}%
  \BibitemOpen
  \bibfield  {author} {\bibinfo {author} {\bibfnamefont {C.~N.}\ \bibnamefont {Yang}}\ and\ \bibinfo {author} {\bibfnamefont {R.~L.}\ \bibnamefont {Mills}},\ }\href {\doibase 10.1103/PhysRev.96.191} {\bibfield  {journal} {\bibinfo  {journal} {Phys. Rev.}\ }\textbf {\bibinfo {volume} {96}},\ \bibinfo {pages} {191} (\bibinfo {year} {1954})}\BibitemShut {NoStop}%
\bibitem [{\citenamefont {Savary}\ and\ \citenamefont {Balents}(2016)}]{Savary2016}%
  \BibitemOpen
  \bibfield  {author} {\bibinfo {author} {\bibfnamefont {L.}~\bibnamefont {Savary}}\ and\ \bibinfo {author} {\bibfnamefont {L.}~\bibnamefont {Balents}},\ }\href {\doibase 10.1088/0034-4885/80/1/016502} {\bibfield  {journal} {\bibinfo  {journal} {Reports on Progress in Physics}\ }\textbf {\bibinfo {volume} {80}},\ \bibinfo {pages} {016502} (\bibinfo {year} {2016})}\BibitemShut {NoStop}%
\bibitem [{\citenamefont {Balents}(2010)}]{Balents2010}%
  \BibitemOpen
  \bibfield  {author} {\bibinfo {author} {\bibfnamefont {L.}~\bibnamefont {Balents}},\ }\href {\doibase 10.1038/nature08917} {\bibfield  {journal} {\bibinfo  {journal} {Nature}\ }\textbf {\bibinfo {volume} {464}},\ \bibinfo {pages} {199} (\bibinfo {year} {2010})}\BibitemShut {NoStop}%
\bibitem [{\citenamefont {Senthil}\ and\ \citenamefont {Fisher}(2000)}]{Senthil2000}%
  \BibitemOpen
  \bibfield  {author} {\bibinfo {author} {\bibfnamefont {T.}~\bibnamefont {Senthil}}\ and\ \bibinfo {author} {\bibfnamefont {M.~P.~A.}\ \bibnamefont {Fisher}},\ }\href {\doibase 10.1103/PhysRevB.62.7850} {\bibfield  {journal} {\bibinfo  {journal} {Phys. Rev. B}\ }\textbf {\bibinfo {volume} {62}},\ \bibinfo {pages} {7850} (\bibinfo {year} {2000})}\BibitemShut {NoStop}%
\bibitem [{\citenamefont {Kleinert}(1989)}]{kleinert_book}%
  \BibitemOpen
  \bibfield  {author} {\bibinfo {author} {\bibfnamefont {H.}~\bibnamefont {Kleinert}},\ }\href {\doibase 10.1142/0356} {\emph {\bibinfo {title} {Gauge Fields in Condensed Matter}}}\ (\bibinfo  {publisher} {World Scientific},\ \bibinfo {year} {1989})\BibitemShut {NoStop}%
\bibitem [{\citenamefont {Sachdev}(2018)}]{Sachdev2019}%
  \BibitemOpen
  \bibfield  {author} {\bibinfo {author} {\bibfnamefont {S.}~\bibnamefont {Sachdev}},\ }\href {\doibase 10.1088/1361-6633/aae110} {\bibfield  {journal} {\bibinfo  {journal} {Reports on Progress in Physics}\ }\textbf {\bibinfo {volume} {82}},\ \bibinfo {pages} {014001} (\bibinfo {year} {2018})}\BibitemShut {NoStop}%
\bibitem [{\citenamefont {Auerbach}(2012)}]{Auerbach_book}%
  \BibitemOpen
  \bibfield  {author} {\bibinfo {author} {\bibfnamefont {A.}~\bibnamefont {Auerbach}},\ }\href {https://books.google.de/books?id=d-sHCAAAQBAJ} {\emph {\bibinfo {title} {Interacting Electrons and Quantum Magnetism}}},\ Graduate Texts in Contemporary Physics\ (\bibinfo  {publisher} {Springer New York},\ \bibinfo {year} {2012})\BibitemShut {NoStop}%
\bibitem [{\citenamefont {Borla}\ \emph {et~al.}(2022)\citenamefont {Borla}, \citenamefont {Jeevanesan}, \citenamefont {Pollmann},\ and\ \citenamefont {Moroz}}]{Borla2022}%
  \BibitemOpen
  \bibfield  {author} {\bibinfo {author} {\bibfnamefont {U.}~\bibnamefont {Borla}}, \bibinfo {author} {\bibfnamefont {B.}~\bibnamefont {Jeevanesan}}, \bibinfo {author} {\bibfnamefont {F.}~\bibnamefont {Pollmann}}, \ and\ \bibinfo {author} {\bibfnamefont {S.}~\bibnamefont {Moroz}},\ }\href {\doibase 10.1103/PhysRevB.105.075132} {\bibfield  {journal} {\bibinfo  {journal} {Phys. Rev. B}\ }\textbf {\bibinfo {volume} {105}},\ \bibinfo {pages} {075132} (\bibinfo {year} {2022})}\BibitemShut {NoStop}%
\bibitem [{\citenamefont {Kogut}\ and\ \citenamefont {Susskind}(1975)}]{Kogut1975}%
  \BibitemOpen
  \bibfield  {author} {\bibinfo {author} {\bibfnamefont {J.}~\bibnamefont {Kogut}}\ and\ \bibinfo {author} {\bibfnamefont {L.}~\bibnamefont {Susskind}},\ }\href {\doibase 10.1103/PhysRevD.11.395} {\bibfield  {journal} {\bibinfo  {journal} {Phys. Rev. D}\ }\textbf {\bibinfo {volume} {11}},\ \bibinfo {pages} {395} (\bibinfo {year} {1975})}\BibitemShut {NoStop}%
\bibitem [{\citenamefont {Gazit}\ \emph {et~al.}(2017)\citenamefont {Gazit}, \citenamefont {Randeria},\ and\ \citenamefont {Vishwanath}}]{Gazit2017}%
  \BibitemOpen
  \bibfield  {author} {\bibinfo {author} {\bibfnamefont {S.}~\bibnamefont {Gazit}}, \bibinfo {author} {\bibfnamefont {M.}~\bibnamefont {Randeria}}, \ and\ \bibinfo {author} {\bibfnamefont {A.}~\bibnamefont {Vishwanath}},\ }\href {\doibase 10.1038/nphys4028} {\bibfield  {journal} {\bibinfo  {journal} {Nature Physics}\ }\textbf {\bibinfo {volume} {13}},\ \bibinfo {pages} {484} (\bibinfo {year} {2017})}\BibitemShut {NoStop}%
\bibitem [{\citenamefont {Creutz}\ \emph {et~al.}(1983)\citenamefont {Creutz}, \citenamefont {Jacobs},\ and\ \citenamefont {Rebbi}}]{Creutz1983}%
  \BibitemOpen
  \bibfield  {author} {\bibinfo {author} {\bibfnamefont {M.}~\bibnamefont {Creutz}}, \bibinfo {author} {\bibfnamefont {L.}~\bibnamefont {Jacobs}}, \ and\ \bibinfo {author} {\bibfnamefont {C.}~\bibnamefont {Rebbi}},\ }\href {\doibase https://doi.org/10.1016/0370-1573(83)90016-9} {\bibfield  {journal} {\bibinfo  {journal} {Physics Reports}\ }\textbf {\bibinfo {volume} {95}},\ \bibinfo {pages} {201} (\bibinfo {year} {1983})}\BibitemShut {NoStop}%
\bibitem [{\citenamefont {Bender}\ \emph {et~al.}(2023)\citenamefont {Bender}, \citenamefont {Emonts},\ and\ \citenamefont {Cirac}}]{Bender2023}%
  \BibitemOpen
  \bibfield  {author} {\bibinfo {author} {\bibfnamefont {J.}~\bibnamefont {Bender}}, \bibinfo {author} {\bibfnamefont {P.}~\bibnamefont {Emonts}}, \ and\ \bibinfo {author} {\bibfnamefont {J.~I.}\ \bibnamefont {Cirac}},\ }\href {\doibase 10.1103/PhysRevResearch.5.043128} {\bibfield  {journal} {\bibinfo  {journal} {Phys. Rev. Res.}\ }\textbf {\bibinfo {volume} {5}},\ \bibinfo {pages} {043128} (\bibinfo {year} {2023})}\BibitemShut {NoStop}%
\bibitem [{\citenamefont {Bañuls}\ \emph {et~al.}(2020)\citenamefont {Bañuls}, \citenamefont {Blatt}, \citenamefont {Catani} \emph {et~al.}}]{Banuls2020}%
  \BibitemOpen
  \bibfield  {author} {\bibinfo {author} {\bibfnamefont {M.~C.}\ \bibnamefont {Bañuls}}, \bibinfo {author} {\bibfnamefont {R.}~\bibnamefont {Blatt}}, \bibinfo {author} {\bibfnamefont {J.}~\bibnamefont {Catani}},  \emph {et~al.},\ }\href {\doibase 10.1140/epjd/e2020-100571-8} {\bibfield  {journal} {\bibinfo  {journal} {The European Physical Journal D}\ }\textbf {\bibinfo {volume} {74}},\ \bibinfo {pages} {165} (\bibinfo {year} {2020})}\BibitemShut {NoStop}%
\bibitem [{\citenamefont {Magnifico}\ \emph {et~al.}(2024)\citenamefont {Magnifico}, \citenamefont {Cataldi}, \citenamefont {Rigobello}, \citenamefont {Majcen}, \citenamefont {Jaschke}, \citenamefont {Silvi},\ and\ \citenamefont {Montangero}}]{magnifico2024}%
  \BibitemOpen
  \bibfield  {author} {\bibinfo {author} {\bibfnamefont {G.}~\bibnamefont {Magnifico}}, \bibinfo {author} {\bibfnamefont {G.}~\bibnamefont {Cataldi}}, \bibinfo {author} {\bibfnamefont {M.}~\bibnamefont {Rigobello}}, \bibinfo {author} {\bibfnamefont {P.}~\bibnamefont {Majcen}}, \bibinfo {author} {\bibfnamefont {D.}~\bibnamefont {Jaschke}}, \bibinfo {author} {\bibfnamefont {P.}~\bibnamefont {Silvi}}, \ and\ \bibinfo {author} {\bibfnamefont {S.}~\bibnamefont {Montangero}},\ }\href {https://arxiv.org/abs/2407.03058} {\  (\bibinfo {year} {2024})},\ \Eprint {http://arxiv.org/abs/2407.03058} {arXiv:2407.03058 [hep-lat]} \BibitemShut {NoStop}%
\bibitem [{\citenamefont {Wiese}(2013)}]{Wiese_review}%
  \BibitemOpen
  \bibfield  {author} {\bibinfo {author} {\bibfnamefont {U.~J.}\ \bibnamefont {Wiese}},\ }\href {\doibase 10.1002/andp.201300104} {\bibfield  {journal} {\bibinfo  {journal} {Annalen der Physik}\ }\textbf {\bibinfo {volume} {525}},\ \bibinfo {pages} {777} (\bibinfo {year} {2013})}\BibitemShut {NoStop}%
\bibitem [{\citenamefont {Halimeh}\ \emph {et~al.}(2025)\citenamefont {Halimeh}, \citenamefont {Aidelsburger}, \citenamefont {Grusdt}, \citenamefont {Hauke},\ and\ \citenamefont {Yang}}]{Halimeh2025}%
  \BibitemOpen
  \bibfield  {author} {\bibinfo {author} {\bibfnamefont {J.~C.}\ \bibnamefont {Halimeh}}, \bibinfo {author} {\bibfnamefont {M.}~\bibnamefont {Aidelsburger}}, \bibinfo {author} {\bibfnamefont {F.}~\bibnamefont {Grusdt}}, \bibinfo {author} {\bibfnamefont {P.}~\bibnamefont {Hauke}}, \ and\ \bibinfo {author} {\bibfnamefont {B.}~\bibnamefont {Yang}},\ }\href {\doibase 10.1038/s41567-024-02721-8} {\bibfield  {journal} {\bibinfo  {journal} {Nature Physics}\ }\textbf {\bibinfo {volume} {21}},\ \bibinfo {pages} {25} (\bibinfo {year} {2025})}\BibitemShut {NoStop}%
\bibitem [{\citenamefont {Dalmonte}\ and\ \citenamefont {Montangero}(2016)}]{Dalmonte_review}%
  \BibitemOpen
  \bibfield  {author} {\bibinfo {author} {\bibfnamefont {M.}~\bibnamefont {Dalmonte}}\ and\ \bibinfo {author} {\bibfnamefont {S.}~\bibnamefont {Montangero}},\ }\href {\doibase 10.1080/00107514.2016.1151199} {\bibfield  {journal} {\bibinfo  {journal} {Contemporary Physics}\ }\textbf {\bibinfo {volume} {57}},\ \bibinfo {pages} {388} (\bibinfo {year} {2016})}\BibitemShut {NoStop}%
\bibitem [{\citenamefont {Zohar}\ \emph {et~al.}(2015)\citenamefont {Zohar}, \citenamefont {Cirac},\ and\ \citenamefont {Reznik}}]{Zohar_review}%
  \BibitemOpen
  \bibfield  {author} {\bibinfo {author} {\bibfnamefont {E.}~\bibnamefont {Zohar}}, \bibinfo {author} {\bibfnamefont {J.~I.}\ \bibnamefont {Cirac}}, \ and\ \bibinfo {author} {\bibfnamefont {B.}~\bibnamefont {Reznik}},\ }\href {\doibase 10.1088/0034-4885/79/1/014401} {\bibfield  {journal} {\bibinfo  {journal} {Reports on Progress in Physics}\ }\textbf {\bibinfo {volume} {79}},\ \bibinfo {pages} {014401} (\bibinfo {year} {2015})}\BibitemShut {NoStop}%
\bibitem [{\citenamefont {Aidelsburger}\ \emph {et~al.}(2022)\citenamefont {Aidelsburger}, \citenamefont {Barbiero}, \citenamefont {Bermudez}, \citenamefont {Chanda} \emph {et~al.}}]{aidelsburger2021cold}%
  \BibitemOpen
  \bibfield  {author} {\bibinfo {author} {\bibfnamefont {M.}~\bibnamefont {Aidelsburger}}, \bibinfo {author} {\bibfnamefont {L.}~\bibnamefont {Barbiero}}, \bibinfo {author} {\bibfnamefont {A.}~\bibnamefont {Bermudez}}, \bibinfo {author} {\bibfnamefont {T.}~\bibnamefont {Chanda}},  \emph {et~al.},\ }\href {\doibase 10.1098/rsta.2021.0064} {\bibfield  {journal} {\bibinfo  {journal} {Philosophical Transactions of the Royal Society A: Mathematical, Physical and Engineering Sciences}\ }\textbf {\bibinfo {volume} {380}},\ \bibinfo {pages} {20210064} (\bibinfo {year} {2022})}\BibitemShut {NoStop}%
\bibitem [{\citenamefont {Zohar}(2022)}]{zohar2021quantum}%
  \BibitemOpen
  \bibfield  {author} {\bibinfo {author} {\bibfnamefont {E.}~\bibnamefont {Zohar}},\ }\href {\doibase 10.1098/rsta.2021.0069} {\bibfield  {journal} {\bibinfo  {journal} {Philosophical Transactions of the Royal Society A: Mathematical, Physical and Engineering Sciences}\ }\textbf {\bibinfo {volume} {380}},\ \bibinfo {pages} {20210069} (\bibinfo {year} {2022})}\BibitemShut {NoStop}%
\bibitem [{\citenamefont {Bauer}\ \emph {et~al.}(2023)\citenamefont {Bauer}, \citenamefont {Davoudi}, \citenamefont {Balantekin}, \citenamefont {Bhattacharya} \emph {et~al.}}]{Bauer2023}%
  \BibitemOpen
  \bibfield  {author} {\bibinfo {author} {\bibfnamefont {C.~W.}\ \bibnamefont {Bauer}}, \bibinfo {author} {\bibfnamefont {Z.}~\bibnamefont {Davoudi}}, \bibinfo {author} {\bibfnamefont {A.~B.}\ \bibnamefont {Balantekin}}, \bibinfo {author} {\bibfnamefont {T.}~\bibnamefont {Bhattacharya}},  \emph {et~al.},\ }\href {\doibase 10.1103/PRXQuantum.4.027001} {\bibfield  {journal} {\bibinfo  {journal} {PRX Quantum}\ }\textbf {\bibinfo {volume} {4}},\ \bibinfo {pages} {027001} (\bibinfo {year} {2023})}\BibitemShut {NoStop}%
\bibitem [{\citenamefont {Bernien}\ \emph {et~al.}(2017)\citenamefont {Bernien}, \citenamefont {Schwartz}, \citenamefont {Keesling}, \citenamefont {Levine}, \citenamefont {Omran}, \citenamefont {Pichler}, \citenamefont {Choi}, \citenamefont {Zibrov}, \citenamefont {Endres}, \citenamefont {Greiner}, \citenamefont {Vuletić},\ and\ \citenamefont {Lukin}}]{Bernien2017}%
  \BibitemOpen
  \bibfield  {author} {\bibinfo {author} {\bibfnamefont {H.}~\bibnamefont {Bernien}}, \bibinfo {author} {\bibfnamefont {S.}~\bibnamefont {Schwartz}}, \bibinfo {author} {\bibfnamefont {A.}~\bibnamefont {Keesling}}, \bibinfo {author} {\bibfnamefont {H.}~\bibnamefont {Levine}}, \bibinfo {author} {\bibfnamefont {A.}~\bibnamefont {Omran}}, \bibinfo {author} {\bibfnamefont {H.}~\bibnamefont {Pichler}}, \bibinfo {author} {\bibfnamefont {S.}~\bibnamefont {Choi}}, \bibinfo {author} {\bibfnamefont {A.~S.}\ \bibnamefont {Zibrov}}, \bibinfo {author} {\bibfnamefont {M.}~\bibnamefont {Endres}}, \bibinfo {author} {\bibfnamefont {M.}~\bibnamefont {Greiner}}, \bibinfo {author} {\bibfnamefont {V.}~\bibnamefont {Vuletić}}, \ and\ \bibinfo {author} {\bibfnamefont {M.~D.}\ \bibnamefont {Lukin}},\ }\href {\doibase 10.1038/nature24622} {\bibfield  {journal} {\bibinfo  {journal} {Nature}\ }\textbf {\bibinfo {volume} {551}},\ \bibinfo {pages} {579} (\bibinfo {year} {2017})}\BibitemShut {NoStop}%
\bibitem [{\citenamefont {Gyawali}\ \emph {et~al.}(2024)\citenamefont {Gyawali}, \citenamefont {Cochran}, \citenamefont {Lensky}, \citenamefont {Rosenberg} \emph {et~al.}}]{gyawali2024}%
  \BibitemOpen
  \bibfield  {author} {\bibinfo {author} {\bibfnamefont {G.}~\bibnamefont {Gyawali}}, \bibinfo {author} {\bibfnamefont {T.}~\bibnamefont {Cochran}}, \bibinfo {author} {\bibfnamefont {Y.}~\bibnamefont {Lensky}}, \bibinfo {author} {\bibfnamefont {E.}~\bibnamefont {Rosenberg}},  \emph {et~al.},\ }\href {https://arxiv.org/abs/2410.06557} {\  (\bibinfo {year} {2024})},\ \Eprint {http://arxiv.org/abs/2410.06557} {arXiv:2410.06557 [quant-ph]} \BibitemShut {NoStop}%
\bibitem [{\citenamefont {Gonzalez-Cuadra}\ \emph {et~al.}(2025)\citenamefont {Gonzalez-Cuadra}, \citenamefont {Hamdan}, \citenamefont {Zache} \emph {et~al.}}]{gonzalezcuadra2024}%
  \BibitemOpen
  \bibfield  {author} {\bibinfo {author} {\bibfnamefont {D.}~\bibnamefont {Gonzalez-Cuadra}}, \bibinfo {author} {\bibfnamefont {M.}~\bibnamefont {Hamdan}}, \bibinfo {author} {\bibfnamefont {T.~V.}\ \bibnamefont {Zache}},  \emph {et~al.},\ }\href {https://doi.org/10.1038/s41586-025-09051-6} {\bibfield  {journal} {\bibinfo  {journal} {Nature}\ } (\bibinfo {year} {2025})}\BibitemShut {NoStop}%
\bibitem [{\citenamefont {Di~Meglio}\ \emph {et~al.}(2024)\citenamefont {Di~Meglio}, \citenamefont {Jansen}, \citenamefont {Tavernelli}, \citenamefont {Alexandrou}, \citenamefont {Arunachalam}, \citenamefont {Bauer}, \citenamefont {Borras}, \citenamefont {Carrazza}, \citenamefont {Crippa} \emph {et~al.}}]{DiMeglio2024}%
  \BibitemOpen
  \bibfield  {author} {\bibinfo {author} {\bibfnamefont {A.}~\bibnamefont {Di~Meglio}}, \bibinfo {author} {\bibfnamefont {K.}~\bibnamefont {Jansen}}, \bibinfo {author} {\bibfnamefont {I.}~\bibnamefont {Tavernelli}}, \bibinfo {author} {\bibfnamefont {C.}~\bibnamefont {Alexandrou}}, \bibinfo {author} {\bibfnamefont {S.}~\bibnamefont {Arunachalam}}, \bibinfo {author} {\bibfnamefont {C.~W.}\ \bibnamefont {Bauer}}, \bibinfo {author} {\bibfnamefont {K.}~\bibnamefont {Borras}}, \bibinfo {author} {\bibfnamefont {S.}~\bibnamefont {Carrazza}}, \bibinfo {author} {\bibfnamefont {A.}~\bibnamefont {Crippa}},  \emph {et~al.},\ }\href {\doibase 10.1103/PRXQuantum.5.037001} {\bibfield  {journal} {\bibinfo  {journal} {PRX Quantum}\ }\textbf {\bibinfo {volume} {5}},\ \bibinfo {pages} {037001} (\bibinfo {year} {2024})}\BibitemShut {NoStop}%
\bibitem [{\citenamefont {Mil}\ \emph {et~al.}(2020)\citenamefont {Mil}, \citenamefont {Zache}, \citenamefont {Hegde}, \citenamefont {Xia}, \citenamefont {Bhatt}, \citenamefont {Oberthaler}, \citenamefont {Hauke}, \citenamefont {Berges},\ and\ \citenamefont {Jendrzejewski}}]{Mil2020}%
  \BibitemOpen
  \bibfield  {author} {\bibinfo {author} {\bibfnamefont {A.}~\bibnamefont {Mil}}, \bibinfo {author} {\bibfnamefont {T.~V.}\ \bibnamefont {Zache}}, \bibinfo {author} {\bibfnamefont {A.}~\bibnamefont {Hegde}}, \bibinfo {author} {\bibfnamefont {A.}~\bibnamefont {Xia}}, \bibinfo {author} {\bibfnamefont {R.~P.}\ \bibnamefont {Bhatt}}, \bibinfo {author} {\bibfnamefont {M.~K.}\ \bibnamefont {Oberthaler}}, \bibinfo {author} {\bibfnamefont {P.}~\bibnamefont {Hauke}}, \bibinfo {author} {\bibfnamefont {J.}~\bibnamefont {Berges}}, \ and\ \bibinfo {author} {\bibfnamefont {F.}~\bibnamefont {Jendrzejewski}},\ }\href {\doibase 10.1126/science.aaz5312} {\bibfield  {journal} {\bibinfo  {journal} {Science}\ }\textbf {\bibinfo {volume} {367}},\ \bibinfo {pages} {1128} (\bibinfo {year} {2020})}\BibitemShut {NoStop}%
\bibitem [{\citenamefont {Yang}\ \emph {et~al.}(2020)\citenamefont {Yang}, \citenamefont {Sun}, \citenamefont {Ott}, \citenamefont {Wang}, \citenamefont {Zache}, \citenamefont {Halimeh}, \citenamefont {Yuan}, \citenamefont {Hauke},\ and\ \citenamefont {Pan}}]{Yang2020}%
  \BibitemOpen
  \bibfield  {author} {\bibinfo {author} {\bibfnamefont {B.}~\bibnamefont {Yang}}, \bibinfo {author} {\bibfnamefont {H.}~\bibnamefont {Sun}}, \bibinfo {author} {\bibfnamefont {R.}~\bibnamefont {Ott}}, \bibinfo {author} {\bibfnamefont {H.-Y.}\ \bibnamefont {Wang}}, \bibinfo {author} {\bibfnamefont {T.~V.}\ \bibnamefont {Zache}}, \bibinfo {author} {\bibfnamefont {J.~C.}\ \bibnamefont {Halimeh}}, \bibinfo {author} {\bibfnamefont {Z.-S.}\ \bibnamefont {Yuan}}, \bibinfo {author} {\bibfnamefont {P.}~\bibnamefont {Hauke}}, \ and\ \bibinfo {author} {\bibfnamefont {J.-W.}\ \bibnamefont {Pan}},\ }\href {\doibase 10.1038/s41586-020-2910-8} {\bibfield  {journal} {\bibinfo  {journal} {Nature}\ }\textbf {\bibinfo {volume} {587}},\ \bibinfo {pages} {392} (\bibinfo {year} {2020})}\BibitemShut {NoStop}%
\bibitem [{\citenamefont {Zhou}\ \emph {et~al.}(2022)\citenamefont {Zhou}, \citenamefont {Su}, \citenamefont {Halimeh}, \citenamefont {Ott}, \citenamefont {Sun}, \citenamefont {Hauke}, \citenamefont {Yang}, \citenamefont {Yuan}, \citenamefont {Berges},\ and\ \citenamefont {Pan}}]{Zhou2022}%
  \BibitemOpen
  \bibfield  {author} {\bibinfo {author} {\bibfnamefont {Z.-Y.}\ \bibnamefont {Zhou}}, \bibinfo {author} {\bibfnamefont {G.-X.}\ \bibnamefont {Su}}, \bibinfo {author} {\bibfnamefont {J.~C.}\ \bibnamefont {Halimeh}}, \bibinfo {author} {\bibfnamefont {R.}~\bibnamefont {Ott}}, \bibinfo {author} {\bibfnamefont {H.}~\bibnamefont {Sun}}, \bibinfo {author} {\bibfnamefont {P.}~\bibnamefont {Hauke}}, \bibinfo {author} {\bibfnamefont {B.}~\bibnamefont {Yang}}, \bibinfo {author} {\bibfnamefont {Z.-S.}\ \bibnamefont {Yuan}}, \bibinfo {author} {\bibfnamefont {J.}~\bibnamefont {Berges}}, \ and\ \bibinfo {author} {\bibfnamefont {J.-W.}\ \bibnamefont {Pan}},\ }\href {\doibase 10.1126/science.abl6277} {\bibfield  {journal} {\bibinfo  {journal} {Science}\ }\textbf {\bibinfo {volume} {377}},\ \bibinfo {pages} {311} (\bibinfo {year} {2022})}\BibitemShut {NoStop}%
\bibitem [{\citenamefont {Surace}\ \emph {et~al.}(2020)\citenamefont {Surace}, \citenamefont {Mazza}, \citenamefont {Giudici}, \citenamefont {Lerose}, \citenamefont {Gambassi},\ and\ \citenamefont {Dalmonte}}]{Surace2020}%
  \BibitemOpen
  \bibfield  {author} {\bibinfo {author} {\bibfnamefont {F.~M.}\ \bibnamefont {Surace}}, \bibinfo {author} {\bibfnamefont {P.~P.}\ \bibnamefont {Mazza}}, \bibinfo {author} {\bibfnamefont {G.}~\bibnamefont {Giudici}}, \bibinfo {author} {\bibfnamefont {A.}~\bibnamefont {Lerose}}, \bibinfo {author} {\bibfnamefont {A.}~\bibnamefont {Gambassi}}, \ and\ \bibinfo {author} {\bibfnamefont {M.}~\bibnamefont {Dalmonte}},\ }\href {\doibase 10.1103/PhysRevX.10.021041} {\bibfield  {journal} {\bibinfo  {journal} {Phys. Rev. X}\ }\textbf {\bibinfo {volume} {10}},\ \bibinfo {pages} {021041} (\bibinfo {year} {2020})}\BibitemShut {NoStop}%
\bibitem [{\citenamefont {Wang}\ \emph {et~al.}(2023)\citenamefont {Wang}, \citenamefont {Zhang}, \citenamefont {Yao}, \citenamefont {Liu}, \citenamefont {Zhu}, \citenamefont {Zheng}, \citenamefont {Wang}, \citenamefont {Zhai}, \citenamefont {Yuan},\ and\ \citenamefont {Pan}}]{Wang2023}%
  \BibitemOpen
  \bibfield  {author} {\bibinfo {author} {\bibfnamefont {H.-Y.}\ \bibnamefont {Wang}}, \bibinfo {author} {\bibfnamefont {W.-Y.}\ \bibnamefont {Zhang}}, \bibinfo {author} {\bibfnamefont {Z.}~\bibnamefont {Yao}}, \bibinfo {author} {\bibfnamefont {Y.}~\bibnamefont {Liu}}, \bibinfo {author} {\bibfnamefont {Z.-H.}\ \bibnamefont {Zhu}}, \bibinfo {author} {\bibfnamefont {Y.-G.}\ \bibnamefont {Zheng}}, \bibinfo {author} {\bibfnamefont {X.-K.}\ \bibnamefont {Wang}}, \bibinfo {author} {\bibfnamefont {H.}~\bibnamefont {Zhai}}, \bibinfo {author} {\bibfnamefont {Z.-S.}\ \bibnamefont {Yuan}}, \ and\ \bibinfo {author} {\bibfnamefont {J.-W.}\ \bibnamefont {Pan}},\ }\href {\doibase 10.1103/PhysRevLett.131.050401} {\bibfield  {journal} {\bibinfo  {journal} {Phys. Rev. Lett.}\ }\textbf {\bibinfo {volume} {131}},\ \bibinfo {pages} {050401} (\bibinfo {year} {2023})}\BibitemShut {NoStop}%
\bibitem [{\citenamefont {Schweizer}\ \emph {et~al.}(2019)\citenamefont {Schweizer}, \citenamefont {Grusdt}, \citenamefont {Berngruber}, \citenamefont {Barbiero}, \citenamefont {Demler}, \citenamefont {Goldman},\ and\ \citenamefont {Bloch}}]{Schweizer2019}%
  \BibitemOpen
  \bibfield  {author} {\bibinfo {author} {\bibfnamefont {C.}~\bibnamefont {Schweizer}}, \bibinfo {author} {\bibfnamefont {F.}~\bibnamefont {Grusdt}}, \bibinfo {author} {\bibfnamefont {M.}~\bibnamefont {Berngruber}}, \bibinfo {author} {\bibfnamefont {L.}~\bibnamefont {Barbiero}}, \bibinfo {author} {\bibfnamefont {E.}~\bibnamefont {Demler}}, \bibinfo {author} {\bibfnamefont {N.}~\bibnamefont {Goldman}}, \ and\ \bibinfo {author} {\bibfnamefont {I.}~\bibnamefont {Bloch}},\ }\href {\doibase 10.1038/s41567-019-0649-7} {\bibfield  {journal} {\bibinfo  {journal} {Nature Physics}\ }\textbf {\bibinfo {volume} {15}},\ \bibinfo {pages} {1168} (\bibinfo {year} {2019})}\BibitemShut {NoStop}%
\bibitem [{\citenamefont {Zhang}\ \emph {et~al.}(2025)\citenamefont {Zhang}, \citenamefont {Liu}, \citenamefont {Cheng} \emph {et~al.}}]{Zhang2025}%
  \BibitemOpen
  \bibfield  {author} {\bibinfo {author} {\bibfnamefont {W.-Y.}\ \bibnamefont {Zhang}}, \bibinfo {author} {\bibfnamefont {Y.}~\bibnamefont {Liu}}, \bibinfo {author} {\bibfnamefont {Y.}~\bibnamefont {Cheng}},  \emph {et~al.},\ }\href {\doibase 10.1038/s41567-024-02702-x} {\bibfield  {journal} {\bibinfo  {journal} {Nature Physics}\ }\textbf {\bibinfo {volume} {21}},\ \bibinfo {pages} {155} (\bibinfo {year} {2025})}\BibitemShut {NoStop}%
\bibitem [{\citenamefont {Su}\ \emph {et~al.}(2023)\citenamefont {Su}, \citenamefont {Sun}, \citenamefont {Hudomal}, \citenamefont {Desaules}, \citenamefont {Zhou}, \citenamefont {Yang}, \citenamefont {Halimeh}, \citenamefont {Yuan}, \citenamefont {Papi\ifmmode~\acute{c}\else \'{c}\fi{}},\ and\ \citenamefont {Pan}}]{Su2023}%
  \BibitemOpen
  \bibfield  {author} {\bibinfo {author} {\bibfnamefont {G.-X.}\ \bibnamefont {Su}}, \bibinfo {author} {\bibfnamefont {H.}~\bibnamefont {Sun}}, \bibinfo {author} {\bibfnamefont {A.}~\bibnamefont {Hudomal}}, \bibinfo {author} {\bibfnamefont {J.-Y.}\ \bibnamefont {Desaules}}, \bibinfo {author} {\bibfnamefont {Z.-Y.}\ \bibnamefont {Zhou}}, \bibinfo {author} {\bibfnamefont {B.}~\bibnamefont {Yang}}, \bibinfo {author} {\bibfnamefont {J.~C.}\ \bibnamefont {Halimeh}}, \bibinfo {author} {\bibfnamefont {Z.-S.}\ \bibnamefont {Yuan}}, \bibinfo {author} {\bibfnamefont {Z.}~\bibnamefont {Papi\ifmmode~\acute{c}\else \'{c}\fi{}}}, \ and\ \bibinfo {author} {\bibfnamefont {J.-W.}\ \bibnamefont {Pan}},\ }\href {\doibase 10.1103/PhysRevResearch.5.023010} {\bibfield  {journal} {\bibinfo  {journal} {Phys. Rev. Res.}\ }\textbf {\bibinfo {volume} {5}},\ \bibinfo {pages} {023010} (\bibinfo {year} {2023})}\BibitemShut {NoStop}%
\bibitem [{\citenamefont {Zhu}\ \emph {et~al.}(2024)\citenamefont {Zhu}, \citenamefont {Liu}, \citenamefont {Lagnese}, \citenamefont {Surace}, \citenamefont {Zhang}, \citenamefont {He}, \citenamefont {Halimeh}, \citenamefont {Dalmonte}, \citenamefont {Morampudi}, \citenamefont {Wilczek}, \citenamefont {Yuan},\ and\ \citenamefont {Pan}}]{Zhu2024}%
  \BibitemOpen
  \bibfield  {author} {\bibinfo {author} {\bibfnamefont {Z.-H.}\ \bibnamefont {Zhu}}, \bibinfo {author} {\bibfnamefont {Y.}~\bibnamefont {Liu}}, \bibinfo {author} {\bibfnamefont {G.}~\bibnamefont {Lagnese}}, \bibinfo {author} {\bibfnamefont {F.~M.}\ \bibnamefont {Surace}}, \bibinfo {author} {\bibfnamefont {W.-Y.}\ \bibnamefont {Zhang}}, \bibinfo {author} {\bibfnamefont {M.-G.}\ \bibnamefont {He}}, \bibinfo {author} {\bibfnamefont {J.~C.}\ \bibnamefont {Halimeh}}, \bibinfo {author} {\bibfnamefont {M.}~\bibnamefont {Dalmonte}}, \bibinfo {author} {\bibfnamefont {S.~C.}\ \bibnamefont {Morampudi}}, \bibinfo {author} {\bibfnamefont {F.}~\bibnamefont {Wilczek}}, \bibinfo {author} {\bibfnamefont {Z.-S.}\ \bibnamefont {Yuan}}, \ and\ \bibinfo {author} {\bibfnamefont {J.-W.}\ \bibnamefont {Pan}},\ }\href {https://arxiv.org/abs/2411.12565} {\  (\bibinfo {year} {2024})},\ \Eprint {http://arxiv.org/abs/2411.12565} {arXiv:2411.12565 [cond-mat.quant-gas]} \BibitemShut {NoStop}%
\bibitem [{\citenamefont {Kapit}\ and\ \citenamefont {Mueller}(2011)}]{Kapit2011}%
  \BibitemOpen
  \bibfield  {author} {\bibinfo {author} {\bibfnamefont {E.}~\bibnamefont {Kapit}}\ and\ \bibinfo {author} {\bibfnamefont {E.}~\bibnamefont {Mueller}},\ }\href {\doibase 10.1103/PhysRevA.83.033625} {\bibfield  {journal} {\bibinfo  {journal} {Phys. Rev. A}\ }\textbf {\bibinfo {volume} {83}},\ \bibinfo {pages} {033625} (\bibinfo {year} {2011})}\BibitemShut {NoStop}%
\bibitem [{\citenamefont {B\"uchler}\ \emph {et~al.}(2005)\citenamefont {B\"uchler}, \citenamefont {Hermele}, \citenamefont {Huber}, \citenamefont {Fisher},\ and\ \citenamefont {Zoller}}]{Buechler2005}%
  \BibitemOpen
  \bibfield  {author} {\bibinfo {author} {\bibfnamefont {H.~P.}\ \bibnamefont {B\"uchler}}, \bibinfo {author} {\bibfnamefont {M.}~\bibnamefont {Hermele}}, \bibinfo {author} {\bibfnamefont {S.~D.}\ \bibnamefont {Huber}}, \bibinfo {author} {\bibfnamefont {M.~P.~A.}\ \bibnamefont {Fisher}}, \ and\ \bibinfo {author} {\bibfnamefont {P.}~\bibnamefont {Zoller}},\ }\href {\doibase 10.1103/PhysRevLett.95.040402} {\bibfield  {journal} {\bibinfo  {journal} {Phys. Rev. Lett.}\ }\textbf {\bibinfo {volume} {95}},\ \bibinfo {pages} {040402} (\bibinfo {year} {2005})}\BibitemShut {NoStop}%
\bibitem [{\citenamefont {Zohar}\ and\ \citenamefont {Reznik}(2011)}]{Zohar2011}%
  \BibitemOpen
  \bibfield  {author} {\bibinfo {author} {\bibfnamefont {E.}~\bibnamefont {Zohar}}\ and\ \bibinfo {author} {\bibfnamefont {B.}~\bibnamefont {Reznik}},\ }\href {\doibase 10.1103/PhysRevLett.107.275301} {\bibfield  {journal} {\bibinfo  {journal} {Phys. Rev. Lett.}\ }\textbf {\bibinfo {volume} {107}},\ \bibinfo {pages} {275301} (\bibinfo {year} {2011})}\BibitemShut {NoStop}%
\bibitem [{\citenamefont {Feldmeier}\ \emph {et~al.}(2024)\citenamefont {Feldmeier}, \citenamefont {Maskara}, \citenamefont {Köylüoğlu},\ and\ \citenamefont {Lukin}}]{feldmeier2024}%
  \BibitemOpen
  \bibfield  {author} {\bibinfo {author} {\bibfnamefont {J.}~\bibnamefont {Feldmeier}}, \bibinfo {author} {\bibfnamefont {N.}~\bibnamefont {Maskara}}, \bibinfo {author} {\bibfnamefont {N.~U.}\ \bibnamefont {Köylüoğlu}}, \ and\ \bibinfo {author} {\bibfnamefont {M.~D.}\ \bibnamefont {Lukin}},\ }\href {https://arxiv.org/abs/2408.02733} {\  (\bibinfo {year} {2024})},\ \Eprint {http://arxiv.org/abs/2408.02733} {arXiv:2408.02733 [quant-ph]} \BibitemShut {NoStop}%
\bibitem [{\citenamefont {Weimer}\ \emph {et~al.}(2010)\citenamefont {Weimer}, \citenamefont {Müller}, \citenamefont {Lesanovsky}, \citenamefont {Zoller},\ and\ \citenamefont {Büchler}}]{Weimer2010}%
  \BibitemOpen
  \bibfield  {author} {\bibinfo {author} {\bibfnamefont {H.}~\bibnamefont {Weimer}}, \bibinfo {author} {\bibfnamefont {M.}~\bibnamefont {Müller}}, \bibinfo {author} {\bibfnamefont {I.}~\bibnamefont {Lesanovsky}}, \bibinfo {author} {\bibfnamefont {P.}~\bibnamefont {Zoller}}, \ and\ \bibinfo {author} {\bibfnamefont {H.~P.}\ \bibnamefont {Büchler}},\ }\href {\doibase 10.1038/nphys1614} {\bibfield  {journal} {\bibinfo  {journal} {Nature Physics}\ }\textbf {\bibinfo {volume} {6}} (\bibinfo {year} {2010}),\ 10.1038/nphys1614}\BibitemShut {NoStop}%
\bibitem [{\citenamefont {Marcos}\ \emph {et~al.}(2013)\citenamefont {Marcos}, \citenamefont {Rabl}, \citenamefont {Rico},\ and\ \citenamefont {Zoller}}]{Marcos2013}%
  \BibitemOpen
  \bibfield  {author} {\bibinfo {author} {\bibfnamefont {D.}~\bibnamefont {Marcos}}, \bibinfo {author} {\bibfnamefont {P.}~\bibnamefont {Rabl}}, \bibinfo {author} {\bibfnamefont {E.}~\bibnamefont {Rico}}, \ and\ \bibinfo {author} {\bibfnamefont {P.}~\bibnamefont {Zoller}},\ }\href {\doibase 10.1103/PhysRevLett.111.110504} {\bibfield  {journal} {\bibinfo  {journal} {Phys. Rev. Lett.}\ }\textbf {\bibinfo {volume} {111}},\ \bibinfo {pages} {110504} (\bibinfo {year} {2013})}\BibitemShut {NoStop}%
\bibitem [{\citenamefont {Marcos}\ \emph {et~al.}(2014)\citenamefont {Marcos}, \citenamefont {Widmer}, \citenamefont {Rico}, \citenamefont {Hafezi}, \citenamefont {Rabl}, \citenamefont {Wiese},\ and\ \citenamefont {Zoller}}]{Marcos2014}%
  \BibitemOpen
  \bibfield  {author} {\bibinfo {author} {\bibfnamefont {D.}~\bibnamefont {Marcos}}, \bibinfo {author} {\bibfnamefont {P.}~\bibnamefont {Widmer}}, \bibinfo {author} {\bibfnamefont {E.}~\bibnamefont {Rico}}, \bibinfo {author} {\bibfnamefont {M.}~\bibnamefont {Hafezi}}, \bibinfo {author} {\bibfnamefont {P.}~\bibnamefont {Rabl}}, \bibinfo {author} {\bibfnamefont {U.-J.}\ \bibnamefont {Wiese}}, \ and\ \bibinfo {author} {\bibfnamefont {P.}~\bibnamefont {Zoller}},\ }\href {\doibase https://doi.org/10.1016/j.aop.2014.09.011} {\bibfield  {journal} {\bibinfo  {journal} {Annals of Physics}\ }\textbf {\bibinfo {volume} {351}},\ \bibinfo {pages} {634} (\bibinfo {year} {2014})}\BibitemShut {NoStop}%
\bibitem [{\citenamefont {Hayata}\ \emph {et~al.}(2024)\citenamefont {Hayata}, \citenamefont {Seki},\ and\ \citenamefont {Yamamoto}}]{Hayata2024}%
  \BibitemOpen
  \bibfield  {author} {\bibinfo {author} {\bibfnamefont {T.}~\bibnamefont {Hayata}}, \bibinfo {author} {\bibfnamefont {K.}~\bibnamefont {Seki}}, \ and\ \bibinfo {author} {\bibfnamefont {A.}~\bibnamefont {Yamamoto}},\ }\href {\doibase 10.1103/PhysRevD.110.114503} {\bibfield  {journal} {\bibinfo  {journal} {Phys. Rev. D}\ }\textbf {\bibinfo {volume} {110}},\ \bibinfo {pages} {114503} (\bibinfo {year} {2024})}\BibitemShut {NoStop}%
\bibitem [{\citenamefont {Ge}\ \emph {et~al.}(2022)\citenamefont {Ge}, \citenamefont {Huang}, \citenamefont {Meng},\ and\ \citenamefont {Fan}}]{Ge2022}%
  \BibitemOpen
  \bibfield  {author} {\bibinfo {author} {\bibfnamefont {Z.-Y.}\ \bibnamefont {Ge}}, \bibinfo {author} {\bibfnamefont {R.-Z.}\ \bibnamefont {Huang}}, \bibinfo {author} {\bibfnamefont {Z.-Y.}\ \bibnamefont {Meng}}, \ and\ \bibinfo {author} {\bibfnamefont {H.}~\bibnamefont {Fan}},\ }\href {\doibase 10.1088/1674-1056/ac380e} {\bibfield  {journal} {\bibinfo  {journal} {Chinese Physics B}\ }\textbf {\bibinfo {volume} {31}},\ \bibinfo {pages} {020304} (\bibinfo {year} {2022})}\BibitemShut {NoStop}%
\bibitem [{\citenamefont {Martinez}\ \emph {et~al.}(2016)\citenamefont {Martinez}, \citenamefont {Muschik}, \citenamefont {Schindler}, \citenamefont {Nigg}, \citenamefont {Erhard}, \citenamefont {Heyl}, \citenamefont {Hauke}, \citenamefont {Dalmonte}, \citenamefont {Monz}, \citenamefont {Zoller},\ and\ \citenamefont {Blatt}}]{martinez2016real}%
  \BibitemOpen
  \bibfield  {author} {\bibinfo {author} {\bibfnamefont {E.~A.}\ \bibnamefont {Martinez}}, \bibinfo {author} {\bibfnamefont {C.~A.}\ \bibnamefont {Muschik}}, \bibinfo {author} {\bibfnamefont {P.}~\bibnamefont {Schindler}}, \bibinfo {author} {\bibfnamefont {D.}~\bibnamefont {Nigg}}, \bibinfo {author} {\bibfnamefont {A.}~\bibnamefont {Erhard}}, \bibinfo {author} {\bibfnamefont {M.}~\bibnamefont {Heyl}}, \bibinfo {author} {\bibfnamefont {P.}~\bibnamefont {Hauke}}, \bibinfo {author} {\bibfnamefont {M.}~\bibnamefont {Dalmonte}}, \bibinfo {author} {\bibfnamefont {T.}~\bibnamefont {Monz}}, \bibinfo {author} {\bibfnamefont {P.}~\bibnamefont {Zoller}}, \ and\ \bibinfo {author} {\bibfnamefont {R.}~\bibnamefont {Blatt}},\ }\href {\doibase 10.1038/nature18318} {\bibfield  {journal} {\bibinfo  {journal} {Nature}\ }\textbf {\bibinfo {volume} {534}},\ \bibinfo {pages} {516} (\bibinfo {year} {2016})}\BibitemShut {NoStop}%
\bibitem [{\citenamefont {Muschik}\ \emph {et~al.}(2017)\citenamefont {Muschik}, \citenamefont {Heyl}, \citenamefont {Martinez}, \citenamefont {Monz}, \citenamefont {Schindler}, \citenamefont {Vogell}, \citenamefont {Dalmonte}, \citenamefont {Hauke}, \citenamefont {Blatt},\ and\ \citenamefont {Zoller}}]{Muschik2017}%
  \BibitemOpen
  \bibfield  {author} {\bibinfo {author} {\bibfnamefont {C.}~\bibnamefont {Muschik}}, \bibinfo {author} {\bibfnamefont {M.}~\bibnamefont {Heyl}}, \bibinfo {author} {\bibfnamefont {E.}~\bibnamefont {Martinez}}, \bibinfo {author} {\bibfnamefont {T.}~\bibnamefont {Monz}}, \bibinfo {author} {\bibfnamefont {P.}~\bibnamefont {Schindler}}, \bibinfo {author} {\bibfnamefont {B.}~\bibnamefont {Vogell}}, \bibinfo {author} {\bibfnamefont {M.}~\bibnamefont {Dalmonte}}, \bibinfo {author} {\bibfnamefont {P.}~\bibnamefont {Hauke}}, \bibinfo {author} {\bibfnamefont {R.}~\bibnamefont {Blatt}}, \ and\ \bibinfo {author} {\bibfnamefont {P.}~\bibnamefont {Zoller}},\ }\href {\doibase 10.1088/1367-2630/aa89ab} {\bibfield  {journal} {\bibinfo  {journal} {New Journal of Physics}\ }\textbf {\bibinfo {volume} {19}},\ \bibinfo {pages} {103020} (\bibinfo {year} {2017})}\BibitemShut {NoStop}%
\bibitem [{\citenamefont {Nguyen}\ \emph {et~al.}(2022)\citenamefont {Nguyen}, \citenamefont {Tran}, \citenamefont {Zhu}, \citenamefont {Green}, \citenamefont {Alderete}, \citenamefont {Davoudi},\ and\ \citenamefont {Linke}}]{Nguyen2022}%
  \BibitemOpen
  \bibfield  {author} {\bibinfo {author} {\bibfnamefont {N.~H.}\ \bibnamefont {Nguyen}}, \bibinfo {author} {\bibfnamefont {M.~C.}\ \bibnamefont {Tran}}, \bibinfo {author} {\bibfnamefont {Y.}~\bibnamefont {Zhu}}, \bibinfo {author} {\bibfnamefont {A.~M.}\ \bibnamefont {Green}}, \bibinfo {author} {\bibfnamefont {C.~H.}\ \bibnamefont {Alderete}}, \bibinfo {author} {\bibfnamefont {Z.}~\bibnamefont {Davoudi}}, \ and\ \bibinfo {author} {\bibfnamefont {N.~M.}\ \bibnamefont {Linke}},\ }\href {\doibase 10.1103/PRXQuantum.3.020324} {\bibfield  {journal} {\bibinfo  {journal} {PRX Quantum}\ }\textbf {\bibinfo {volume} {3}},\ \bibinfo {pages} {020324} (\bibinfo {year} {2022})}\BibitemShut {NoStop}%
\bibitem [{\citenamefont {Kokail}\ \emph {et~al.}(2019)\citenamefont {Kokail}, \citenamefont {Maier}, \citenamefont {van Bijnen}, \citenamefont {Brydges}, \citenamefont {Joshi}, \citenamefont {Jurcevic}, \citenamefont {Muschik}, \citenamefont {Silvi}, \citenamefont {Blatt}, \citenamefont {Roos},\ and\ \citenamefont {Zoller}}]{Kokail2019}%
  \BibitemOpen
  \bibfield  {author} {\bibinfo {author} {\bibfnamefont {C.}~\bibnamefont {Kokail}}, \bibinfo {author} {\bibfnamefont {C.}~\bibnamefont {Maier}}, \bibinfo {author} {\bibfnamefont {R.}~\bibnamefont {van Bijnen}}, \bibinfo {author} {\bibfnamefont {T.}~\bibnamefont {Brydges}}, \bibinfo {author} {\bibfnamefont {M.~K.}\ \bibnamefont {Joshi}}, \bibinfo {author} {\bibfnamefont {P.}~\bibnamefont {Jurcevic}}, \bibinfo {author} {\bibfnamefont {C.~A.}\ \bibnamefont {Muschik}}, \bibinfo {author} {\bibfnamefont {P.}~\bibnamefont {Silvi}}, \bibinfo {author} {\bibfnamefont {R.}~\bibnamefont {Blatt}}, \bibinfo {author} {\bibfnamefont {C.~F.}\ \bibnamefont {Roos}}, \ and\ \bibinfo {author} {\bibfnamefont {P.}~\bibnamefont {Zoller}},\ }\href {\doibase 10.1038/s41586-019-1177-4} {\bibfield  {journal} {\bibinfo  {journal} {Nature}\ }\textbf {\bibinfo {volume} {569}},\ \bibinfo {pages} {355} (\bibinfo {year} {2019})}\BibitemShut {NoStop}%
\bibitem [{\citenamefont {Klco}\ \emph {et~al.}(2018)\citenamefont {Klco}, \citenamefont {Dumitrescu}, \citenamefont {McCaskey}, \citenamefont {Morris}, \citenamefont {Pooser}, \citenamefont {Sanz}, \citenamefont {Solano}, \citenamefont {Lougovski},\ and\ \citenamefont {Savage}}]{Klco2018}%
  \BibitemOpen
  \bibfield  {author} {\bibinfo {author} {\bibfnamefont {N.}~\bibnamefont {Klco}}, \bibinfo {author} {\bibfnamefont {E.~F.}\ \bibnamefont {Dumitrescu}}, \bibinfo {author} {\bibfnamefont {A.~J.}\ \bibnamefont {McCaskey}}, \bibinfo {author} {\bibfnamefont {T.~D.}\ \bibnamefont {Morris}}, \bibinfo {author} {\bibfnamefont {R.~C.}\ \bibnamefont {Pooser}}, \bibinfo {author} {\bibfnamefont {M.}~\bibnamefont {Sanz}}, \bibinfo {author} {\bibfnamefont {E.}~\bibnamefont {Solano}}, \bibinfo {author} {\bibfnamefont {P.}~\bibnamefont {Lougovski}}, \ and\ \bibinfo {author} {\bibfnamefont {M.~J.}\ \bibnamefont {Savage}},\ }\href {\doibase 10.1103/PhysRevA.98.032331} {\bibfield  {journal} {\bibinfo  {journal} {Phys. Rev. A}\ }\textbf {\bibinfo {volume} {98}},\ \bibinfo {pages} {032331} (\bibinfo {year} {2018})}\BibitemShut {NoStop}%
\bibitem [{\citenamefont {Mildenberger}\ \emph {et~al.}(2025)\citenamefont {Mildenberger}, \citenamefont {Mruczkiewicz}, \citenamefont {Halimeh}, \citenamefont {Jiang},\ and\ \citenamefont {Hauke}}]{Mildenberger2025}%
  \BibitemOpen
  \bibfield  {author} {\bibinfo {author} {\bibfnamefont {J.}~\bibnamefont {Mildenberger}}, \bibinfo {author} {\bibfnamefont {W.}~\bibnamefont {Mruczkiewicz}}, \bibinfo {author} {\bibfnamefont {J.~C.}\ \bibnamefont {Halimeh}}, \bibinfo {author} {\bibfnamefont {Z.}~\bibnamefont {Jiang}}, \ and\ \bibinfo {author} {\bibfnamefont {P.}~\bibnamefont {Hauke}},\ }\href {\doibase 10.1038/s41567-024-02723-6} {\bibfield  {journal} {\bibinfo  {journal} {Nature Physics}\ }\textbf {\bibinfo {volume} {21}},\ \bibinfo {pages} {312–317} (\bibinfo {year} {2025})}\BibitemShut {NoStop}%
\bibitem [{\citenamefont {Wang}\ \emph {et~al.}(2022)\citenamefont {Wang}, \citenamefont {Ge}, \citenamefont {Xiang}, \citenamefont {Song}, \citenamefont {Huang}, \citenamefont {Song}, \citenamefont {Guo}, \citenamefont {Su}, \citenamefont {Xu}, \citenamefont {Zheng},\ and\ \citenamefont {Fan}}]{Wang2021}%
  \BibitemOpen
  \bibfield  {author} {\bibinfo {author} {\bibfnamefont {Z.}~\bibnamefont {Wang}}, \bibinfo {author} {\bibfnamefont {Z.-Y.}\ \bibnamefont {Ge}}, \bibinfo {author} {\bibfnamefont {Z.}~\bibnamefont {Xiang}}, \bibinfo {author} {\bibfnamefont {X.}~\bibnamefont {Song}}, \bibinfo {author} {\bibfnamefont {R.-Z.}\ \bibnamefont {Huang}}, \bibinfo {author} {\bibfnamefont {P.}~\bibnamefont {Song}}, \bibinfo {author} {\bibfnamefont {X.-Y.}\ \bibnamefont {Guo}}, \bibinfo {author} {\bibfnamefont {L.}~\bibnamefont {Su}}, \bibinfo {author} {\bibfnamefont {K.}~\bibnamefont {Xu}}, \bibinfo {author} {\bibfnamefont {D.}~\bibnamefont {Zheng}}, \ and\ \bibinfo {author} {\bibfnamefont {H.}~\bibnamefont {Fan}},\ }\href {\doibase 10.1103/PhysRevResearch.4.L022060} {\bibfield  {journal} {\bibinfo  {journal} {Phys. Rev. Res.}\ }\textbf {\bibinfo {volume} {4}},\ \bibinfo {pages} {L022060} (\bibinfo {year} {2022})}\BibitemShut {NoStop}%
\bibitem [{\citenamefont {Charles}\ \emph {et~al.}(2024)\citenamefont {Charles}, \citenamefont {Gustafson}, \citenamefont {Hardt}, \citenamefont {Herren}, \citenamefont {Hogan}, \citenamefont {Lamm}, \citenamefont {Starecheski}, \citenamefont {Van~de Water},\ and\ \citenamefont {Wagman}}]{charles2023}%
  \BibitemOpen
  \bibfield  {author} {\bibinfo {author} {\bibfnamefont {C.}~\bibnamefont {Charles}}, \bibinfo {author} {\bibfnamefont {E.~J.}\ \bibnamefont {Gustafson}}, \bibinfo {author} {\bibfnamefont {E.}~\bibnamefont {Hardt}}, \bibinfo {author} {\bibfnamefont {F.}~\bibnamefont {Herren}}, \bibinfo {author} {\bibfnamefont {N.}~\bibnamefont {Hogan}}, \bibinfo {author} {\bibfnamefont {H.}~\bibnamefont {Lamm}}, \bibinfo {author} {\bibfnamefont {S.}~\bibnamefont {Starecheski}}, \bibinfo {author} {\bibfnamefont {R.~S.}\ \bibnamefont {Van~de Water}}, \ and\ \bibinfo {author} {\bibfnamefont {M.~L.}\ \bibnamefont {Wagman}},\ }\href {\doibase 10.1103/PhysRevE.109.015307} {\bibfield  {journal} {\bibinfo  {journal} {Phys. Rev. E}\ }\textbf {\bibinfo {volume} {109}},\ \bibinfo {pages} {015307} (\bibinfo {year} {2024})}\BibitemShut {NoStop}%
\bibitem [{\citenamefont {Surace}\ \emph {et~al.}(2023)\citenamefont {Surace}, \citenamefont {Fromholz}, \citenamefont {Oppong}, \citenamefont {Dalmonte},\ and\ \citenamefont {Aidelsburger}}]{surace2023}%
  \BibitemOpen
  \bibfield  {author} {\bibinfo {author} {\bibfnamefont {F.~M.}\ \bibnamefont {Surace}}, \bibinfo {author} {\bibfnamefont {P.}~\bibnamefont {Fromholz}}, \bibinfo {author} {\bibfnamefont {N.~D.}\ \bibnamefont {Oppong}}, \bibinfo {author} {\bibfnamefont {M.}~\bibnamefont {Dalmonte}}, \ and\ \bibinfo {author} {\bibfnamefont {M.}~\bibnamefont {Aidelsburger}},\ }\href {\doibase 10.1103/PRXQuantum.4.020330} {\bibfield  {journal} {\bibinfo  {journal} {PRX Quantum}\ }\textbf {\bibinfo {volume} {4}},\ \bibinfo {pages} {020330} (\bibinfo {year} {2023})}\BibitemShut {NoStop}%
\bibitem [{\citenamefont {Cochran}\ \emph {et~al.}(2025)\citenamefont {Cochran}, \citenamefont {Jobst}, \citenamefont {Rosenberg} \emph {et~al.}}]{cochran2024}%
  \BibitemOpen
  \bibfield  {author} {\bibinfo {author} {\bibfnamefont {T.~A.}\ \bibnamefont {Cochran}}, \bibinfo {author} {\bibfnamefont {B.}~\bibnamefont {Jobst}}, \bibinfo {author} {\bibfnamefont {E.}~\bibnamefont {Rosenberg}},  \emph {et~al.},\ }\href {https://doi.org/10.1038/s41586-025-08999-9} {\bibfield  {journal} {\bibinfo  {journal} {Nature}\ } (\bibinfo {year} {2025})}\BibitemShut {NoStop}%
\bibitem [{\citenamefont {De}\ \emph {et~al.}(2024)\citenamefont {De}, \citenamefont {Lerose}, \citenamefont {Luo}, \citenamefont {Surace}, \citenamefont {Schuckert}, \citenamefont {Bennewitz}, \citenamefont {Ware}, \citenamefont {Morong}, \citenamefont {Collins}, \citenamefont {Davoudi}, \citenamefont {Gorshkov}, \citenamefont {Katz},\ and\ \citenamefont {Monroe}}]{de2024}%
  \BibitemOpen
  \bibfield  {author} {\bibinfo {author} {\bibfnamefont {A.}~\bibnamefont {De}}, \bibinfo {author} {\bibfnamefont {A.}~\bibnamefont {Lerose}}, \bibinfo {author} {\bibfnamefont {D.}~\bibnamefont {Luo}}, \bibinfo {author} {\bibfnamefont {F.~M.}\ \bibnamefont {Surace}}, \bibinfo {author} {\bibfnamefont {A.}~\bibnamefont {Schuckert}}, \bibinfo {author} {\bibfnamefont {E.~R.}\ \bibnamefont {Bennewitz}}, \bibinfo {author} {\bibfnamefont {B.}~\bibnamefont {Ware}}, \bibinfo {author} {\bibfnamefont {W.}~\bibnamefont {Morong}}, \bibinfo {author} {\bibfnamefont {K.~S.}\ \bibnamefont {Collins}}, \bibinfo {author} {\bibfnamefont {Z.}~\bibnamefont {Davoudi}}, \bibinfo {author} {\bibfnamefont {A.~V.}\ \bibnamefont {Gorshkov}}, \bibinfo {author} {\bibfnamefont {O.}~\bibnamefont {Katz}}, \ and\ \bibinfo {author} {\bibfnamefont {C.}~\bibnamefont {Monroe}},\ }\href {https://arxiv.org/abs/2410.13815} {\  (\bibinfo {year} {2024})},\ \Eprint {http://arxiv.org/abs/2410.13815} {arXiv:2410.13815 [quant-ph]} \BibitemShut {NoStop}%
\bibitem [{\citenamefont {Atas}\ \emph {et~al.}(2021)\citenamefont {Atas}, \citenamefont {Zhang}, \citenamefont {Lewis}, \citenamefont {Jahanpour}, \citenamefont {Haase},\ and\ \citenamefont {Muschik}}]{Atas2021}%
  \BibitemOpen
  \bibfield  {author} {\bibinfo {author} {\bibfnamefont {Y.~Y.}\ \bibnamefont {Atas}}, \bibinfo {author} {\bibfnamefont {J.}~\bibnamefont {Zhang}}, \bibinfo {author} {\bibfnamefont {R.}~\bibnamefont {Lewis}}, \bibinfo {author} {\bibfnamefont {A.}~\bibnamefont {Jahanpour}}, \bibinfo {author} {\bibfnamefont {J.~F.}\ \bibnamefont {Haase}}, \ and\ \bibinfo {author} {\bibfnamefont {C.~A.}\ \bibnamefont {Muschik}},\ }\href {\doibase 10.1038/s41467-021-26825-4} {\bibfield  {journal} {\bibinfo  {journal} {Nature Communications}\ }\textbf {\bibinfo {volume} {12}},\ \bibinfo {pages} {6499} (\bibinfo {year} {2021})}\BibitemShut {NoStop}%
\bibitem [{\citenamefont {Banerjee}\ \emph {et~al.}(2013)\citenamefont {Banerjee}, \citenamefont {B\"ogli}, \citenamefont {Dalmonte}, \citenamefont {Rico}, \citenamefont {Stebler}, \citenamefont {Wiese},\ and\ \citenamefont {Zoller}}]{banerjee2013}%
  \BibitemOpen
  \bibfield  {author} {\bibinfo {author} {\bibfnamefont {D.}~\bibnamefont {Banerjee}}, \bibinfo {author} {\bibfnamefont {M.}~\bibnamefont {B\"ogli}}, \bibinfo {author} {\bibfnamefont {M.}~\bibnamefont {Dalmonte}}, \bibinfo {author} {\bibfnamefont {E.}~\bibnamefont {Rico}}, \bibinfo {author} {\bibfnamefont {P.}~\bibnamefont {Stebler}}, \bibinfo {author} {\bibfnamefont {U.-J.}\ \bibnamefont {Wiese}}, \ and\ \bibinfo {author} {\bibfnamefont {P.}~\bibnamefont {Zoller}},\ }\href {\doibase 10.1103/PhysRevLett.110.125303} {\bibfield  {journal} {\bibinfo  {journal} {Phys. Rev. Lett.}\ }\textbf {\bibinfo {volume} {110}},\ \bibinfo {pages} {125303} (\bibinfo {year} {2013})}\BibitemShut {NoStop}%
\bibitem [{\citenamefont {Klco}\ \emph {et~al.}(2020)\citenamefont {Klco}, \citenamefont {Savage},\ and\ \citenamefont {Stryker}}]{Klco2020}%
  \BibitemOpen
  \bibfield  {author} {\bibinfo {author} {\bibfnamefont {N.}~\bibnamefont {Klco}}, \bibinfo {author} {\bibfnamefont {M.~J.}\ \bibnamefont {Savage}}, \ and\ \bibinfo {author} {\bibfnamefont {J.~R.}\ \bibnamefont {Stryker}},\ }\href {\doibase 10.1103/PhysRevD.101.074512} {\bibfield  {journal} {\bibinfo  {journal} {Phys. Rev. D}\ }\textbf {\bibinfo {volume} {101}},\ \bibinfo {pages} {074512} (\bibinfo {year} {2020})}\BibitemShut {NoStop}%
\bibitem [{\citenamefont {Surace}\ \emph {et~al.}(2024)\citenamefont {Surace}, \citenamefont {Fromholz}, \citenamefont {Scazza},\ and\ \citenamefont {Dalmonte}}]{surace2024}%
  \BibitemOpen
  \bibfield  {author} {\bibinfo {author} {\bibfnamefont {F.~M.}\ \bibnamefont {Surace}}, \bibinfo {author} {\bibfnamefont {P.}~\bibnamefont {Fromholz}}, \bibinfo {author} {\bibfnamefont {F.}~\bibnamefont {Scazza}}, \ and\ \bibinfo {author} {\bibfnamefont {M.}~\bibnamefont {Dalmonte}},\ }\href {\doibase 10.22331/q-2024-05-23-1359} {\bibfield  {journal} {\bibinfo  {journal} {Quantum}\ }\textbf {\bibinfo {volume} {8}},\ \bibinfo {pages} {1359} (\bibinfo {year} {2024})}\BibitemShut {NoStop}%
\bibitem [{\citenamefont {Halimeh}\ \emph {et~al.}(2024)\citenamefont {Halimeh}, \citenamefont {Homeier}, \citenamefont {Bohrdt},\ and\ \citenamefont {Grusdt}}]{halimeh2024}%
  \BibitemOpen
  \bibfield  {author} {\bibinfo {author} {\bibfnamefont {J.~C.}\ \bibnamefont {Halimeh}}, \bibinfo {author} {\bibfnamefont {L.}~\bibnamefont {Homeier}}, \bibinfo {author} {\bibfnamefont {A.}~\bibnamefont {Bohrdt}}, \ and\ \bibinfo {author} {\bibfnamefont {F.}~\bibnamefont {Grusdt}},\ }\href {\doibase 10.1103/PRXQuantum.5.030358} {\bibfield  {journal} {\bibinfo  {journal} {PRX Quantum}\ }\textbf {\bibinfo {volume} {5}},\ \bibinfo {pages} {030358} (\bibinfo {year} {2024})}\BibitemShut {NoStop}%
\bibitem [{\citenamefont {Zohar}\ and\ \citenamefont {Cirac}(2019)}]{Zohar2019}%
  \BibitemOpen
  \bibfield  {author} {\bibinfo {author} {\bibfnamefont {E.}~\bibnamefont {Zohar}}\ and\ \bibinfo {author} {\bibfnamefont {J.~I.}\ \bibnamefont {Cirac}},\ }\href {\doibase 10.1103/PhysRevD.99.114511} {\bibfield  {journal} {\bibinfo  {journal} {Phys. Rev. D}\ }\textbf {\bibinfo {volume} {99}},\ \bibinfo {pages} {114511} (\bibinfo {year} {2019})}\BibitemShut {NoStop}%
\bibitem [{\citenamefont {Halimeh}\ and\ \citenamefont {Hauke}(2020)}]{halimeh2020reliability}%
  \BibitemOpen
  \bibfield  {author} {\bibinfo {author} {\bibfnamefont {J.~C.}\ \bibnamefont {Halimeh}}\ and\ \bibinfo {author} {\bibfnamefont {P.}~\bibnamefont {Hauke}},\ }\href {\doibase 10.1103/PhysRevLett.125.030503} {\bibfield  {journal} {\bibinfo  {journal} {Phys. Rev. Lett.}\ }\textbf {\bibinfo {volume} {125}},\ \bibinfo {pages} {030503} (\bibinfo {year} {2020})}\BibitemShut {NoStop}%
\bibitem [{\citenamefont {Halimeh}\ \emph {et~al.}(2022{\natexlab{a}})\citenamefont {Halimeh}, \citenamefont {Lang},\ and\ \citenamefont {Hauke}}]{halimeh2022gaugeprotection}%
  \BibitemOpen
  \bibfield  {author} {\bibinfo {author} {\bibfnamefont {J.~C.}\ \bibnamefont {Halimeh}}, \bibinfo {author} {\bibfnamefont {H.}~\bibnamefont {Lang}}, \ and\ \bibinfo {author} {\bibfnamefont {P.}~\bibnamefont {Hauke}},\ }\href {\doibase 10.1088/1367-2630/ac5564} {\bibfield  {journal} {\bibinfo  {journal} {New Journal of Physics}\ }\textbf {\bibinfo {volume} {24}},\ \bibinfo {pages} {033015} (\bibinfo {year} {2022}{\natexlab{a}})}\BibitemShut {NoStop}%
\bibitem [{\citenamefont {Halimeh}\ and\ \citenamefont {Hauke}(2022)}]{halimeh2022stabilizinggaugetheoriesquantum}%
  \BibitemOpen
  \bibfield  {author} {\bibinfo {author} {\bibfnamefont {J.~C.}\ \bibnamefont {Halimeh}}\ and\ \bibinfo {author} {\bibfnamefont {P.}~\bibnamefont {Hauke}},\ }\href {https://arxiv.org/abs/2204.13709} {\  (\bibinfo {year} {2022})},\ \Eprint {http://arxiv.org/abs/2204.13709} {arXiv:2204.13709 [cond-mat.quant-gas]} \BibitemShut {NoStop}%
\bibitem [{\citenamefont {Tagliacozzo}\ \emph {et~al.}(2013)\citenamefont {Tagliacozzo}, \citenamefont {Celi}, \citenamefont {Orland}, \citenamefont {Mitchell},\ and\ \citenamefont {Lewenstein}}]{tagliacozzo2013}%
  \BibitemOpen
  \bibfield  {author} {\bibinfo {author} {\bibfnamefont {L.}~\bibnamefont {Tagliacozzo}}, \bibinfo {author} {\bibfnamefont {A.}~\bibnamefont {Celi}}, \bibinfo {author} {\bibfnamefont {P.}~\bibnamefont {Orland}}, \bibinfo {author} {\bibfnamefont {M.~W.}\ \bibnamefont {Mitchell}}, \ and\ \bibinfo {author} {\bibfnamefont {M.}~\bibnamefont {Lewenstein}},\ }\href {\doibase 10.1038/ncomms3615} {\bibfield  {journal} {\bibinfo  {journal} {Nature Communications}\ }\textbf {\bibinfo {volume} {4}},\ \bibinfo {pages} {2615} (\bibinfo {year} {2013})}\BibitemShut {NoStop}%
\bibitem [{\citenamefont {Zohar}\ \emph {et~al.}(2013)\citenamefont {Zohar}, \citenamefont {Cirac},\ and\ \citenamefont {Reznik}}]{Zohar2013coldatom}%
  \BibitemOpen
  \bibfield  {author} {\bibinfo {author} {\bibfnamefont {E.}~\bibnamefont {Zohar}}, \bibinfo {author} {\bibfnamefont {J.~I.}\ \bibnamefont {Cirac}}, \ and\ \bibinfo {author} {\bibfnamefont {B.}~\bibnamefont {Reznik}},\ }\href {\doibase 10.1103/PhysRevLett.110.125304} {\bibfield  {journal} {\bibinfo  {journal} {Phys. Rev. Lett.}\ }\textbf {\bibinfo {volume} {110}},\ \bibinfo {pages} {125304} (\bibinfo {year} {2013})}\BibitemShut {NoStop}%
\bibitem [{\citenamefont {Brower}\ \emph {et~al.}(1999)\citenamefont {Brower}, \citenamefont {Chandrasekharan},\ and\ \citenamefont {Wiese}}]{Brower1999}%
  \BibitemOpen
  \bibfield  {author} {\bibinfo {author} {\bibfnamefont {R.}~\bibnamefont {Brower}}, \bibinfo {author} {\bibfnamefont {S.}~\bibnamefont {Chandrasekharan}}, \ and\ \bibinfo {author} {\bibfnamefont {U.-J.}\ \bibnamefont {Wiese}},\ }\href {\doibase 10.1103/PhysRevD.60.094502} {\bibfield  {journal} {\bibinfo  {journal} {Phys. Rev. D}\ }\textbf {\bibinfo {volume} {60}},\ \bibinfo {pages} {094502} (\bibinfo {year} {1999})}\BibitemShut {NoStop}%
\bibitem [{\citenamefont {Chandrasekharan}\ and\ \citenamefont {Wiese}(1997)}]{Chan_Wiese_1997}%
  \BibitemOpen
  \bibfield  {author} {\bibinfo {author} {\bibfnamefont {S.}~\bibnamefont {Chandrasekharan}}\ and\ \bibinfo {author} {\bibfnamefont {U.-J.}\ \bibnamefont {Wiese}},\ }\href {\doibase 10.1016/s0550-3213(97)80041-7} {\bibfield  {journal} {\bibinfo  {journal} {Nuclear Physics B}\ }\textbf {\bibinfo {volume} {492}},\ \bibinfo {pages} {455–471} (\bibinfo {year} {1997})}\BibitemShut {NoStop}%
\bibitem [{\citenamefont {Bar}\ \emph {et~al.}(2002)\citenamefont {Bar}, \citenamefont {Brower}, \citenamefont {Schlittgen},\ and\ \citenamefont {Wiese}}]{Bar2001}%
  \BibitemOpen
  \bibfield  {author} {\bibinfo {author} {\bibfnamefont {O.}~\bibnamefont {Bar}}, \bibinfo {author} {\bibfnamefont {R.}~\bibnamefont {Brower}}, \bibinfo {author} {\bibfnamefont {B.}~\bibnamefont {Schlittgen}}, \ and\ \bibinfo {author} {\bibfnamefont {U.~J.}\ \bibnamefont {Wiese}},\ }\href {\doibase 10.1016/S0920-5632(01)01916-8} {\bibfield  {journal} {\bibinfo  {journal} {Nucl. Phys. B Proc. Suppl.}\ }\textbf {\bibinfo {volume} {106}},\ \bibinfo {pages} {1019} (\bibinfo {year} {2002})}\BibitemShut {NoStop}%
\bibitem [{\citenamefont {Wang}\ and\ \citenamefont {Pollet}(2025)}]{Wang2025}%
  \BibitemOpen
  \bibfield  {author} {\bibinfo {author} {\bibfnamefont {Z.}~\bibnamefont {Wang}}\ and\ \bibinfo {author} {\bibfnamefont {L.}~\bibnamefont {Pollet}},\ }\href {https://doi.org/10.1103/physrevlett.134.086601} {\bibfield  {journal} {\bibinfo  {journal} {Physical Review Letters}\ }\textbf {\bibinfo {volume} {134}},\ \bibinfo {pages} {086601} (\bibinfo {year} {2025})}\BibitemShut {NoStop}%
\bibitem [{\citenamefont {Homeier}\ \emph {et~al.}(2021)\citenamefont {Homeier}, \citenamefont {Schweizer}, \citenamefont {Aidelsburger}, \citenamefont {Fedorov},\ and\ \citenamefont {Grusdt}}]{homeier2021}%
  \BibitemOpen
  \bibfield  {author} {\bibinfo {author} {\bibfnamefont {L.}~\bibnamefont {Homeier}}, \bibinfo {author} {\bibfnamefont {C.}~\bibnamefont {Schweizer}}, \bibinfo {author} {\bibfnamefont {M.}~\bibnamefont {Aidelsburger}}, \bibinfo {author} {\bibfnamefont {A.}~\bibnamefont {Fedorov}}, \ and\ \bibinfo {author} {\bibfnamefont {F.}~\bibnamefont {Grusdt}},\ }\href {\doibase 10.1103/PhysRevB.104.085138} {\bibfield  {journal} {\bibinfo  {journal} {Phys. Rev. B}\ }\textbf {\bibinfo {volume} {104}},\ \bibinfo {pages} {085138} (\bibinfo {year} {2021})}\BibitemShut {NoStop}%
\bibitem [{\citenamefont {Dai}\ \emph {et~al.}(2017)\citenamefont {Dai}, \citenamefont {Yang}, \citenamefont {Reingruber} \emph {et~al.}}]{Dai2017}%
  \BibitemOpen
  \bibfield  {author} {\bibinfo {author} {\bibfnamefont {H.-N.}\ \bibnamefont {Dai}}, \bibinfo {author} {\bibfnamefont {B.}~\bibnamefont {Yang}}, \bibinfo {author} {\bibfnamefont {A.}~\bibnamefont {Reingruber}},  \emph {et~al.},\ }\href {\doibase 10.1038/nphys4243} {\bibfield  {journal} {\bibinfo  {journal} {Nature Physics}\ }\textbf {\bibinfo {volume} {13}},\ \bibinfo {pages} {1195} (\bibinfo {year} {2017})}\BibitemShut {NoStop}%
\bibitem [{\citenamefont {Halimeh}\ \emph {et~al.}(2021)\citenamefont {Halimeh}, \citenamefont {Lang}, \citenamefont {Mildenberger}, \citenamefont {Jiang},\ and\ \citenamefont {Hauke}}]{Halimeh2020single_bodies}%
  \BibitemOpen
  \bibfield  {author} {\bibinfo {author} {\bibfnamefont {J.~C.}\ \bibnamefont {Halimeh}}, \bibinfo {author} {\bibfnamefont {H.}~\bibnamefont {Lang}}, \bibinfo {author} {\bibfnamefont {J.}~\bibnamefont {Mildenberger}}, \bibinfo {author} {\bibfnamefont {Z.}~\bibnamefont {Jiang}}, \ and\ \bibinfo {author} {\bibfnamefont {P.}~\bibnamefont {Hauke}},\ }\href {\doibase 10.1103/PRXQuantum.2.040311} {\bibfield  {journal} {\bibinfo  {journal} {PRX Quantum}\ }\textbf {\bibinfo {volume} {2}},\ \bibinfo {pages} {040311} (\bibinfo {year} {2021})}\BibitemShut {NoStop}%
\bibitem [{\citenamefont {Scazza}\ \emph {et~al.}(2014)\citenamefont {Scazza}, \citenamefont {Hofrichter}, \citenamefont {Höfer}, \citenamefont {De~Groot}, \citenamefont {Bloch},\ and\ \citenamefont {Fölling}}]{scazza2014observation}%
  \BibitemOpen
  \bibfield  {author} {\bibinfo {author} {\bibfnamefont {F.}~\bibnamefont {Scazza}}, \bibinfo {author} {\bibfnamefont {C.}~\bibnamefont {Hofrichter}}, \bibinfo {author} {\bibfnamefont {M.}~\bibnamefont {Höfer}}, \bibinfo {author} {\bibfnamefont {P.~C.}\ \bibnamefont {De~Groot}}, \bibinfo {author} {\bibfnamefont {I.}~\bibnamefont {Bloch}}, \ and\ \bibinfo {author} {\bibfnamefont {S.}~\bibnamefont {Fölling}},\ }\href {\doibase 10.1038/nphys3061} {\bibfield  {journal} {\bibinfo  {journal} {Nature Physics}\ }\textbf {\bibinfo {volume} {10}},\ \bibinfo {pages} {779–784} (\bibinfo {year} {2014})}\BibitemShut {NoStop}%
\bibitem [{\citenamefont {Cappellini}\ \emph {et~al.}(2014)\citenamefont {Cappellini}, \citenamefont {Mancini}, \citenamefont {Pagano}, \citenamefont {Lombardi}, \citenamefont {Livi}, \citenamefont {Siciliani~de Cumis}, \citenamefont {Cancio}, \citenamefont {Pizzocaro}, \citenamefont {Calonico}, \citenamefont {Levi}, \citenamefont {Sias}, \citenamefont {Catani}, \citenamefont {Inguscio},\ and\ \citenamefont {Fallani}}]{Cappellini2014}%
  \BibitemOpen
  \bibfield  {author} {\bibinfo {author} {\bibfnamefont {G.}~\bibnamefont {Cappellini}}, \bibinfo {author} {\bibfnamefont {M.}~\bibnamefont {Mancini}}, \bibinfo {author} {\bibfnamefont {G.}~\bibnamefont {Pagano}}, \bibinfo {author} {\bibfnamefont {P.}~\bibnamefont {Lombardi}}, \bibinfo {author} {\bibfnamefont {L.}~\bibnamefont {Livi}}, \bibinfo {author} {\bibfnamefont {M.}~\bibnamefont {Siciliani~de Cumis}}, \bibinfo {author} {\bibfnamefont {P.}~\bibnamefont {Cancio}}, \bibinfo {author} {\bibfnamefont {M.}~\bibnamefont {Pizzocaro}}, \bibinfo {author} {\bibfnamefont {D.}~\bibnamefont {Calonico}}, \bibinfo {author} {\bibfnamefont {F.}~\bibnamefont {Levi}}, \bibinfo {author} {\bibfnamefont {C.}~\bibnamefont {Sias}}, \bibinfo {author} {\bibfnamefont {J.}~\bibnamefont {Catani}}, \bibinfo {author} {\bibfnamefont {M.}~\bibnamefont {Inguscio}}, \ and\ \bibinfo {author} {\bibfnamefont {L.}~\bibnamefont {Fallani}},\ }\href {\doibase 10.1103/PhysRevLett.113.120402} {\bibfield  {journal} {\bibinfo  {journal} {Phys. Rev.
  Lett.}\ }\textbf {\bibinfo {volume} {113}},\ \bibinfo {pages} {120402} (\bibinfo {year} {2014})}\BibitemShut {NoStop}%
\bibitem [{\citenamefont {Gorshkov}\ \emph {et~al.}(2010)\citenamefont {Gorshkov}, \citenamefont {Hermele}, \citenamefont {Gurarie}, \citenamefont {Xu}, \citenamefont {Julienne}, \citenamefont {Ye}, \citenamefont {Zoller}, \citenamefont {Demler}, \citenamefont {Lukin},\ and\ \citenamefont {Rey}}]{gorshkov2010}%
  \BibitemOpen
  \bibfield  {author} {\bibinfo {author} {\bibfnamefont {A.~V.}\ \bibnamefont {Gorshkov}}, \bibinfo {author} {\bibfnamefont {M.}~\bibnamefont {Hermele}}, \bibinfo {author} {\bibfnamefont {V.}~\bibnamefont {Gurarie}}, \bibinfo {author} {\bibfnamefont {C.}~\bibnamefont {Xu}}, \bibinfo {author} {\bibfnamefont {P.~S.}\ \bibnamefont {Julienne}}, \bibinfo {author} {\bibfnamefont {J.}~\bibnamefont {Ye}}, \bibinfo {author} {\bibfnamefont {P.}~\bibnamefont {Zoller}}, \bibinfo {author} {\bibfnamefont {E.}~\bibnamefont {Demler}}, \bibinfo {author} {\bibfnamefont {M.~D.}\ \bibnamefont {Lukin}}, \ and\ \bibinfo {author} {\bibfnamefont {A.~M.}\ \bibnamefont {Rey}},\ }\href {\doibase 10.1038/nphys1535} {\bibfield  {journal} {\bibinfo  {journal} {Nature Physics}\ }\textbf {\bibinfo {volume} {6}},\ \bibinfo {pages} {289} (\bibinfo {year} {2010})}\BibitemShut {NoStop}%
\bibitem [{\citenamefont {Zhang}\ \emph {et~al.}(2014)\citenamefont {Zhang}, \citenamefont {Bishof}, \citenamefont {Bromley}, \citenamefont {Kraus}, \citenamefont {Safronova}, \citenamefont {Zoller}, \citenamefont {Rey},\ and\ \citenamefont {Ye}}]{Zhang2014}%
  \BibitemOpen
  \bibfield  {author} {\bibinfo {author} {\bibfnamefont {X.}~\bibnamefont {Zhang}}, \bibinfo {author} {\bibfnamefont {M.}~\bibnamefont {Bishof}}, \bibinfo {author} {\bibfnamefont {S.~L.}\ \bibnamefont {Bromley}}, \bibinfo {author} {\bibfnamefont {C.~V.}\ \bibnamefont {Kraus}}, \bibinfo {author} {\bibfnamefont {M.~S.}\ \bibnamefont {Safronova}}, \bibinfo {author} {\bibfnamefont {P.}~\bibnamefont {Zoller}}, \bibinfo {author} {\bibfnamefont {A.~M.}\ \bibnamefont {Rey}}, \ and\ \bibinfo {author} {\bibfnamefont {J.}~\bibnamefont {Ye}},\ }\href {\doibase 10.1126/science.1254978} {\bibfield  {journal} {\bibinfo  {journal} {Science}\ }\textbf {\bibinfo {volume} {345}},\ \bibinfo {pages} {1467} (\bibinfo {year} {2014})}\BibitemShut {NoStop}%
\bibitem [{\citenamefont {Mezzacapo}\ \emph {et~al.}(2015)\citenamefont {Mezzacapo}, \citenamefont {Rico}, \citenamefont {Sabín}, \citenamefont {Egusquiza}, \citenamefont {Lamata},\ and\ \citenamefont {Solano}}]{Mezzacapo2015}%
  \BibitemOpen
  \bibfield  {author} {\bibinfo {author} {\bibfnamefont {A.}~\bibnamefont {Mezzacapo}}, \bibinfo {author} {\bibfnamefont {E.}~\bibnamefont {Rico}}, \bibinfo {author} {\bibfnamefont {C.}~\bibnamefont {Sabín}}, \bibinfo {author} {\bibfnamefont {I.}~\bibnamefont {Egusquiza}}, \bibinfo {author} {\bibfnamefont {L.}~\bibnamefont {Lamata}}, \ and\ \bibinfo {author} {\bibfnamefont {E.}~\bibnamefont {Solano}},\ }\href {https://doi.org/10.1103/physrevlett.115.240502} {\bibfield  {journal} {\bibinfo  {journal} {Physical Review Letters}\ }\textbf {\bibinfo {volume} {115}},\ \bibinfo {pages} {240502} (\bibinfo {year} {2015})}\BibitemShut {NoStop}%
\bibitem [{\citenamefont {González-Cuadra}\ \emph {et~al.}(2022)\citenamefont {González-Cuadra}, \citenamefont {Zache}, \citenamefont {Carrasco}, \citenamefont {Kraus},\ and\ \citenamefont {Zoller}}]{GonzalezCuadra2022}%
  \BibitemOpen
  \bibfield  {author} {\bibinfo {author} {\bibfnamefont {D.}~\bibnamefont {González-Cuadra}}, \bibinfo {author} {\bibfnamefont {T.~V.}\ \bibnamefont {Zache}}, \bibinfo {author} {\bibfnamefont {J.}~\bibnamefont {Carrasco}}, \bibinfo {author} {\bibfnamefont {B.}~\bibnamefont {Kraus}}, \ and\ \bibinfo {author} {\bibfnamefont {P.}~\bibnamefont {Zoller}},\ }\href {https://doi.org/10.1103/physrevlett.129.160501} {\bibfield  {journal} {\bibinfo  {journal} {Physical Review Letters}\ }\textbf {\bibinfo {volume} {129}},\ \bibinfo {pages} {160501} (\bibinfo {year} {2022})}\BibitemShut {NoStop}%
\bibitem [{\citenamefont {Gaz}\ \emph {et~al.}(2025)\citenamefont {Gaz}, \citenamefont {Popov}, \citenamefont {Pardo}, \citenamefont {Lewenstein}, \citenamefont {Hauke},\ and\ \citenamefont {Zohar}}]{Gaz2025}%
  \BibitemOpen
  \bibfield  {author} {\bibinfo {author} {\bibfnamefont {E.}~\bibnamefont {Gaz}}, \bibinfo {author} {\bibfnamefont {P.~P.}\ \bibnamefont {Popov}}, \bibinfo {author} {\bibfnamefont {G.}~\bibnamefont {Pardo}}, \bibinfo {author} {\bibfnamefont {M.}~\bibnamefont {Lewenstein}}, \bibinfo {author} {\bibfnamefont {P.}~\bibnamefont {Hauke}}, \ and\ \bibinfo {author} {\bibfnamefont {E.}~\bibnamefont {Zohar}},\ }\href {https://arxiv.org/pdf/2501.17863.pdf} {\enquote {\bibinfo {title} {Quantum simulation of non-abelian lattice gauge theories: a variational approach to $\mathbb{D}_8$},}\ } (\bibinfo {year} {2025}),\ \Eprint {http://arxiv.org/abs/2501.17863} {arXiv:2501.17863} \BibitemShut {NoStop}%
\bibitem [{\citenamefont {Bettermann}\ \emph {et~al.}(2023)\citenamefont {Bettermann}, \citenamefont {Darkwah~Oppong}, \citenamefont {Pasqualetti}, \citenamefont {Riegger}, \citenamefont {Bloch},\ and\ \citenamefont {F\"olling}}]{Bettermann2023}%
  \BibitemOpen
  \bibfield  {author} {\bibinfo {author} {\bibfnamefont {O.}~\bibnamefont {Bettermann}}, \bibinfo {author} {\bibfnamefont {N.}~\bibnamefont {Darkwah~Oppong}}, \bibinfo {author} {\bibfnamefont {G.}~\bibnamefont {Pasqualetti}}, \bibinfo {author} {\bibfnamefont {L.}~\bibnamefont {Riegger}}, \bibinfo {author} {\bibfnamefont {I.}~\bibnamefont {Bloch}}, \ and\ \bibinfo {author} {\bibfnamefont {S.}~\bibnamefont {F\"olling}},\ }\href {\doibase 10.1103/PhysRevA.108.L041302} {\bibfield  {journal} {\bibinfo  {journal} {Phys. Rev. A}\ }\textbf {\bibinfo {volume} {108}},\ \bibinfo {pages} {L041302} (\bibinfo {year} {2023})}\BibitemShut {NoStop}%
\bibitem [{\citenamefont {Pagano}\ \emph {et~al.}(2015)\citenamefont {Pagano}, \citenamefont {Mancini}, \citenamefont {Cappellini}, \citenamefont {Livi}, \citenamefont {Sias}, \citenamefont {Catani}, \citenamefont {Inguscio},\ and\ \citenamefont {Fallani}}]{Pagano2015}%
  \BibitemOpen
  \bibfield  {author} {\bibinfo {author} {\bibfnamefont {G.}~\bibnamefont {Pagano}}, \bibinfo {author} {\bibfnamefont {M.}~\bibnamefont {Mancini}}, \bibinfo {author} {\bibfnamefont {G.}~\bibnamefont {Cappellini}}, \bibinfo {author} {\bibfnamefont {L.}~\bibnamefont {Livi}}, \bibinfo {author} {\bibfnamefont {C.}~\bibnamefont {Sias}}, \bibinfo {author} {\bibfnamefont {J.}~\bibnamefont {Catani}}, \bibinfo {author} {\bibfnamefont {M.}~\bibnamefont {Inguscio}}, \ and\ \bibinfo {author} {\bibfnamefont {L.}~\bibnamefont {Fallani}},\ }\href {\doibase 10.1103/PhysRevLett.115.265301} {\bibfield  {journal} {\bibinfo  {journal} {Phys. Rev. Lett.}\ }\textbf {\bibinfo {volume} {115}},\ \bibinfo {pages} {265301} (\bibinfo {year} {2015})}\BibitemShut {NoStop}%
\bibitem [{\citenamefont {H\"ofer}\ \emph {et~al.}(2015)\citenamefont {H\"ofer}, \citenamefont {Riegger}, \citenamefont {Scazza}, \citenamefont {Hofrichter}, \citenamefont {Fernandes}, \citenamefont {Parish}, \citenamefont {Levinsen}, \citenamefont {Bloch},\ and\ \citenamefont {F\"olling}}]{Hofer2015}%
  \BibitemOpen
  \bibfield  {author} {\bibinfo {author} {\bibfnamefont {M.}~\bibnamefont {H\"ofer}}, \bibinfo {author} {\bibfnamefont {L.}~\bibnamefont {Riegger}}, \bibinfo {author} {\bibfnamefont {F.}~\bibnamefont {Scazza}}, \bibinfo {author} {\bibfnamefont {C.}~\bibnamefont {Hofrichter}}, \bibinfo {author} {\bibfnamefont {D.~R.}\ \bibnamefont {Fernandes}}, \bibinfo {author} {\bibfnamefont {M.~M.}\ \bibnamefont {Parish}}, \bibinfo {author} {\bibfnamefont {J.}~\bibnamefont {Levinsen}}, \bibinfo {author} {\bibfnamefont {I.}~\bibnamefont {Bloch}}, \ and\ \bibinfo {author} {\bibfnamefont {S.}~\bibnamefont {F\"olling}},\ }\href {\doibase 10.1103/PhysRevLett.115.265302} {\bibfield  {journal} {\bibinfo  {journal} {Phys. Rev. Lett.}\ }\textbf {\bibinfo {volume} {115}},\ \bibinfo {pages} {265302} (\bibinfo {year} {2015})}\BibitemShut {NoStop}%
\bibitem [{\citenamefont {Halimeh}\ \emph {et~al.}(2022{\natexlab{b}})\citenamefont {Halimeh}, \citenamefont {Homeier}, \citenamefont {Schweizer}, \citenamefont {Aidelsburger}, \citenamefont {Hauke},\ and\ \citenamefont {Grusdt}}]{Halimeh2022stabilizing}%
  \BibitemOpen
  \bibfield  {author} {\bibinfo {author} {\bibfnamefont {J.~C.}\ \bibnamefont {Halimeh}}, \bibinfo {author} {\bibfnamefont {L.}~\bibnamefont {Homeier}}, \bibinfo {author} {\bibfnamefont {C.}~\bibnamefont {Schweizer}}, \bibinfo {author} {\bibfnamefont {M.}~\bibnamefont {Aidelsburger}}, \bibinfo {author} {\bibfnamefont {P.}~\bibnamefont {Hauke}}, \ and\ \bibinfo {author} {\bibfnamefont {F.}~\bibnamefont {Grusdt}},\ }\href {\doibase 10.1103/PhysRevResearch.4.033120} {\bibfield  {journal} {\bibinfo  {journal} {Phys. Rev. Res.}\ }\textbf {\bibinfo {volume} {4}},\ \bibinfo {pages} {033120} (\bibinfo {year} {2022}{\natexlab{b}})}\BibitemShut {NoStop}%
\bibitem [{\citenamefont {Homeier}\ \emph {et~al.}(2023)\citenamefont {Homeier}, \citenamefont {Bohrdt}, \citenamefont {Linsel}, \citenamefont {Demler}, \citenamefont {Halimeh},\ and\ \citenamefont {Grusdt}}]{homeier2023Z2}%
  \BibitemOpen
  \bibfield  {author} {\bibinfo {author} {\bibfnamefont {L.}~\bibnamefont {Homeier}}, \bibinfo {author} {\bibfnamefont {A.}~\bibnamefont {Bohrdt}}, \bibinfo {author} {\bibfnamefont {S.}~\bibnamefont {Linsel}}, \bibinfo {author} {\bibfnamefont {E.}~\bibnamefont {Demler}}, \bibinfo {author} {\bibfnamefont {J.~C.}\ \bibnamefont {Halimeh}}, \ and\ \bibinfo {author} {\bibfnamefont {F.}~\bibnamefont {Grusdt}},\ }\href {\doibase 10.1038/s42005-023-01237-6} {\bibfield  {journal} {\bibinfo  {journal} {Communications Physics}\ }\textbf {\bibinfo {volume} {6}},\ \bibinfo {pages} {127} (\bibinfo {year} {2023})}\BibitemShut {NoStop}%
\bibitem [{\citenamefont {Trotzky}\ \emph {et~al.}(2008)\citenamefont {Trotzky}, \citenamefont {Cheinet}, \citenamefont {Fölling}, \citenamefont {Feld}, \citenamefont {Schnorrberger}, \citenamefont {Rey}, \citenamefont {Polkovnikov}, \citenamefont {Demler}, \citenamefont {Lukin},\ and\ \citenamefont {Bloch}}]{Trotzky2008}%
  \BibitemOpen
  \bibfield  {author} {\bibinfo {author} {\bibfnamefont {S.}~\bibnamefont {Trotzky}}, \bibinfo {author} {\bibfnamefont {P.}~\bibnamefont {Cheinet}}, \bibinfo {author} {\bibfnamefont {S.}~\bibnamefont {Fölling}}, \bibinfo {author} {\bibfnamefont {M.}~\bibnamefont {Feld}}, \bibinfo {author} {\bibfnamefont {U.}~\bibnamefont {Schnorrberger}}, \bibinfo {author} {\bibfnamefont {A.~M.}\ \bibnamefont {Rey}}, \bibinfo {author} {\bibfnamefont {A.}~\bibnamefont {Polkovnikov}}, \bibinfo {author} {\bibfnamefont {E.~A.}\ \bibnamefont {Demler}}, \bibinfo {author} {\bibfnamefont {M.~D.}\ \bibnamefont {Lukin}}, \ and\ \bibinfo {author} {\bibfnamefont {I.}~\bibnamefont {Bloch}},\ }\href {\doibase 10.1126/science.1150841} {\bibfield  {journal} {\bibinfo  {journal} {Science}\ }\textbf {\bibinfo {volume} {319}},\ \bibinfo {pages} {295} (\bibinfo {year} {2008})}\BibitemShut {NoStop}%
\bibitem [{\citenamefont {Duan}\ \emph {et~al.}(2003)\citenamefont {Duan}, \citenamefont {Demler},\ and\ \citenamefont {Lukin}}]{Duan2003}%
  \BibitemOpen
  \bibfield  {author} {\bibinfo {author} {\bibfnamefont {L.-M.}\ \bibnamefont {Duan}}, \bibinfo {author} {\bibfnamefont {E.}~\bibnamefont {Demler}}, \ and\ \bibinfo {author} {\bibfnamefont {M.~D.}\ \bibnamefont {Lukin}},\ }\href {\doibase 10.1103/PhysRevLett.91.090402} {\bibfield  {journal} {\bibinfo  {journal} {Phys. Rev. Lett.}\ }\textbf {\bibinfo {volume} {91}},\ \bibinfo {pages} {090402} (\bibinfo {year} {2003})}\BibitemShut {NoStop}%
\bibitem [{\citenamefont {Scazza}(2015)}]{ediss18159}%
  \BibitemOpen
  \bibfield  {author} {\bibinfo {author} {\bibfnamefont {F.}~\bibnamefont {Scazza}},\ }\href {http://nbn-resolving.de/urn:nbn:de:bvb:19-181593} {\enquote {\bibinfo {title} {Probing su(n)-symmetric orbital interactions with ytterbium fermi gases in optical lattices},}\ } (\bibinfo {year} {2015})\BibitemShut {NoStop}%
\bibitem [{\citenamefont {Santos}\ \emph {et~al.}(2000)\citenamefont {Santos}, \citenamefont {Shlyapnikov}, \citenamefont {Zoller},\ and\ \citenamefont {Lewenstein}}]{Santos2000}%
  \BibitemOpen
  \bibfield  {author} {\bibinfo {author} {\bibfnamefont {L.}~\bibnamefont {Santos}}, \bibinfo {author} {\bibfnamefont {G.~V.}\ \bibnamefont {Shlyapnikov}}, \bibinfo {author} {\bibfnamefont {P.}~\bibnamefont {Zoller}}, \ and\ \bibinfo {author} {\bibfnamefont {M.}~\bibnamefont {Lewenstein}},\ }\href {\doibase 10.1103/PhysRevLett.85.1791} {\bibfield  {journal} {\bibinfo  {journal} {Phys. Rev. Lett.}\ }\textbf {\bibinfo {volume} {85}},\ \bibinfo {pages} {1791} (\bibinfo {year} {2000})}\BibitemShut {NoStop}%
\bibitem [{\citenamefont {Nishad}\ \emph {et~al.}(2023)\citenamefont {Nishad}, \citenamefont {Keselman}, \citenamefont {Lahaye}, \citenamefont {Browaeys},\ and\ \citenamefont {Tsesses}}]{Nishad2023}%
  \BibitemOpen
  \bibfield  {author} {\bibinfo {author} {\bibfnamefont {N.}~\bibnamefont {Nishad}}, \bibinfo {author} {\bibfnamefont {A.}~\bibnamefont {Keselman}}, \bibinfo {author} {\bibfnamefont {T.}~\bibnamefont {Lahaye}}, \bibinfo {author} {\bibfnamefont {A.}~\bibnamefont {Browaeys}}, \ and\ \bibinfo {author} {\bibfnamefont {S.}~\bibnamefont {Tsesses}},\ }\href {\doibase 10.1103/PhysRevA.108.053318} {\bibfield  {journal} {\bibinfo  {journal} {Phys. Rev. A}\ }\textbf {\bibinfo {volume} {108}},\ \bibinfo {pages} {053318} (\bibinfo {year} {2023})}\BibitemShut {NoStop}%
\bibitem [{\citenamefont {Zhao}\ \emph {et~al.}(2023)\citenamefont {Zhao}, \citenamefont {Lee}, \citenamefont {Aliyu},\ and\ \citenamefont {Loh}}]{Zhao2023}%
  \BibitemOpen
  \bibfield  {author} {\bibinfo {author} {\bibfnamefont {L.}~\bibnamefont {Zhao}}, \bibinfo {author} {\bibfnamefont {M.~D.~K.}\ \bibnamefont {Lee}}, \bibinfo {author} {\bibfnamefont {M.~M.}\ \bibnamefont {Aliyu}}, \ and\ \bibinfo {author} {\bibfnamefont {H.}~\bibnamefont {Loh}},\ }\href {\doibase 10.1038/s41467-023-42899-8} {\bibfield  {journal} {\bibinfo  {journal} {Nature Communications}\ }\textbf {\bibinfo {volume} {14}},\ \bibinfo {pages} {7128} (\bibinfo {year} {2023})}\BibitemShut {NoStop}%
\bibitem [{\citenamefont {Weckesser}\ \emph {et~al.}(2024)\citenamefont {Weckesser}, \citenamefont {Srakaew}, \citenamefont {Blatz}, \citenamefont {Wei}, \citenamefont {Adler}, \citenamefont {Agrawal}, \citenamefont {Bohrdt}, \citenamefont {Bloch},\ and\ \citenamefont {Zeiher}}]{Weckesser2024}%
  \BibitemOpen
  \bibfield  {author} {\bibinfo {author} {\bibfnamefont {P.}~\bibnamefont {Weckesser}}, \bibinfo {author} {\bibfnamefont {K.}~\bibnamefont {Srakaew}}, \bibinfo {author} {\bibfnamefont {T.}~\bibnamefont {Blatz}}, \bibinfo {author} {\bibfnamefont {D.}~\bibnamefont {Wei}}, \bibinfo {author} {\bibfnamefont {D.}~\bibnamefont {Adler}}, \bibinfo {author} {\bibfnamefont {S.}~\bibnamefont {Agrawal}}, \bibinfo {author} {\bibfnamefont {A.}~\bibnamefont {Bohrdt}}, \bibinfo {author} {\bibfnamefont {I.}~\bibnamefont {Bloch}}, \ and\ \bibinfo {author} {\bibfnamefont {J.}~\bibnamefont {Zeiher}},\ }\href {https://arxiv.org/abs/2405.20128} {\  (\bibinfo {year} {2024})}\BibitemShut {NoStop}%
\bibitem [{\citenamefont {Young}\ \emph {et~al.}(2022)\citenamefont {Young}, \citenamefont {Eckner}, \citenamefont {Schine}, \citenamefont {Childs},\ and\ \citenamefont {Kaufman}}]{Young2022}%
  \BibitemOpen
  \bibfield  {author} {\bibinfo {author} {\bibfnamefont {A.~W.}\ \bibnamefont {Young}}, \bibinfo {author} {\bibfnamefont {W.~J.}\ \bibnamefont {Eckner}}, \bibinfo {author} {\bibfnamefont {N.}~\bibnamefont {Schine}}, \bibinfo {author} {\bibfnamefont {A.~M.}\ \bibnamefont {Childs}}, \ and\ \bibinfo {author} {\bibfnamefont {A.~M.}\ \bibnamefont {Kaufman}},\ }\href {\doibase 10.1126/science.abo0608} {\bibfield  {journal} {\bibinfo  {journal} {Science}\ }\textbf {\bibinfo {volume} {377}},\ \bibinfo {pages} {885} (\bibinfo {year} {2022})}\BibitemShut {NoStop}%
\bibitem [{\citenamefont {Stuart}\ and\ \citenamefont {Kuhn}(2018)}]{Stuart_2018}%
  \BibitemOpen
  \bibfield  {author} {\bibinfo {author} {\bibfnamefont {D.}~\bibnamefont {Stuart}}\ and\ \bibinfo {author} {\bibfnamefont {A.}~\bibnamefont {Kuhn}},\ }\href {\doibase 10.1088/1367-2630/aaa634} {\bibfield  {journal} {\bibinfo  {journal} {New Journal of Physics}\ }\textbf {\bibinfo {volume} {20}},\ \bibinfo {pages} {023013} (\bibinfo {year} {2018})}\BibitemShut {NoStop}%
\bibitem [{\citenamefont {Keating}\ \emph {et~al.}(2016)\citenamefont {Keating}, \citenamefont {Baldwin}, \citenamefont {Jau}, \citenamefont {Lee}, \citenamefont {Biedermann},\ and\ \citenamefont {Deutsch}}]{Keating2016}%
  \BibitemOpen
  \bibfield  {author} {\bibinfo {author} {\bibfnamefont {T.}~\bibnamefont {Keating}}, \bibinfo {author} {\bibfnamefont {C.~H.}\ \bibnamefont {Baldwin}}, \bibinfo {author} {\bibfnamefont {Y.-Y.}\ \bibnamefont {Jau}}, \bibinfo {author} {\bibfnamefont {J.}~\bibnamefont {Lee}}, \bibinfo {author} {\bibfnamefont {G.~W.}\ \bibnamefont {Biedermann}}, \ and\ \bibinfo {author} {\bibfnamefont {I.~H.}\ \bibnamefont {Deutsch}},\ }\href {https://doi.org/10.1103/physrevlett.117.213601} {\bibfield  {journal} {\bibinfo  {journal} {Physical Review Letters}\ }\textbf {\bibinfo {volume} {117}},\ \bibinfo {pages} {213601} (\bibinfo {year} {2016})}\BibitemShut {NoStop}%
\bibitem [{\citenamefont {Browaeys}\ and\ \citenamefont {Lahaye}(2020)}]{Browaeys2020}%
  \BibitemOpen
  \bibfield  {author} {\bibinfo {author} {\bibfnamefont {A.}~\bibnamefont {Browaeys}}\ and\ \bibinfo {author} {\bibfnamefont {T.}~\bibnamefont {Lahaye}},\ }\href {\doibase 10.1038/s41567-019-0733-z} {\bibfield  {journal} {\bibinfo  {journal} {Nature Physics}\ }\textbf {\bibinfo {volume} {16}},\ \bibinfo {pages} {132} (\bibinfo {year} {2020})}\BibitemShut {NoStop}%
\bibitem [{\citenamefont {Peper}\ \emph {et~al.}(2025)\citenamefont {Peper}, \citenamefont {Li}, \citenamefont {Knapp}, \citenamefont {Bileska}, \citenamefont {Ma}, \citenamefont {Liu}, \citenamefont {Peng}, \citenamefont {Zhang}, \citenamefont {Horvath}, \citenamefont {Burgers},\ and\ \citenamefont {Thompson}}]{peper2024}%
  \BibitemOpen
  \bibfield  {author} {\bibinfo {author} {\bibfnamefont {M.}~\bibnamefont {Peper}}, \bibinfo {author} {\bibfnamefont {Y.}~\bibnamefont {Li}}, \bibinfo {author} {\bibfnamefont {D.~Y.}\ \bibnamefont {Knapp}}, \bibinfo {author} {\bibfnamefont {M.}~\bibnamefont {Bileska}}, \bibinfo {author} {\bibfnamefont {S.}~\bibnamefont {Ma}}, \bibinfo {author} {\bibfnamefont {G.}~\bibnamefont {Liu}}, \bibinfo {author} {\bibfnamefont {P.}~\bibnamefont {Peng}}, \bibinfo {author} {\bibfnamefont {B.}~\bibnamefont {Zhang}}, \bibinfo {author} {\bibfnamefont {S.~P.}\ \bibnamefont {Horvath}}, \bibinfo {author} {\bibfnamefont {A.~P.}\ \bibnamefont {Burgers}}, \ and\ \bibinfo {author} {\bibfnamefont {J.~D.}\ \bibnamefont {Thompson}},\ }\href {\doibase 10.1103/PhysRevX.15.011009} {\bibfield  {journal} {\bibinfo  {journal} {Phys. Rev. X}\ }\textbf {\bibinfo {volume} {15}},\ \bibinfo {pages} {011009} (\bibinfo {year} {2025})}\BibitemShut {NoStop}%
\bibitem [{\citenamefont {Robicheaux}\ \emph {et~al.}(2018)\citenamefont {Robicheaux}, \citenamefont {Booth},\ and\ \citenamefont {Saffman}}]{Robicheaux2018}%
  \BibitemOpen
  \bibfield  {author} {\bibinfo {author} {\bibfnamefont {F.}~\bibnamefont {Robicheaux}}, \bibinfo {author} {\bibfnamefont {D.~W.}\ \bibnamefont {Booth}}, \ and\ \bibinfo {author} {\bibfnamefont {M.}~\bibnamefont {Saffman}},\ }\href {\doibase 10.1103/PhysRevA.97.022508} {\bibfield  {journal} {\bibinfo  {journal} {Phys. Rev. A}\ }\textbf {\bibinfo {volume} {97}},\ \bibinfo {pages} {022508} (\bibinfo {year} {2018})}\BibitemShut {NoStop}%
\bibitem [{\citenamefont {Wu}\ \emph {et~al.}(2022)\citenamefont {Wu}, \citenamefont {Yang}, \citenamefont {Yang}, \citenamefont {M\o{}lmer}, \citenamefont {Pohl}, \citenamefont {Tey},\ and\ \citenamefont {You}}]{Wu2022}%
  \BibitemOpen
  \bibfield  {author} {\bibinfo {author} {\bibfnamefont {X.}~\bibnamefont {Wu}}, \bibinfo {author} {\bibfnamefont {F.}~\bibnamefont {Yang}}, \bibinfo {author} {\bibfnamefont {S.}~\bibnamefont {Yang}}, \bibinfo {author} {\bibfnamefont {K.}~\bibnamefont {M\o{}lmer}}, \bibinfo {author} {\bibfnamefont {T.}~\bibnamefont {Pohl}}, \bibinfo {author} {\bibfnamefont {M.~K.}\ \bibnamefont {Tey}}, \ and\ \bibinfo {author} {\bibfnamefont {L.}~\bibnamefont {You}},\ }\href {\doibase 10.1103/PhysRevResearch.4.L032046} {\bibfield  {journal} {\bibinfo  {journal} {Phys. Rev. Res.}\ }\textbf {\bibinfo {volume} {4}},\ \bibinfo {pages} {L032046} (\bibinfo {year} {2022})}\BibitemShut {NoStop}%
\bibitem [{\citenamefont {Tang}\ \emph {et~al.}(2024)\citenamefont {Tang}, \citenamefont {Li}, \citenamefont {You},\ and\ \citenamefont {Shao}}]{Tang_2024}%
  \BibitemOpen
  \bibfield  {author} {\bibinfo {author} {\bibfnamefont {H.-Y.}\ \bibnamefont {Tang}}, \bibinfo {author} {\bibfnamefont {X.-X.}\ \bibnamefont {Li}}, \bibinfo {author} {\bibfnamefont {J.-B.}\ \bibnamefont {You}}, \ and\ \bibinfo {author} {\bibfnamefont {X.-Q.}\ \bibnamefont {Shao}},\ }\href {\doibase 10.1063/5.0211177} {\bibfield  {journal} {\bibinfo  {journal} {APL Quantum}\ }\textbf {\bibinfo {volume} {1}} (\bibinfo {year} {2024}),\ 10.1063/5.0211177}\BibitemShut {NoStop}%
\bibitem [{\citenamefont {Mitra}\ \emph {et~al.}(2020)\citenamefont {Mitra}, \citenamefont {Martin}, \citenamefont {Biedermann}, \citenamefont {Marino}, \citenamefont {Poggi},\ and\ \citenamefont {Deutsch}}]{Mitra2020}%
  \BibitemOpen
  \bibfield  {author} {\bibinfo {author} {\bibfnamefont {A.}~\bibnamefont {Mitra}}, \bibinfo {author} {\bibfnamefont {M.~J.}\ \bibnamefont {Martin}}, \bibinfo {author} {\bibfnamefont {G.~W.}\ \bibnamefont {Biedermann}}, \bibinfo {author} {\bibfnamefont {A.~M.}\ \bibnamefont {Marino}}, \bibinfo {author} {\bibfnamefont {P.~M.}\ \bibnamefont {Poggi}}, \ and\ \bibinfo {author} {\bibfnamefont {I.~H.}\ \bibnamefont {Deutsch}},\ }\href {\doibase 10.1103/PhysRevA.101.030301} {\bibfield  {journal} {\bibinfo  {journal} {Phys. Rev. A}\ }\textbf {\bibinfo {volume} {101}},\ \bibinfo {pages} {030301} (\bibinfo {year} {2020})}\BibitemShut {NoStop}%
\bibitem [{\citenamefont {Jaksch}\ \emph {et~al.}(2000)\citenamefont {Jaksch}, \citenamefont {Cirac}, \citenamefont {Zoller}, \citenamefont {Rolston}, \citenamefont {C\^ot\'e},\ and\ \citenamefont {Lukin}}]{Jaksch2000}%
  \BibitemOpen
  \bibfield  {author} {\bibinfo {author} {\bibfnamefont {D.}~\bibnamefont {Jaksch}}, \bibinfo {author} {\bibfnamefont {J.~I.}\ \bibnamefont {Cirac}}, \bibinfo {author} {\bibfnamefont {P.}~\bibnamefont {Zoller}}, \bibinfo {author} {\bibfnamefont {S.~L.}\ \bibnamefont {Rolston}}, \bibinfo {author} {\bibfnamefont {R.}~\bibnamefont {C\^ot\'e}}, \ and\ \bibinfo {author} {\bibfnamefont {M.~D.}\ \bibnamefont {Lukin}},\ }\href {\doibase 10.1103/PhysRevLett.85.2208} {\bibfield  {journal} {\bibinfo  {journal} {Phys. Rev. Lett.}\ }\textbf {\bibinfo {volume} {85}},\ \bibinfo {pages} {2208} (\bibinfo {year} {2000})}\BibitemShut {NoStop}%
\bibitem [{\citenamefont {Conolly}\ \emph {et~al.}(1989)\citenamefont {Conolly}, \citenamefont {Nishimura},\ and\ \citenamefont {Macovski}}]{Conolly1989}%
  \BibitemOpen
  \bibfield  {author} {\bibinfo {author} {\bibfnamefont {S.}~\bibnamefont {Conolly}}, \bibinfo {author} {\bibfnamefont {D.}~\bibnamefont {Nishimura}}, \ and\ \bibinfo {author} {\bibfnamefont {A.}~\bibnamefont {Macovski}},\ }\href {\doibase https://doi.org/10.1016/0022-2364(89)90194-7} {\bibfield  {journal} {\bibinfo  {journal} {Journal of Magnetic Resonance (1969)}\ }\textbf {\bibinfo {volume} {83}},\ \bibinfo {pages} {324} (\bibinfo {year} {1989})}\BibitemShut {NoStop}%
\bibitem [{\citenamefont {Gregefalk}\ and\ \citenamefont {Sj\"oqvist}(2022)}]{Gregefalk2022}%
  \BibitemOpen
  \bibfield  {author} {\bibinfo {author} {\bibfnamefont {A.}~\bibnamefont {Gregefalk}}\ and\ \bibinfo {author} {\bibfnamefont {E.}~\bibnamefont {Sj\"oqvist}},\ }\href {\doibase 10.1103/PhysRevApplied.17.024012} {\bibfield  {journal} {\bibinfo  {journal} {Phys. Rev. Appl.}\ }\textbf {\bibinfo {volume} {17}},\ \bibinfo {pages} {024012} (\bibinfo {year} {2022})}\BibitemShut {NoStop}%
\bibitem [{\citenamefont {Goldman}\ and\ \citenamefont {Dalibard}(2014)}]{Goldman2014}%
  \BibitemOpen
  \bibfield  {author} {\bibinfo {author} {\bibfnamefont {N.}~\bibnamefont {Goldman}}\ and\ \bibinfo {author} {\bibfnamefont {J.}~\bibnamefont {Dalibard}},\ }\href {\doibase 10.1103/PhysRevX.4.031027} {\bibfield  {journal} {\bibinfo  {journal} {Phys. Rev. X}\ }\textbf {\bibinfo {volume} {4}},\ \bibinfo {pages} {031027} (\bibinfo {year} {2014})}\BibitemShut {NoStop}%
\bibitem [{\citenamefont {Chinni}\ \emph {et~al.}(2022)\citenamefont {Chinni}, \citenamefont {Mu\~noz Arias}, \citenamefont {Deutsch},\ and\ \citenamefont {Poggi}}]{Chinni2022}%
  \BibitemOpen
  \bibfield  {author} {\bibinfo {author} {\bibfnamefont {K.}~\bibnamefont {Chinni}}, \bibinfo {author} {\bibfnamefont {M.~H.}\ \bibnamefont {Mu\~noz Arias}}, \bibinfo {author} {\bibfnamefont {I.~H.}\ \bibnamefont {Deutsch}}, \ and\ \bibinfo {author} {\bibfnamefont {P.~M.}\ \bibnamefont {Poggi}},\ }\href {\doibase 10.1103/PRXQuantum.3.010351} {\bibfield  {journal} {\bibinfo  {journal} {PRX Quantum}\ }\textbf {\bibinfo {volume} {3}},\ \bibinfo {pages} {010351} (\bibinfo {year} {2022})}\BibitemShut {NoStop}%
\bibitem [{\citenamefont {Levine}\ \emph {et~al.}(2019)\citenamefont {Levine}, \citenamefont {Keesling}, \citenamefont {Semeghini}, \citenamefont {Omran}, \citenamefont {Wang}, \citenamefont {Ebadi}, \citenamefont {Bernien}, \citenamefont {Greiner}, \citenamefont {Vuleti\ifmmode~\acute{c}\else \'{c}\fi{}}, \citenamefont {Pichler},\ and\ \citenamefont {Lukin}}]{Levine2019}%
  \BibitemOpen
  \bibfield  {author} {\bibinfo {author} {\bibfnamefont {H.}~\bibnamefont {Levine}}, \bibinfo {author} {\bibfnamefont {A.}~\bibnamefont {Keesling}}, \bibinfo {author} {\bibfnamefont {G.}~\bibnamefont {Semeghini}}, \bibinfo {author} {\bibfnamefont {A.}~\bibnamefont {Omran}}, \bibinfo {author} {\bibfnamefont {T.~T.}\ \bibnamefont {Wang}}, \bibinfo {author} {\bibfnamefont {S.}~\bibnamefont {Ebadi}}, \bibinfo {author} {\bibfnamefont {H.}~\bibnamefont {Bernien}}, \bibinfo {author} {\bibfnamefont {M.}~\bibnamefont {Greiner}}, \bibinfo {author} {\bibfnamefont {V.}~\bibnamefont {Vuleti\ifmmode~\acute{c}\else \'{c}\fi{}}}, \bibinfo {author} {\bibfnamefont {H.}~\bibnamefont {Pichler}}, \ and\ \bibinfo {author} {\bibfnamefont {M.~D.}\ \bibnamefont {Lukin}},\ }\href {\doibase 10.1103/PhysRevLett.123.170503} {\bibfield  {journal} {\bibinfo  {journal} {Phys. Rev. Lett.}\ }\textbf {\bibinfo {volume} {123}},\ \bibinfo {pages} {170503} (\bibinfo {year} {2019})}\BibitemShut {NoStop}%
\bibitem [{\citenamefont {Zhang}\ \emph {et~al.}(2011)\citenamefont {Zhang}, \citenamefont {Robicheaux},\ and\ \citenamefont {Saffman}}]{Zhang2011}%
  \BibitemOpen
  \bibfield  {author} {\bibinfo {author} {\bibfnamefont {S.}~\bibnamefont {Zhang}}, \bibinfo {author} {\bibfnamefont {F.}~\bibnamefont {Robicheaux}}, \ and\ \bibinfo {author} {\bibfnamefont {M.}~\bibnamefont {Saffman}},\ }\href {\doibase 10.1103/PhysRevA.84.043408} {\bibfield  {journal} {\bibinfo  {journal} {Phys. Rev. A}\ }\textbf {\bibinfo {volume} {84}},\ \bibinfo {pages} {043408} (\bibinfo {year} {2011})}\BibitemShut {NoStop}%
\bibitem [{\citenamefont {Ahlheit}\ \emph {et~al.}(2025)\citenamefont {Ahlheit}, \citenamefont {Nill}, \citenamefont {Svirskiy}, \citenamefont {de~Haan}, \citenamefont {Schroers}, \citenamefont {Alt}, \citenamefont {Stiesdal}, \citenamefont {Lesanovsky},\ and\ \citenamefont {Hofferberth}}]{Ahlheit2025}%
  \BibitemOpen
  \bibfield  {author} {\bibinfo {author} {\bibfnamefont {L.}~\bibnamefont {Ahlheit}}, \bibinfo {author} {\bibfnamefont {C.}~\bibnamefont {Nill}}, \bibinfo {author} {\bibfnamefont {D.}~\bibnamefont {Svirskiy}}, \bibinfo {author} {\bibfnamefont {J.}~\bibnamefont {de~Haan}}, \bibinfo {author} {\bibfnamefont {S.}~\bibnamefont {Schroers}}, \bibinfo {author} {\bibfnamefont {W.}~\bibnamefont {Alt}}, \bibinfo {author} {\bibfnamefont {N.}~\bibnamefont {Stiesdal}}, \bibinfo {author} {\bibfnamefont {I.}~\bibnamefont {Lesanovsky}}, \ and\ \bibinfo {author} {\bibfnamefont {S.}~\bibnamefont {Hofferberth}},\ }\href {\doibase 10.1103/PhysRevA.111.013115} {\bibfield  {journal} {\bibinfo  {journal} {Phys. Rev. A}\ }\textbf {\bibinfo {volume} {111}},\ \bibinfo {pages} {013115} (\bibinfo {year} {2025})}\BibitemShut {NoStop}%
\bibitem [{\citenamefont {Darkwah~Oppong}\ \emph {et~al.}(2022)\citenamefont {Darkwah~Oppong}, \citenamefont {Pasqualetti}, \citenamefont {Bettermann}, \citenamefont {Zechmann}, \citenamefont {Knap}, \citenamefont {Bloch},\ and\ \citenamefont {F\"olling}}]{Darkwah_Oppong2022}%
  \BibitemOpen
  \bibfield  {author} {\bibinfo {author} {\bibfnamefont {N.}~\bibnamefont {Darkwah~Oppong}}, \bibinfo {author} {\bibfnamefont {G.}~\bibnamefont {Pasqualetti}}, \bibinfo {author} {\bibfnamefont {O.}~\bibnamefont {Bettermann}}, \bibinfo {author} {\bibfnamefont {P.}~\bibnamefont {Zechmann}}, \bibinfo {author} {\bibfnamefont {M.}~\bibnamefont {Knap}}, \bibinfo {author} {\bibfnamefont {I.}~\bibnamefont {Bloch}}, \ and\ \bibinfo {author} {\bibfnamefont {S.}~\bibnamefont {F\"olling}},\ }\href {\doibase 10.1103/PhysRevX.12.031026} {\bibfield  {journal} {\bibinfo  {journal} {Phys. Rev. X}\ }\textbf {\bibinfo {volume} {12}},\ \bibinfo {pages} {031026} (\bibinfo {year} {2022})}\BibitemShut {NoStop}%
\bibitem [{\citenamefont {Riegger}\ \emph {et~al.}(2018)\citenamefont {Riegger}, \citenamefont {Darkwah~Oppong}, \citenamefont {H\"ofer}, \citenamefont {Fernandes}, \citenamefont {Bloch},\ and\ \citenamefont {F\"olling}}]{Riegger2018}%
  \BibitemOpen
  \bibfield  {author} {\bibinfo {author} {\bibfnamefont {L.}~\bibnamefont {Riegger}}, \bibinfo {author} {\bibfnamefont {N.}~\bibnamefont {Darkwah~Oppong}}, \bibinfo {author} {\bibfnamefont {M.}~\bibnamefont {H\"ofer}}, \bibinfo {author} {\bibfnamefont {D.~R.}\ \bibnamefont {Fernandes}}, \bibinfo {author} {\bibfnamefont {I.}~\bibnamefont {Bloch}}, \ and\ \bibinfo {author} {\bibfnamefont {S.}~\bibnamefont {F\"olling}},\ }\href {\doibase 10.1103/PhysRevLett.120.143601} {\bibfield  {journal} {\bibinfo  {journal} {Phys. Rev. Lett.}\ }\textbf {\bibinfo {volume} {120}},\ \bibinfo {pages} {143601} (\bibinfo {year} {2018})}\BibitemShut {NoStop}%
\bibitem [{\citenamefont {Meth}\ \emph {et~al.}(2025)\citenamefont {Meth}, \citenamefont {Zhang}, \citenamefont {Haase}, \citenamefont {Edmunds}, \citenamefont {Postler}, \citenamefont {Jena}, \citenamefont {Steiner}, \citenamefont {Dellantonio}, \citenamefont {Blatt}, \citenamefont {Zoller}, \citenamefont {Monz}, \citenamefont {Schindler}, \citenamefont {Muschik},\ and\ \citenamefont {Ringbauer}}]{Meth2025}%
  \BibitemOpen
  \bibfield  {author} {\bibinfo {author} {\bibfnamefont {M.}~\bibnamefont {Meth}}, \bibinfo {author} {\bibfnamefont {J.}~\bibnamefont {Zhang}}, \bibinfo {author} {\bibfnamefont {J.~F.}\ \bibnamefont {Haase}}, \bibinfo {author} {\bibfnamefont {C.}~\bibnamefont {Edmunds}}, \bibinfo {author} {\bibfnamefont {L.}~\bibnamefont {Postler}}, \bibinfo {author} {\bibfnamefont {A.~J.}\ \bibnamefont {Jena}}, \bibinfo {author} {\bibfnamefont {A.}~\bibnamefont {Steiner}}, \bibinfo {author} {\bibfnamefont {L.}~\bibnamefont {Dellantonio}}, \bibinfo {author} {\bibfnamefont {R.}~\bibnamefont {Blatt}}, \bibinfo {author} {\bibfnamefont {P.}~\bibnamefont {Zoller}}, \bibinfo {author} {\bibfnamefont {T.}~\bibnamefont {Monz}}, \bibinfo {author} {\bibfnamefont {P.}~\bibnamefont {Schindler}}, \bibinfo {author} {\bibfnamefont {C.}~\bibnamefont {Muschik}}, \ and\ \bibinfo {author} {\bibfnamefont {M.}~\bibnamefont {Ringbauer}},\ }\href {https://doi.org/10.1038/s41567-025-02797-w} {\bibfield  {journal} {\bibinfo  {journal} {Nature
  Physics}\ } (\bibinfo {year} {2025})}\BibitemShut {NoStop}%
\bibitem [{\citenamefont {Banerjee}\ \emph {et~al.}(2018)\citenamefont {Banerjee}, \citenamefont {Jiang}, \citenamefont {Olesen}, \citenamefont {Orland},\ and\ \citenamefont {Wiese}}]{Banerjee2018}%
  \BibitemOpen
  \bibfield  {author} {\bibinfo {author} {\bibfnamefont {D.}~\bibnamefont {Banerjee}}, \bibinfo {author} {\bibfnamefont {F.-J.}\ \bibnamefont {Jiang}}, \bibinfo {author} {\bibfnamefont {T.~Z.}\ \bibnamefont {Olesen}}, \bibinfo {author} {\bibfnamefont {P.}~\bibnamefont {Orland}}, \ and\ \bibinfo {author} {\bibfnamefont {U.-J.}\ \bibnamefont {Wiese}},\ }\href {https://doi.org/10.1126/science.abl6277https://doi.org/10.1103/physrevb.97.205108} {\bibfield  {journal} {\bibinfo  {journal} {Physical Review B}\ }\textbf {\bibinfo {volume} {97}},\ \bibinfo {pages} {205108} (\bibinfo {year} {2018})}\BibitemShut {NoStop}%
\bibitem [{\citenamefont {{L. Homeier et al.}}(2025)}]{Homeier_inprep}%
  \BibitemOpen
  \bibfield  {author} {\bibinfo {author} {\bibnamefont {{L. Homeier et al.}}},\ }\href@noop {} {\enquote {\bibinfo {title} {in preparation},}\ } (\bibinfo {year} {2025})\BibitemShut {NoStop}%
\bibitem [{\citenamefont {Schuhmacher}\ \emph {et~al.}(2025)\citenamefont {Schuhmacher}, \citenamefont {Su}, \citenamefont {Osborne}, \citenamefont {Gandon}, \citenamefont {Halimeh},\ and\ \citenamefont {Tavernelli}}]{schuhmacher2025}%
  \BibitemOpen
  \bibfield  {author} {\bibinfo {author} {\bibfnamefont {J.}~\bibnamefont {Schuhmacher}}, \bibinfo {author} {\bibfnamefont {G.-X.}\ \bibnamefont {Su}}, \bibinfo {author} {\bibfnamefont {J.~J.}\ \bibnamefont {Osborne}}, \bibinfo {author} {\bibfnamefont {A.}~\bibnamefont {Gandon}}, \bibinfo {author} {\bibfnamefont {J.~C.}\ \bibnamefont {Halimeh}}, \ and\ \bibinfo {author} {\bibfnamefont {I.}~\bibnamefont {Tavernelli}},\ }\href {https://arxiv.org/abs/2505.20387} {\  (\bibinfo {year} {2025})}\BibitemShut {NoStop}%
\bibitem [{\citenamefont {Davoudi}\ \emph {et~al.}(2025)\citenamefont {Davoudi}, \citenamefont {Hsieh},\ and\ \citenamefont {Kadam}}]{davoudi2025}%
  \BibitemOpen
  \bibfield  {author} {\bibinfo {author} {\bibfnamefont {Z.}~\bibnamefont {Davoudi}}, \bibinfo {author} {\bibfnamefont {C.-C.}\ \bibnamefont {Hsieh}}, \ and\ \bibinfo {author} {\bibfnamefont {S.~V.}\ \bibnamefont {Kadam}},\ }\href {https://arxiv.org/abs/2505.20408} {\  (\bibinfo {year} {2025})}\BibitemShut {NoStop}%
\bibitem [{\citenamefont {Foss-Feig}\ \emph {et~al.}(2010)\citenamefont {Foss-Feig}, \citenamefont {Hermele},\ and\ \citenamefont {Rey}}]{Foss-Feig2010}%
  \BibitemOpen
  \bibfield  {author} {\bibinfo {author} {\bibfnamefont {M.}~\bibnamefont {Foss-Feig}}, \bibinfo {author} {\bibfnamefont {M.}~\bibnamefont {Hermele}}, \ and\ \bibinfo {author} {\bibfnamefont {A.~M.}\ \bibnamefont {Rey}},\ }\href {\doibase 10.1103/PhysRevA.81.051603} {\bibfield  {journal} {\bibinfo  {journal} {Phys. Rev. A}\ }\textbf {\bibinfo {volume} {81}},\ \bibinfo {pages} {051603} (\bibinfo {year} {2010})}\BibitemShut {NoStop}%
\bibitem [{\citenamefont {Sotnikov}\ \emph {et~al.}(2020)\citenamefont {Sotnikov}, \citenamefont {Darkwah~Oppong}, \citenamefont {Zambrano},\ and\ \citenamefont {Cichy}}]{Sotnikov2020}%
  \BibitemOpen
  \bibfield  {author} {\bibinfo {author} {\bibfnamefont {A.}~\bibnamefont {Sotnikov}}, \bibinfo {author} {\bibfnamefont {N.}~\bibnamefont {Darkwah~Oppong}}, \bibinfo {author} {\bibfnamefont {Y.}~\bibnamefont {Zambrano}}, \ and\ \bibinfo {author} {\bibfnamefont {A.}~\bibnamefont {Cichy}},\ }\href {\doibase 10.1103/PhysRevResearch.2.023188} {\bibfield  {journal} {\bibinfo  {journal} {Phys. Rev. Res.}\ }\textbf {\bibinfo {volume} {2}},\ \bibinfo {pages} {023188} (\bibinfo {year} {2020})}\BibitemShut {NoStop}%
\bibitem [{\citenamefont {García-Ripoll}\ \emph {et~al.}(2009)\citenamefont {García-Ripoll}, \citenamefont {Dürr}, \citenamefont {Syassen}, \citenamefont {Bauer}, \citenamefont {Lettner}, \citenamefont {Rempe},\ and\ \citenamefont {Cirac}}]{garcia-ripoll2009dissipation}%
  \BibitemOpen
  \bibfield  {author} {\bibinfo {author} {\bibfnamefont {J.~J.}\ \bibnamefont {García-Ripoll}}, \bibinfo {author} {\bibfnamefont {S.}~\bibnamefont {Dürr}}, \bibinfo {author} {\bibfnamefont {N.}~\bibnamefont {Syassen}}, \bibinfo {author} {\bibfnamefont {D.~M.}\ \bibnamefont {Bauer}}, \bibinfo {author} {\bibfnamefont {M.}~\bibnamefont {Lettner}}, \bibinfo {author} {\bibfnamefont {G.}~\bibnamefont {Rempe}}, \ and\ \bibinfo {author} {\bibfnamefont {J.~I.}\ \bibnamefont {Cirac}},\ }\href {\doibase 10.1088/1367-2630/11/1/013053} {\bibfield  {journal} {\bibinfo  {journal} {New Journal of Physics}\ }\textbf {\bibinfo {volume} {11}},\ \bibinfo {pages} {013053} (\bibinfo {year} {2009})}\BibitemShut {NoStop}%
\bibitem [{\citenamefont {Jaksch}\ \emph {et~al.}(1998)\citenamefont {Jaksch}, \citenamefont {Bruder}, \citenamefont {Cirac}, \citenamefont {Gardiner},\ and\ \citenamefont {Zoller}}]{Jaksch1998}%
  \BibitemOpen
  \bibfield  {author} {\bibinfo {author} {\bibfnamefont {D.}~\bibnamefont {Jaksch}}, \bibinfo {author} {\bibfnamefont {C.}~\bibnamefont {Bruder}}, \bibinfo {author} {\bibfnamefont {J.~I.}\ \bibnamefont {Cirac}}, \bibinfo {author} {\bibfnamefont {C.~W.}\ \bibnamefont {Gardiner}}, \ and\ \bibinfo {author} {\bibfnamefont {P.}~\bibnamefont {Zoller}},\ }\href {\doibase 10.1103/PhysRevLett.81.3108} {\bibfield  {journal} {\bibinfo  {journal} {Phys. Rev. Lett.}\ }\textbf {\bibinfo {volume} {81}},\ \bibinfo {pages} {3108} (\bibinfo {year} {1998})}\BibitemShut {NoStop}%
\end{thebibliography}

\end{document}